\documentclass[fleqn,usenatbib]{mnras}
\usepackage{newtxtext,newtxmath}
\usepackage[T1]{fontenc}
\DeclareRobustCommand{\VAN}[3]{#2}
\let\VANthebibliography\thebibliography
\def\thebibliography{\DeclareRobustCommand{\VAN}[3]{##3}\VANthebibliography}

\usepackage{graphicx}	
\usepackage{amsmath}	
\usepackage[flushleft]{threeparttable}
\newcommand\kms{km s$^{-1}$}
\newcommand\masyr{mas yr$^{-1}$}
\newcommand\teff{$T_{eff}$}
\newcommand\logg{$\log g$}

\newcommand\msun{M$_\odot$}

\DeclareMathOperator{\atantwo}{atan2}
\usepackage{listings}
\title[ONC dynamics]{A gravitational and dynamical framework of star formation: \\ The Orion Nebula}

\author[M. Kounkel et al.]{
Marina Kounkel,$^{1}$\thanks{E-mail: marina.kounkel@vanderbilt.edu}
Keivan G.\ Stassun,$^{1}$
Kevin Covey$^{2}$
and Lee Hartmann$^{3}$
\\
$^{1}$Department of Physics and Astronomy, Vanderbilt University, VU Station 1807, Nashville, TN 37235, USA\\
$^{2}$Department of Physics and Astronomy, Western Washington University, 516 High St, Bellingham, WA 98225, USA\\
$^{3}$Department of Astronomy, University of Michigan, 1085 S. University Ave., Ann Arbor, MI 48109
}

\date{Accepted XXX. Received YYY; in original form ZZZ}

\pubyear{2022}

\begin{document}
\label{firstpage}
\pagerange{\pageref{firstpage}--\pageref{lastpage}}
\maketitle

\begin{abstract}
The Orion Nebula Cluster (ONC) is the most massive region of active star formation within a kpc of the Sun. Using Gaia DR3 parallaxes and proper motions, we examine the bulk motions of stars radially and tangentially relative to the cluster center. We find an age gradient with distance to the stars in the ONC, from 385~pc for the oldest stars, to 395~pc for the younger stars, indicating that the star forming front is propagating into the cloud. We find an organized signature of rotation of the central cluster, but it is present only in stars younger than 2~Myr. We also observe a net infall of young stars into the center of the ONC's deep gravitational potential well. The infalling sources lie preferentially along the filament, on the other hand, outflowing sources are distributed spherically around the cluster, and they have larger velocity dispersion. We further propose a solution to a long-standing question of why the ONC shows a weak signature of expansion even though the cluster is likely bound: much of this expansion may be driven by unstable N-body interactions among stars, resulting in low-velocity ejections. Though analyzing signatures imprinted on stellar dynamics across different spatial scales, these observation shed new light on the signatures of formation and evolution of young clusters.
\end{abstract}


\section{Introduction}

The gravitational fields of star-forming molecular clouds can influence the dynamics of both stars and gas over significant spatial scales. In the solar neighborhood, the most massive of these star forming regions is the Orion Complex. It contains a number of clusters, the most massive of which is the Orion Nebula (ONC), alone containing well over 4000 stars within the radius $\sim$10 pc of each other, of which $\sim$2000 are concentrated within a radius of $\sim2$ pc within the Trapezium \citep[e.g.,][]{mcbride2019}.

Despite the masses and densities involved in assembling some of the massive star forming regions like Orion, so far there have been only a few significant signatures of gravitational pull of the young populations on their surroundings. \citet{kuhn2019} have analyzed a sample of 28 young clusters; they found that almost none of them appear to have rotational support, and most of them show a preference of expansion with a typical velocity of $\sim0.5$ \kms. Only two clusters, M17 and NGC 6231, have shown some signature of contraction, albeit at a low level of significance. Most importantly, when examining multiple subclusters that are all part of the same population strung together along a filament, they found only a little evidence of convergent motion within them, concluding that they are not involved in a hierarchical assembly that would collect smaller star clumps into a larger cluster.

Over the years, most of the efforts have been concentrated on the dynamical state of the ONC based on its proper motions. The focus has been primarily aimed almost exclusively on the central cluster within it, rather than the full ``head'' of the Orion A molecular cloud. Early work with motions of stars measured from the photographic plates by \citet{jones1988} have concluded that the cluster is most likely not in virial equillibrium, but, as the escape velocity they estimated is comparable or smaller than the velocity dispersion ($\sigma_v$) they measured, they considered the cluster to be only weakly bound. \citet{hillenbrand1998} were able fit the spatial distribution of the stars in the ONC well using the King model profile, and they have argued that while the entire cluster may not necessarily be relaxed, its core is expected to be. They found that the stellar mass within the cluster is only 40\% of what is expected for virial equillibrium, but that the remaining gas may account for the remainder, and that it may yet to form a sufficient number of stars for the cluster to remain bound at later stages in its life. Later, \citet{dzib2017} have used radio interferometry to improve astrometric measurements of stars, and they found no indication of either preferred radial or tangential motion of stars, which would be expected of a virialized population. Finally, since the release of Gaia DR2, several works, such as \citet{kounkel2018a}, \citet{kuhn2019} or \citet{getman2019} have re-evaluated the dynamics of the ONC, with overall conclusions being that the cluster does appear to be bound, although it may show a weak preference towards expansion, mainly along its outer edges.

In this paper we examine the dynamical state of Orion, the gravitational influence that it has on its vicinity, and the expansion of the stars within it. To clarify the terminology, we refer to the entire ``head'' as the ONC, and we would frequently use Trapezium to refer to the central cluster core. 
In Section~\ref{sec:data} we present the data used and the general methods of our analysis. 
Section~\ref{sec:results} presents the primary results of our study.
In Section~\ref{sec:discussion} we discuss the implications of our findings of an expansion signature in the ONC despite the cluster likely being bound.
We conclude with a summary in Section~\ref{sec:concl}.

\section{Data and Methods}\label{sec:data}
\subsection{Data selection}\label{sec:data_onc}

One of the challenges that the previous works have had in analyzing the dynamical state of the ONC is the sheer number of stars contained within it, as it makes it difficult to fully visualize the velocity vectors of stars. As such, although the cluster appears to be virialized on average, some of the substructure may be less apparent when everything is viewed as a whole. More than that, the ONC contains multiple generations of stars, some younger than 1 Myr, some with ages $>$4 Myr. While the distributions of these stars of different ages overlap, they do have their own unique features, and this may influence the kinematics as well \citep{beccari2017}. To mitigate both of these issues, we split up the sample of the stars in the ONC based on their age.

The ages of young stars can commonly be determined from placement of their photometry on the HR diagram. However, some confusion can remain regarding the ages of unresolved binary stars which have a higher combined luminosity in comparison to single stars; without taking this into account, the ages that would be inferred photometrically will be underestimated. Uncertainty in extinction also has a significant effect on the precise age determination. Although it is still possible to observe trends and gradients in age across given populations \citep{beccari2017}, such age bias can often be difficult to quantify.

Instead we use \logg\ as a proxy for age. As a young star evolves and its radius shrinks, its \logg\ increases until it reaches main sequence. Such measurement is a more direct method of determining an age of a star, as it is significantly less biased by multiplicity. To date only a few studies have produced reliable \logg\ values for young stars that have sufficient accuracy and precision to use them as an age proxy However, recently, \citet{olney2020} and \citet{sprague2022} have performed analysis of APOGEE spectra to use data-driven techniques to measure \teff\ and \logg\ that can be compared to the pre-main-sequence isochrones. Ages estimated from \logg\ are not affected by extinction (as \logg\ is derived from the shape of the spectral lines rather than from the luminosity; moreover, APOGEE spectra are in infrared, where the effect of the extinction is significantly smaller than in the optical regime). Similarly, \logg\ measurements appear to be largely unaffected by unresolved binaries. \citet{olney2020} have found that among pre-main sequence stars, there are no systematic difference in \logg\ distribution of spectroscopic binaries (both SB1 and SB2) relative to the total sample of young stars. \citet{Kounkel2021} have found that there is a slight bias in \logg\ distribution of the SB2s found among the more evolved main sequence stars; this bias is $\sim$0.1 dex in magnitude -- comparable to the magnitude of the typical uncertainties. However, even if this bias is present in younger, such as the sample in Orion, given that SB2s are relatively rare in comparison to all unresolved binaries, spectroscopic ages estimates should on average be more robust.

We do note that using \logg\ rather than luminosity does not compensate over some of the other fundamental factors of the systematic uncertainty in deriving stellar ages, including the uncertainty in the absolute calibration of the stellar parameters, as well as the ability of the isochrone models in accurately representing said parameters. This can often be a challenge, particularly among the cooler stars \citep[e.g.,][]{david2019}.

We use the spectroscopic sample of stars observed by APOGEE towards the ONC \citep{kounkel2018a}, and following a crossmatch with Gaia EDR3 \citep{gaia-collaboration2021}, we further restrict it via

\begin{itemize}
\item Spatial cuts in RA \& Decl. $82<\alpha<85^\circ$, $-6.5<\delta<-3.5^\circ$
\item Cut in parallax $2<\pi<3.5$ mas; which translates to a distance cut of 285--500 pc.
\item Proper motions within 6 \masyr\, of the mean of the cluster, with the mean motion in the local standard of rest \footnote{LSR velocity correction is from \citep{schonrich2010}.} $\mu_{\alpha,lsr}=0.8$ \masyr\ and $\mu_{\delta,lsr}=2.8$ \masyr. This translates to a window $\sim$11 \kms\, or $\sim$5-6 times the velocity dispersion of the clusters, excluding any fast dynamically ejected runaways such as those in \citet{mcbride2019}, or the bulk of the potential contamination. The APOGEE sample towards the ONC primarily contains bona-fide members, with only a small fraction of likely field stars, as such, this cut eliminates only 5\% of stars.
\item Spectroscopically determined \teff$<6500$ K and \logg$>3$ dex, restricting the sample to the stars from which spectroscopic ages can be reliably determined.
\end{itemize}
This produces a sample of 1612 stars, of which 893 are found towards the Trapezium. 

The \logg\ values derived via APOGEE Net may still have minor systematic differences relative to various isochrones, such as, e.g., PARSEC \citep{marigo2017}, particularly at the cooler end of \teff$<$3800 K. The isochrones suggest that \logg\ is expected to decrease with decreasing \teff\ for low mass young stars, whereas, in a given cluster of a given age, the measured \logg\ values from \citet{olney2020} may appear to slightly increase with decreasing \teff\, mirroring a relation that is seen among the main sequence stars. As such, applying theoretical isochrones directly results in a significant overestimation of the ages of cool stars in comparison to their hotter counterparts. Without precise empirical isochrones that could be used to characterize this dependence, we approximate a simple monotonic relationship between age and \logg\ without a dependence on \teff, as this offers a better consistency within a given cluster in comparison to the theoretical isochrones.

We roughly separate the sample into 5 bins, and translate these \logg\ cuts into approximate age ranges: \logg$<$3.6 dex ($\lesssim$1 Myr, 125 stars), 3.6 $<$\logg$<$3.8 dex (1$\sim$2 Myr, 265 stars), 3.8$<$\logg$<$4.0 dex (2$\sim$3 Myr, 480 stars), 4.0$<$\logg$<$4.2 dex (3$\sim$5 Myr, 445 stars), and 4.2$<$\logg\ dex ($\gtrsim$5 Myr, 297 stars). The typical precision in the \logg\ in the sample is $\sim$0.1 dex (10th and 90th percentile of the formal error distribution are 0.037 dex and 0.16 dex). We note that these ages are only approximate labels to provide the readers with some intuition regarding the star formation history of the ONC, but fundamentally, these remain to be cuts solely in \logg\ rather than age directly.

There are independent age estimates from photometery for some of the stars against which the spectroscopic ages can be evaluated, e.g., ages of the individual stars from \citet{kounkel2018a} derived using \teff\ and bolometric luminosity relative to the PARSEC isochrones \citep{marigo2017}, or the ages based on Gaia and 2MASS photometery interpolated using Sagitta neural net \citep{mcbride2021}. There are some systematic differences, both between photometric ages estimated form these two approaches, as well as relative to the spectroscopic age estimates used here. However, in both cases, there is a strong trend of older stars having systematically higher \logg. As such, while there may be some cross-contamination of the neighboring bins, there are clear differences between the selection of the oldest and the youngest age bins, regardless of the method used to select them. The overall results presented in this paper are robust against the age metric that is used.

\subsection{Analysis}

To examine the motion of stars in the ONC in the reference frame the center of the cluster, we derive a metric representing the relative radial orientation of motion, $\cos\theta$, where $\theta$ is defined as
\begin{equation}\label{eq:onc}
\begin{split}
\theta=\atantwo(\mu_{\alpha,lsr}-\mu_{\alpha,COM,lsr},\\
\mu_{\delta,lsr}-\mu_{\delta,COM,lsr})- \\
\atantwo(\alpha-\alpha_{COM},\delta-\delta_{COM})
\end{split}
\end{equation}

\noindent where $\atantwo$ is the 2-argument arctangent that is capable of returning values across the full range of a circle, $\alpha$ and $\delta$ are the right ascension and declination (in decimal degrees) of the star in question, $\mu_{\alpha,lsr}$ and $\mu_{\delta,lsr}$ are the star's proper motions in right ascension and declination with respect to the local standard of rest \citep{schonrich2010}, expressed in milli-arcseconds per year,  $\alpha_{COM}$ and $\delta_{COM}$ are the right ascension and declination of the ONC's center of mass ($\alpha_{COM} = 83.8^{\circ}$ and $\delta_{COM} = -5.4^{\circ}$), and $\mu_{\alpha,COM,lsr}$ and $\mu_{\delta,COM,lsr}$ are the mean proper motions of ONC members in right ascension and declination with respect to the local standard of rest ($\mu_{\alpha,COM,lsr} = 0.8 \mathrm{mas\ yr}^{-1}$ and $\mu_{\delta,COM,lsr} = 2.8~\mathrm{mas\ yr}^{-1}$). The center of mass position and the mean velocities are approximations derived based on the sample, and they are used to evaluate intracluster dynamics. The typical uncertainty in $\theta$ from the reported uncertainty in $\mu_\alpha$ and $\mu_\delta$ is $\sim3^\circ$, increasing to $\sim10^\circ$ if we consider a reasonable uncertainty in the reference position and velocity of the cluster. However, the results outlined in this work do not strongly depend on these precise averages, they are robust against the uncertainties, and they remain consistent when other reasonable estimates are used instead.

We then segregate the selected members of the ONC into three categories based on their computed $\cos\theta$ values: $\cos\theta>0.5$ are the sources that are preferentially moving away from the cluster center (566 stars), $\cos\theta<-0.5$ are those moving towards the cluster center (606 stars) and $-0.5<\cos\theta<0.5$ are those moving tangentially around the cluster/rotating around it (440 stars). The latter sources can be further subdivided through tangential orientation, with $\sin\theta<0$ as the sources preferentially moving clockwise (234 stars), and $\sin\theta>0$ as those moving counterclockwise (206 stars).

\section{Results}\label{sec:results}

\begin{figure*}
\includegraphics[width={0.48\textwidth}]{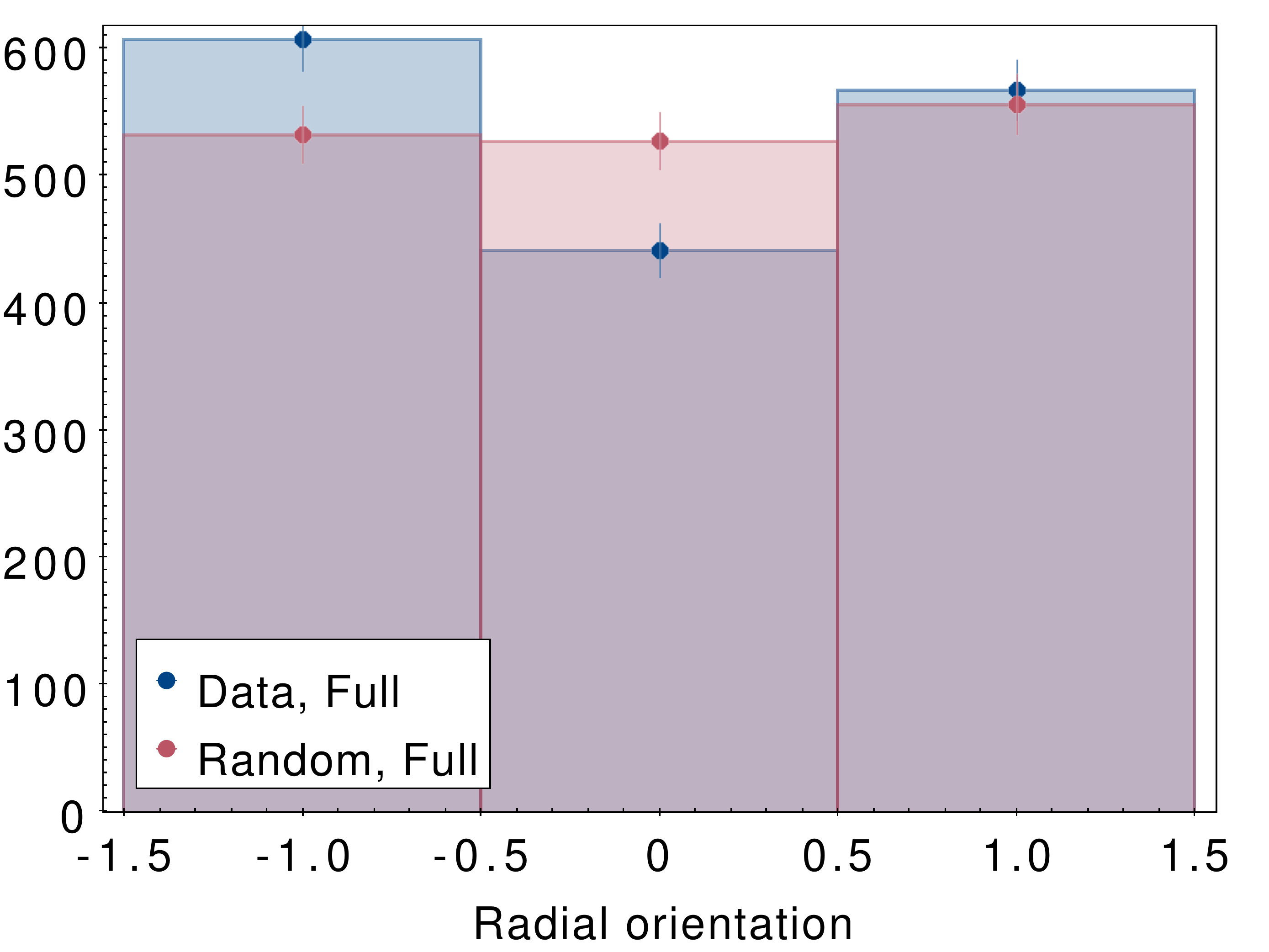}
\includegraphics[width={0.48\textwidth}]{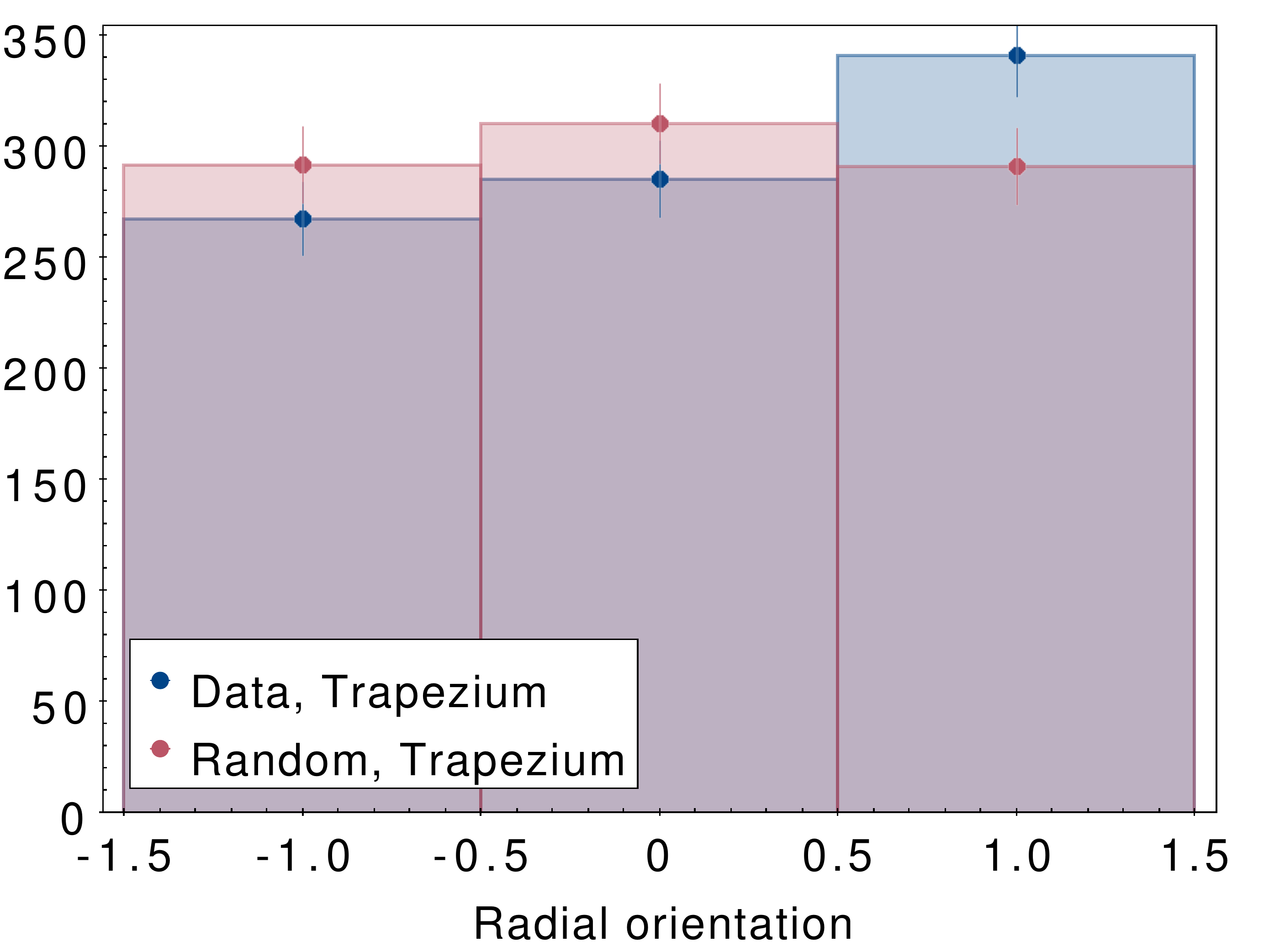}
\caption{Number of sources that are infalling into the ONC ($-1<\cos\theta<-0.5$), those that are preferentially moving away ($0.5<\cos\theta<1$), and those that are moving tangentially around the cluster ($-0.5<\cos\theta<0.5$). Note that although the bin size extends beyond $\pm$1, no objects are found beyond these limits, and the geometry of the cos function is such that these splits at $\pm$0.5 are expected to divide a random sample into three bins of comparable sizes. Left panel shows the data for the entire ONC, right panel just zooms in on the sources found in the inner 0.4$^\circ$ of the central cluster. Blue histograms show the orientation of motion of the real stars, red - stars with randomly generated velocities drawn from a Gaussian distribution consisten with the velocity dispersion of the cluster. The uncertainties are estimated as through Poisson approximation of sqrt(n).
\label{fig:histogram}}
\end{figure*}

To evaluate the distribution of velocities in the ONC, as a comparison, we generated a random sample of proper motions for all of the stars, drawing both $\mu_\alpha$ and $\mu_\delta$ from the $0.9\times1.1$ \masyr\ Gaussian velocity dispersion of the cluster (consistent with the underlying distribution of sources in the real data), preserving the positions of each star. We processed these synthetic velocities in a similar manner, and applied similar cuts. We find that in a purely randomly drawn velocities, there is expected to be an approximately equal number of infalling, outflowing, and rotating stars. In comparison, real data for the entire ONC shows that the fraction of sources that are rotating is suppressed, and, instead, there is an excess in the number of stars that are infalling (Figure \ref{fig:histogram}). Applying two-sided Kolmogorov-Smirnov (KS) test between the full distribution of all alignments between the real and the synthetic data shows that the alignment of stars relative to the cluster center relative to the random is different at $2.5\sigma$ level. Furthermore, zooming in only on the central cluster, within 0.4$^\circ$ of the cluster center, we find there to be an excess in the number of the stars expanding outwards from the center. KS test shows that the distribution of velocities is different from random at $3\sigma$ level. The trends shown in Figure \ref{fig:histogram} are not affected by the precise cut on the $\cos\theta$, e.g, changing the threshold from $\pm0.5$ to other thresholds does not substantively affect the significance of these trends. Similarly, uncertainty in $\theta$ does not skew the underlying distribution.

\begin{figure*}
\includegraphics[width={0.01\textwidth}]{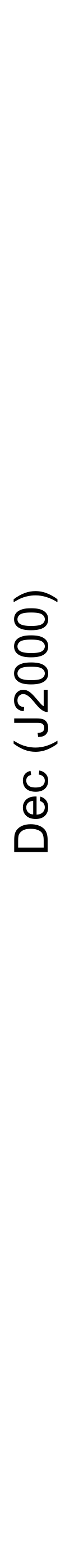}
\includegraphics[width={0.24\textwidth}]{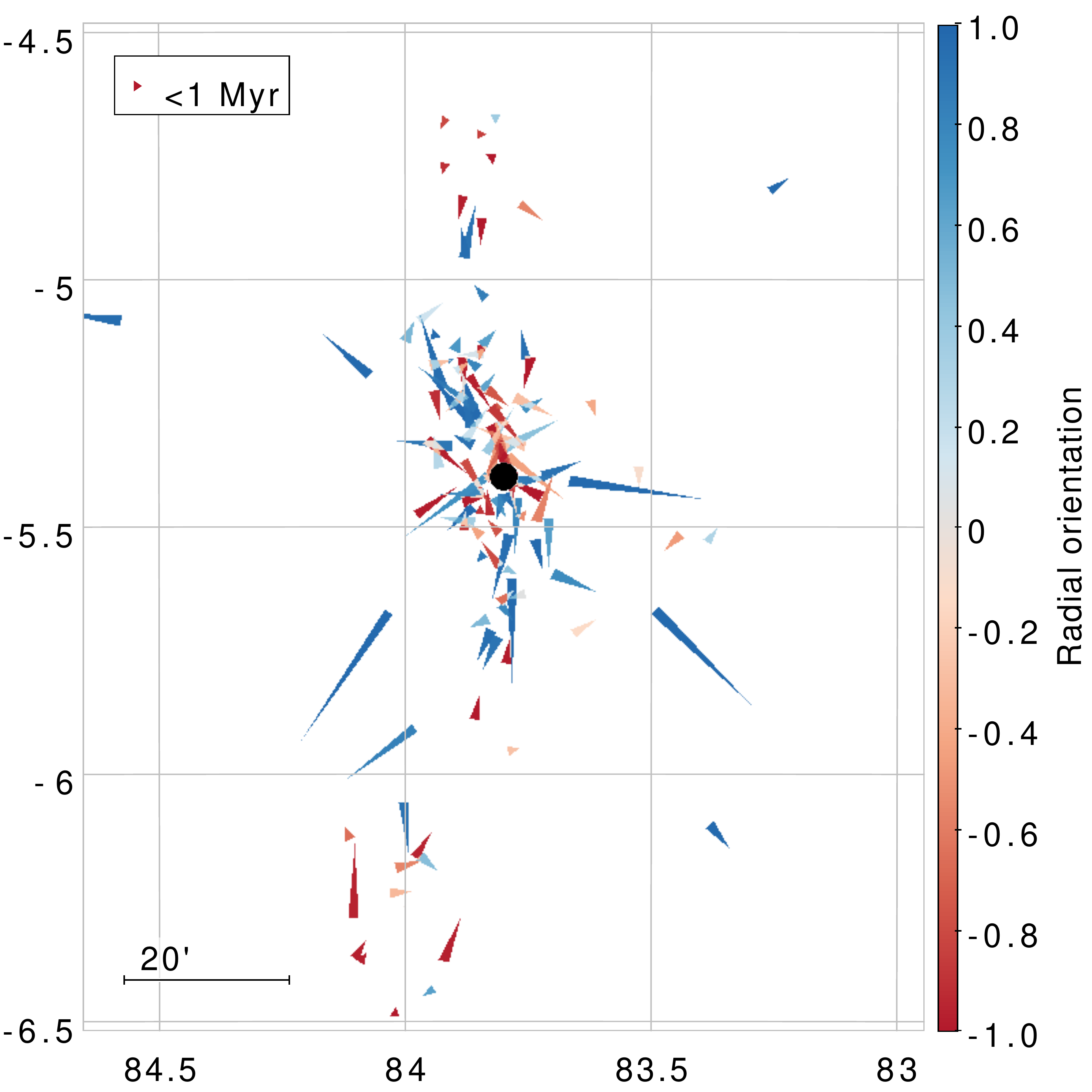}
\includegraphics[width={0.24\textwidth}]{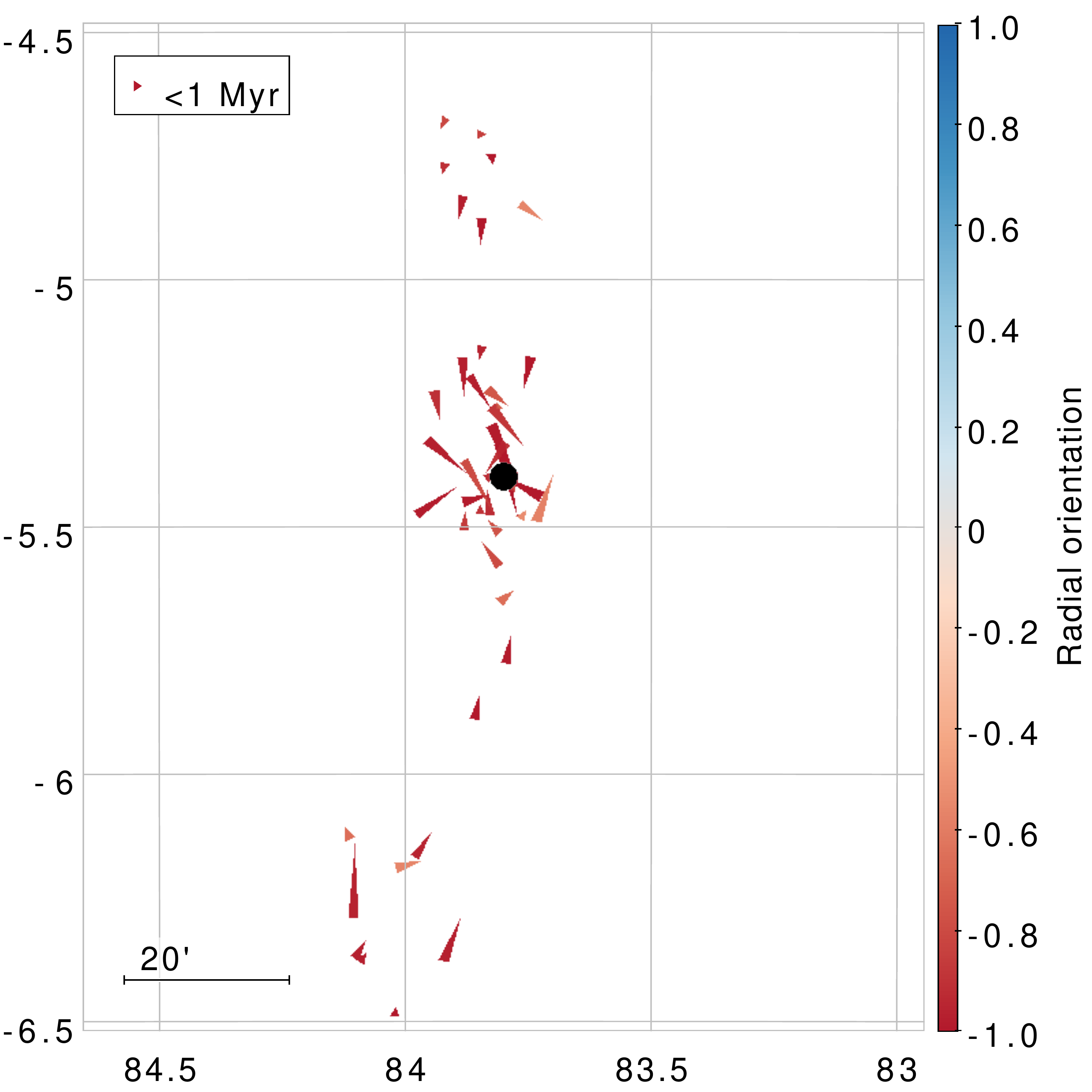}
\includegraphics[width={0.24\textwidth}]{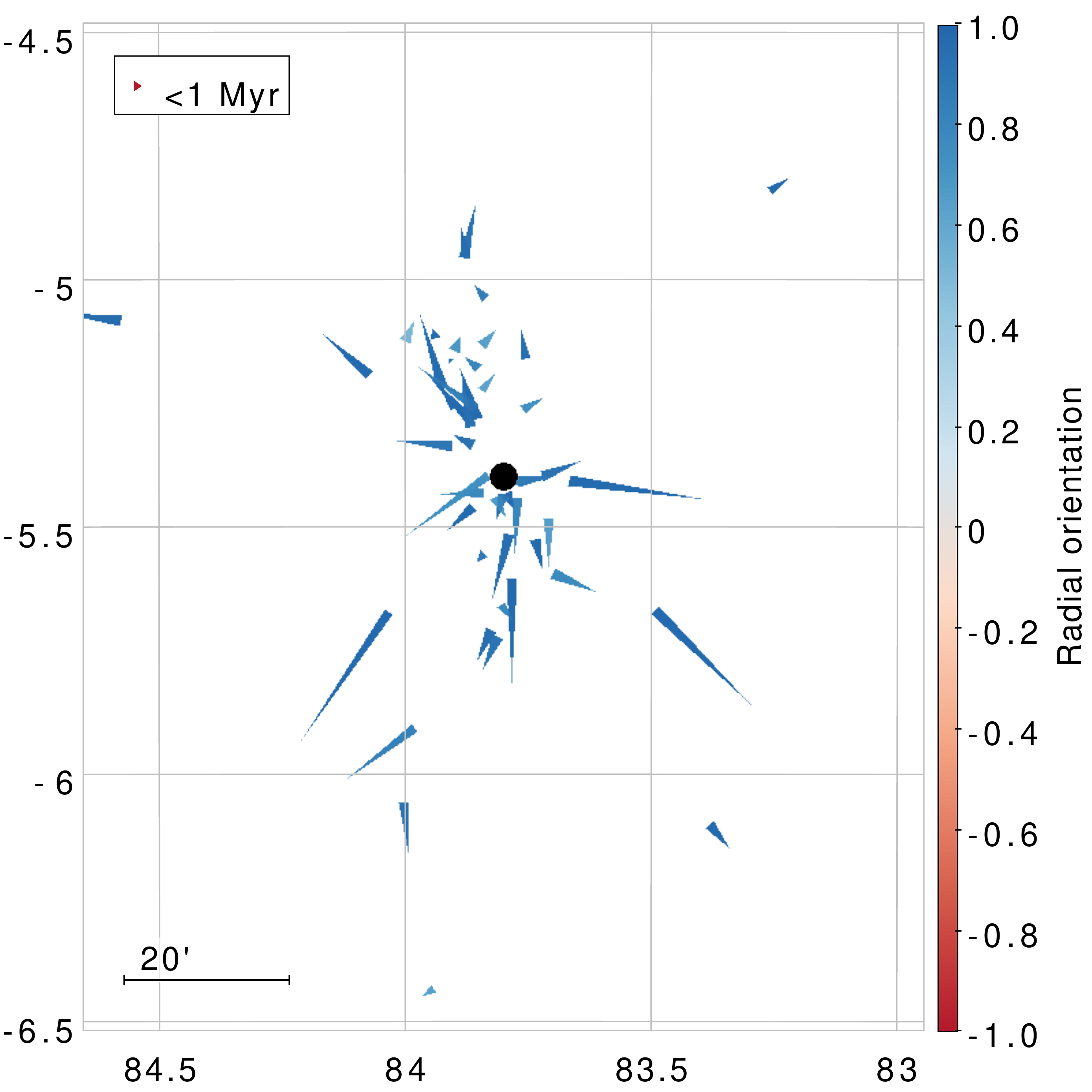}
\includegraphics[width={0.24\textwidth}]{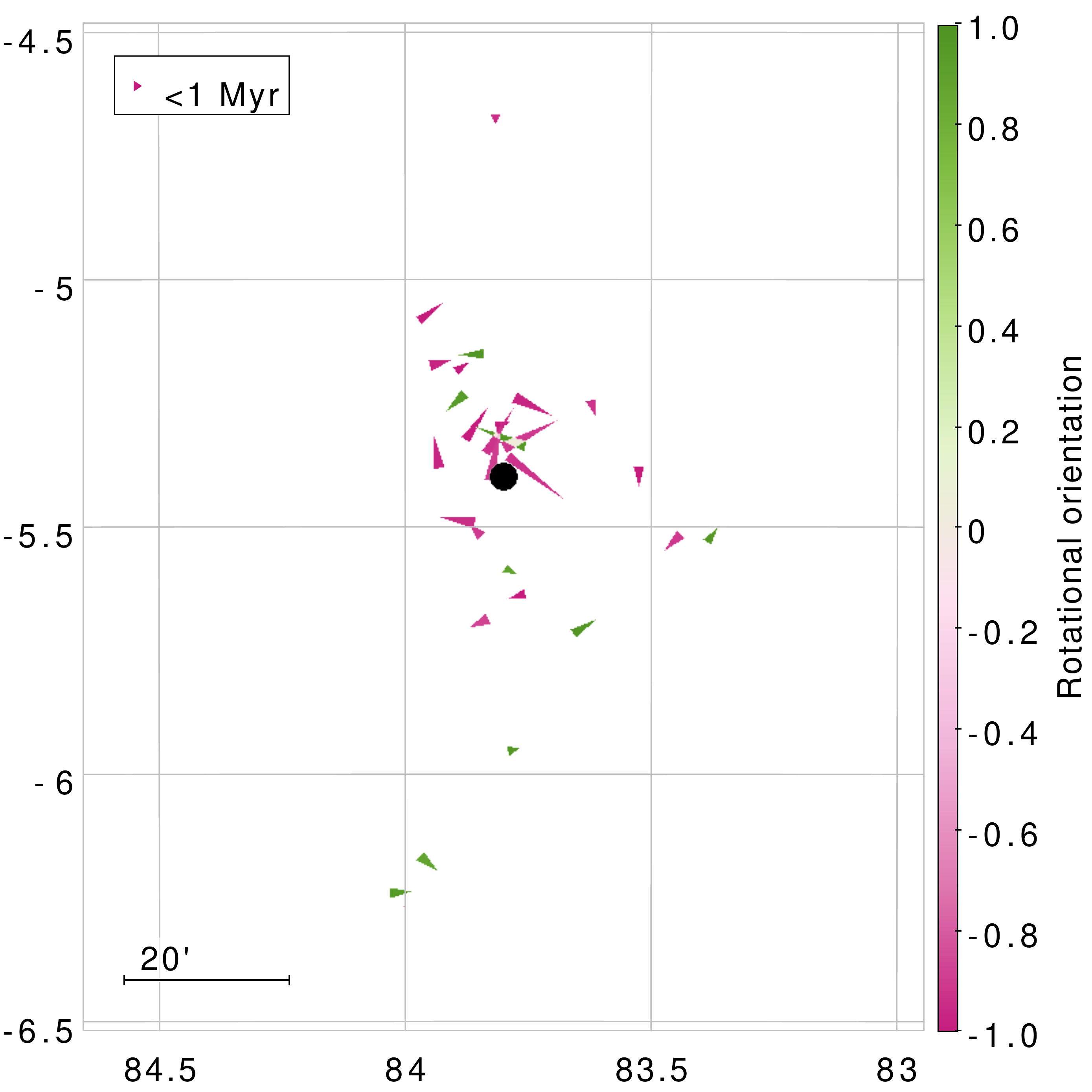}

\includegraphics[width={0.01\textwidth}]{dec.pdf}
\includegraphics[width={0.24\textwidth}]{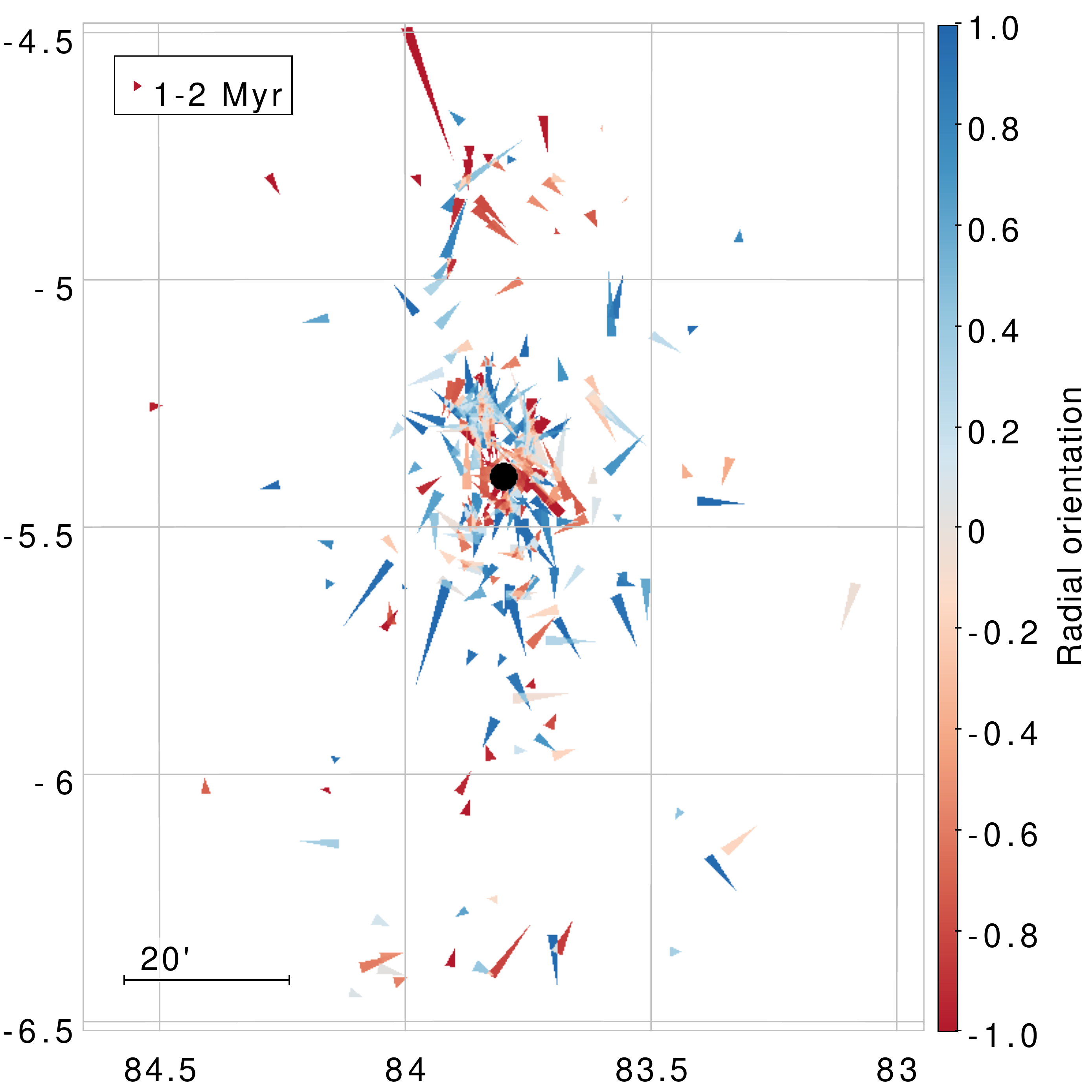}
\includegraphics[width={0.24\textwidth}]{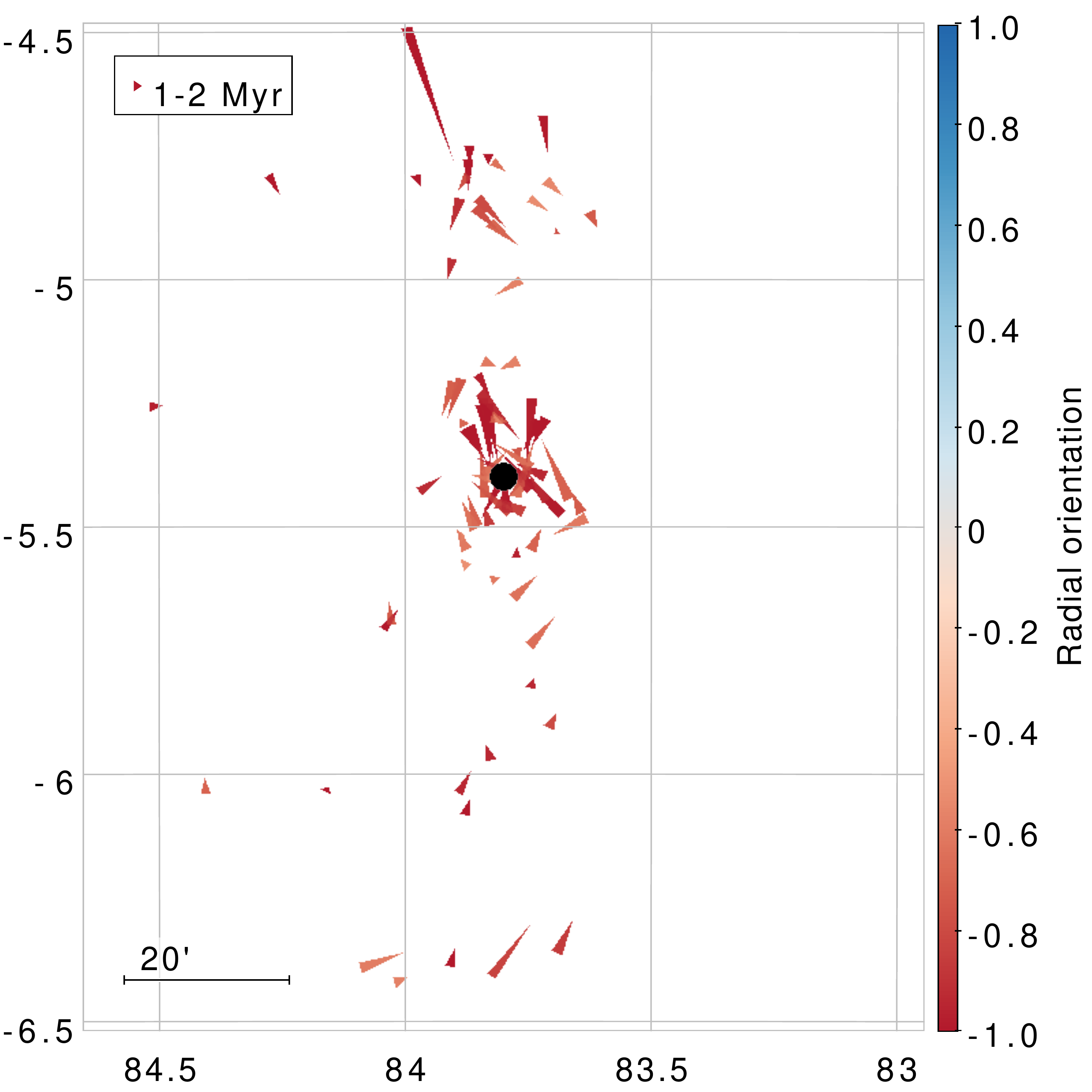}
\includegraphics[width={0.24\textwidth}]{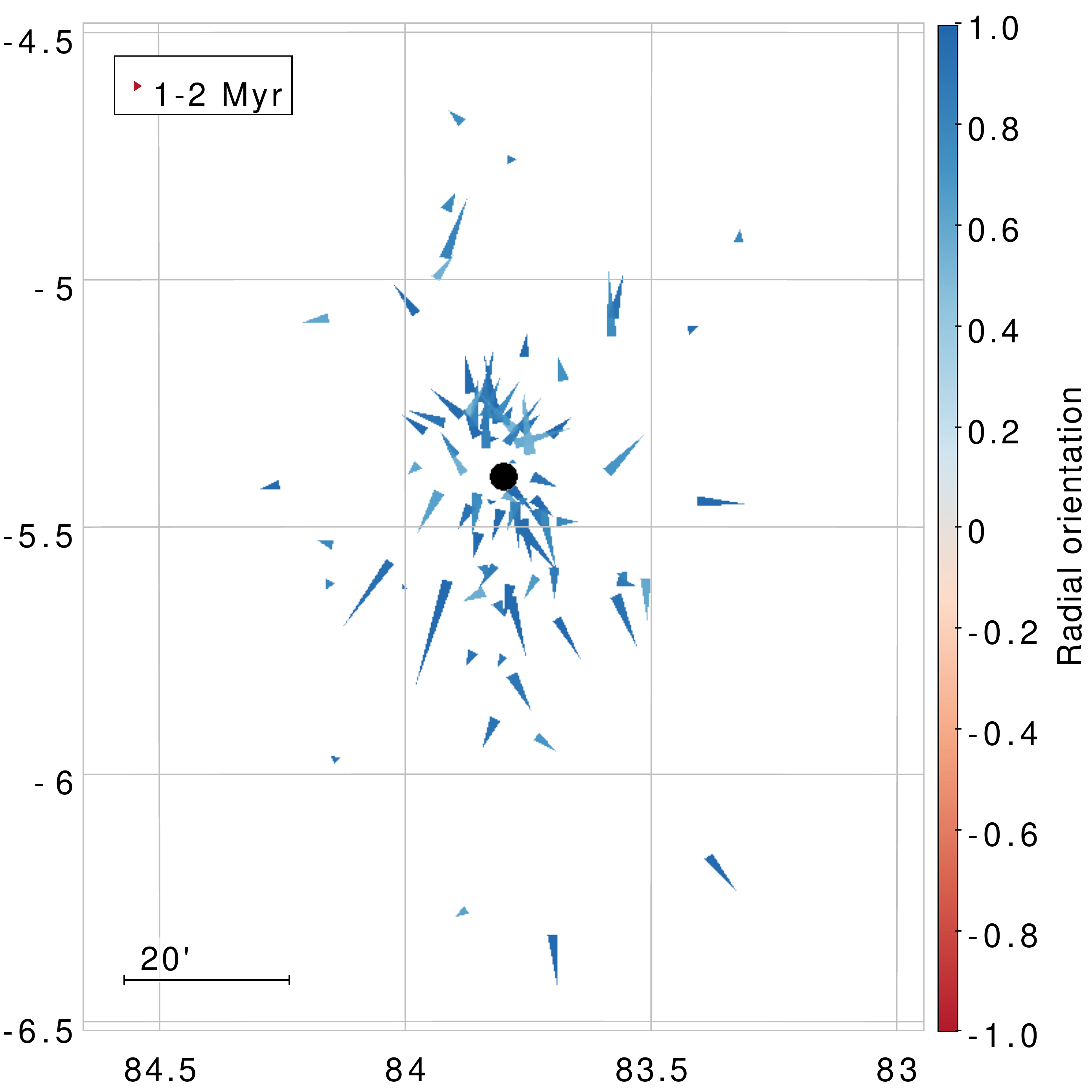}
\includegraphics[width={0.24\textwidth}]{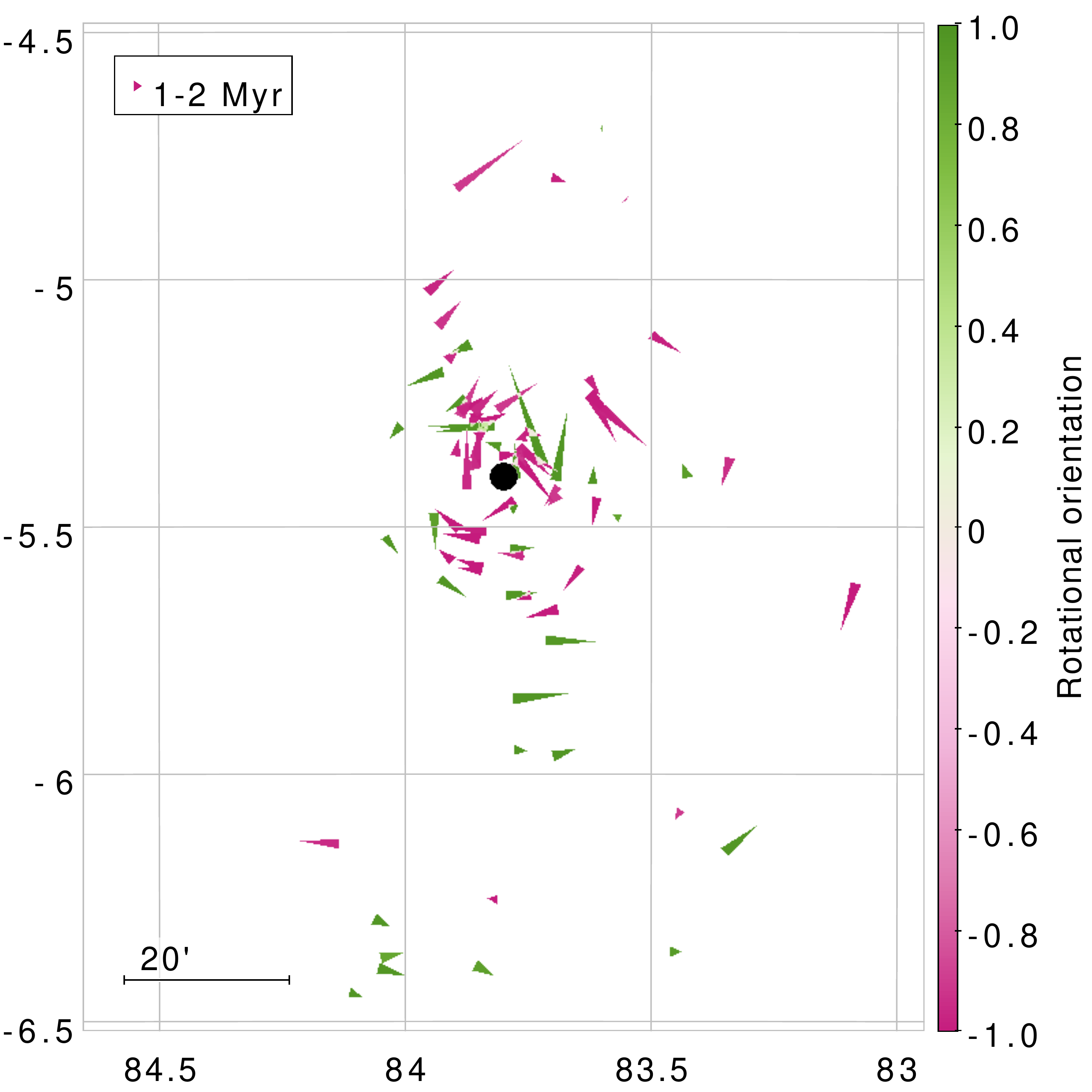}

\includegraphics[width={0.01\textwidth}]{dec.pdf}
\includegraphics[width={0.24\textwidth}]{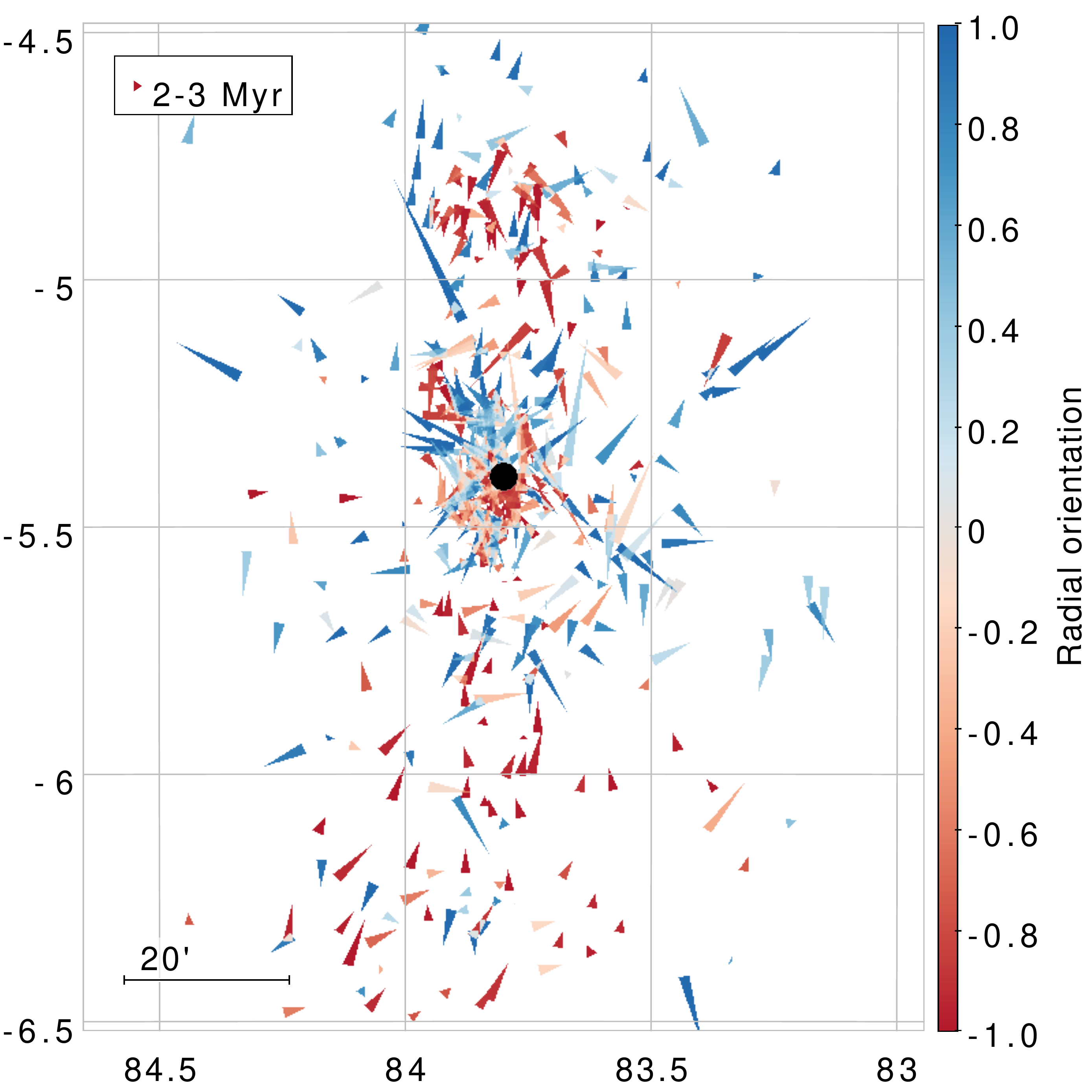}
\includegraphics[width={0.24\textwidth}]{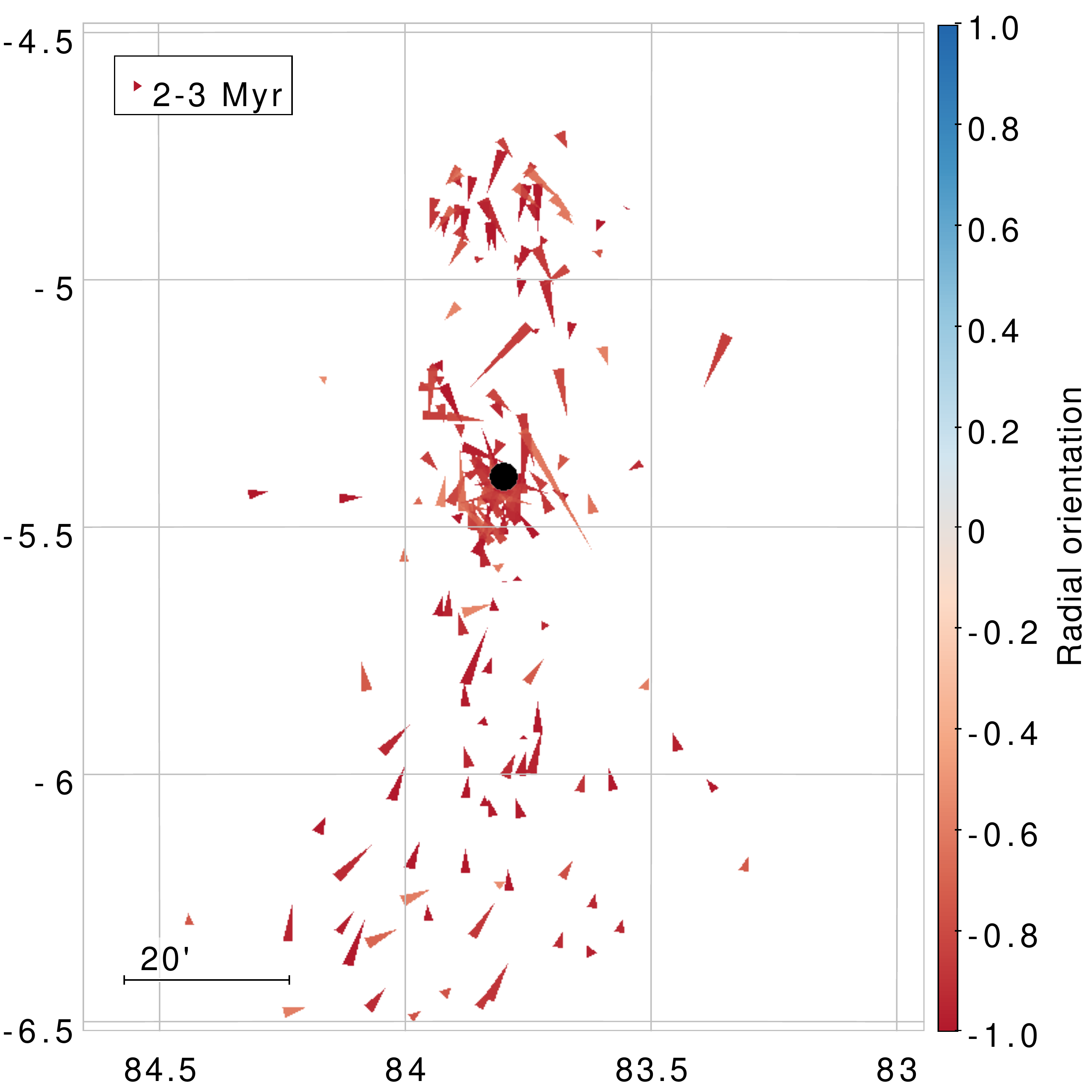}
\includegraphics[width={0.24\textwidth}]{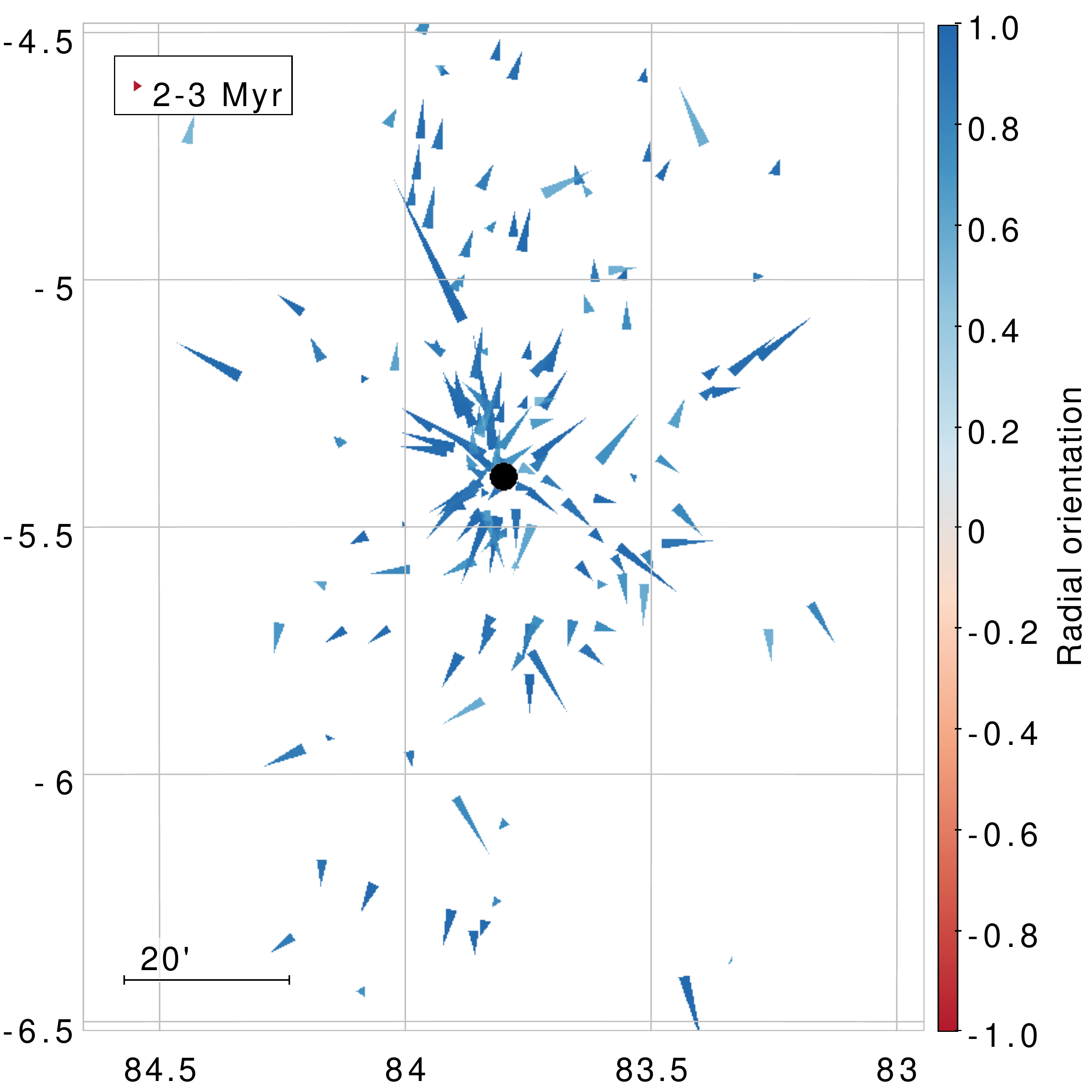}
\includegraphics[width={0.24\textwidth}]{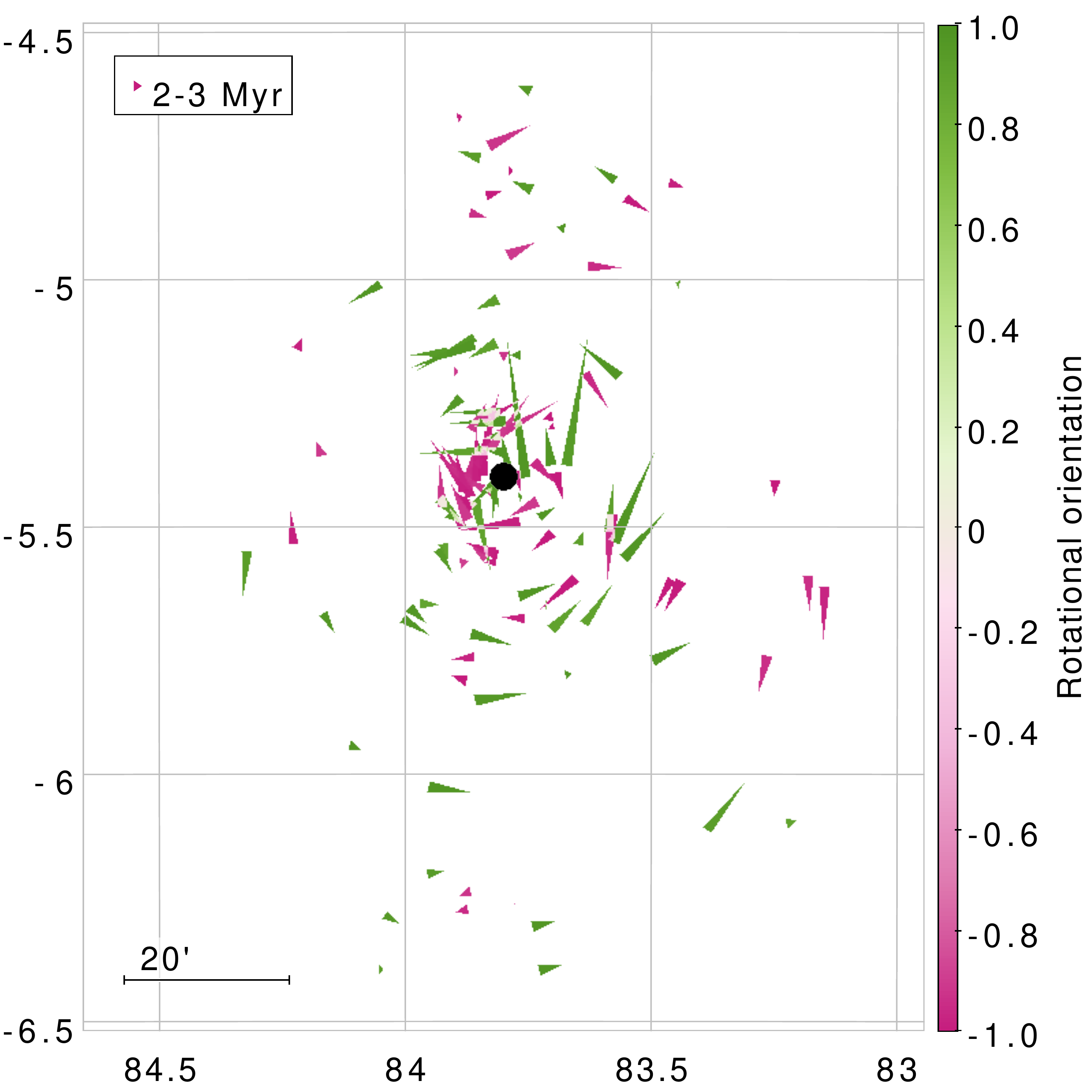}

\includegraphics[width={0.01\textwidth}]{dec.pdf}
\includegraphics[width={0.24\textwidth}]{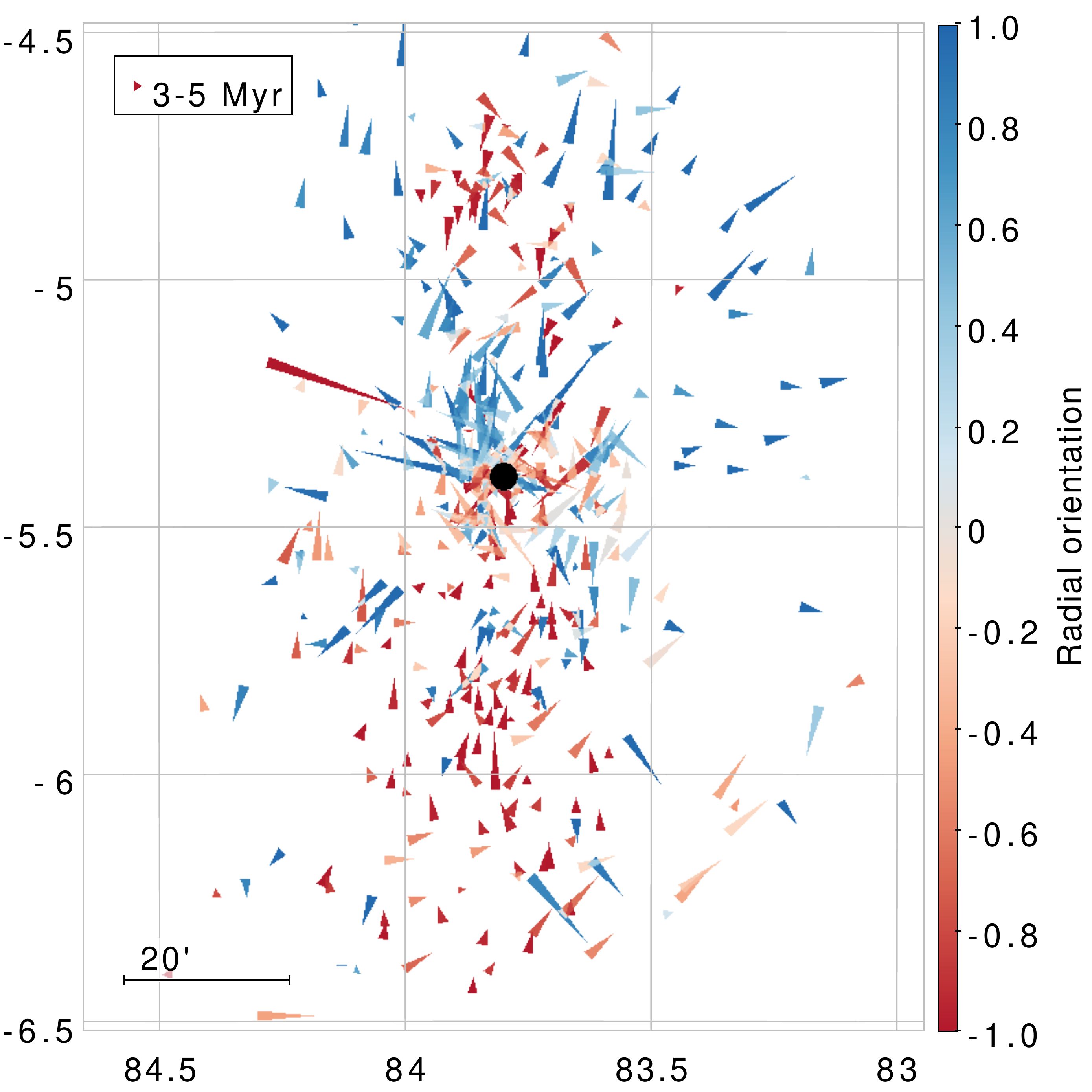}
\includegraphics[width={0.24\textwidth}]{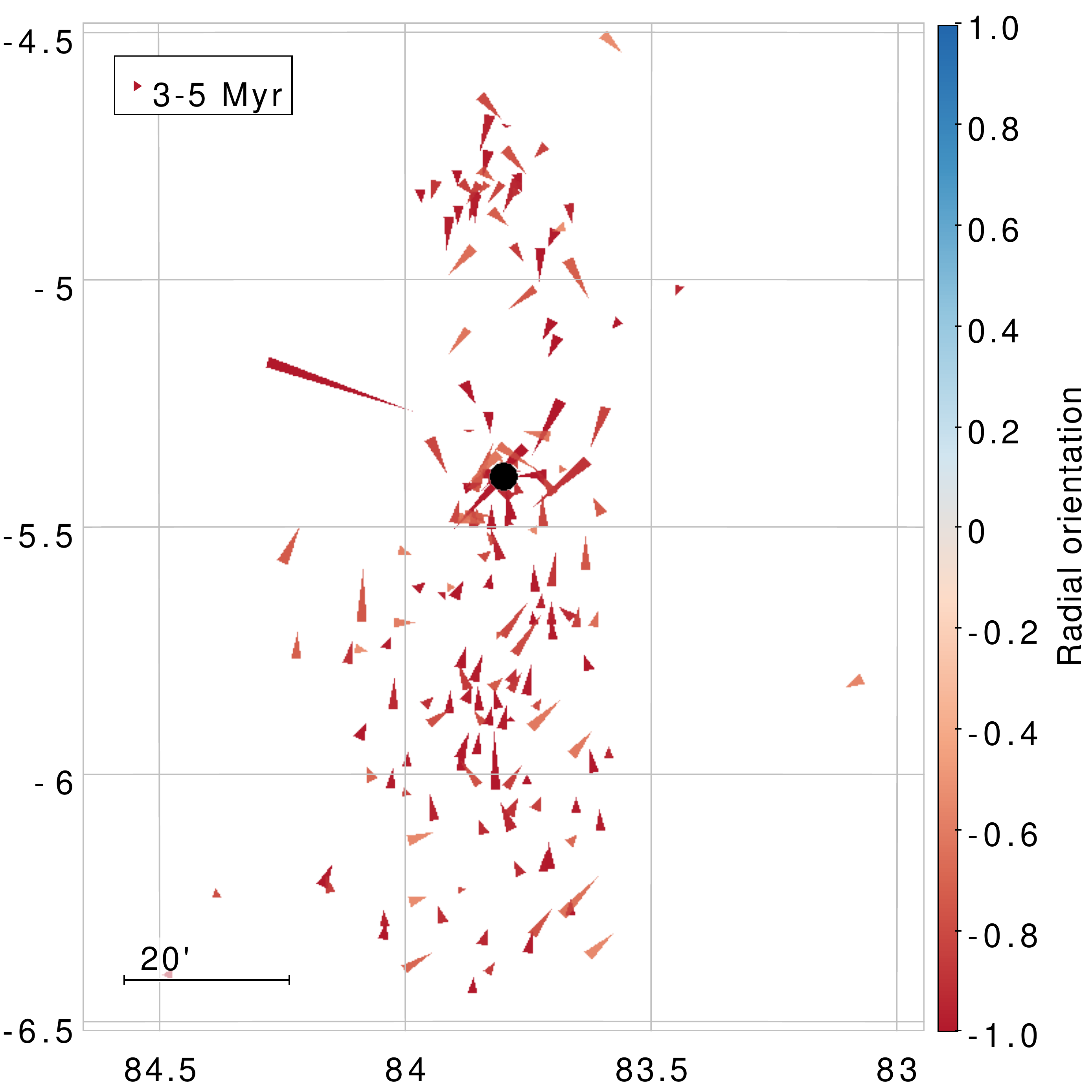}
\includegraphics[width={0.24\textwidth}]{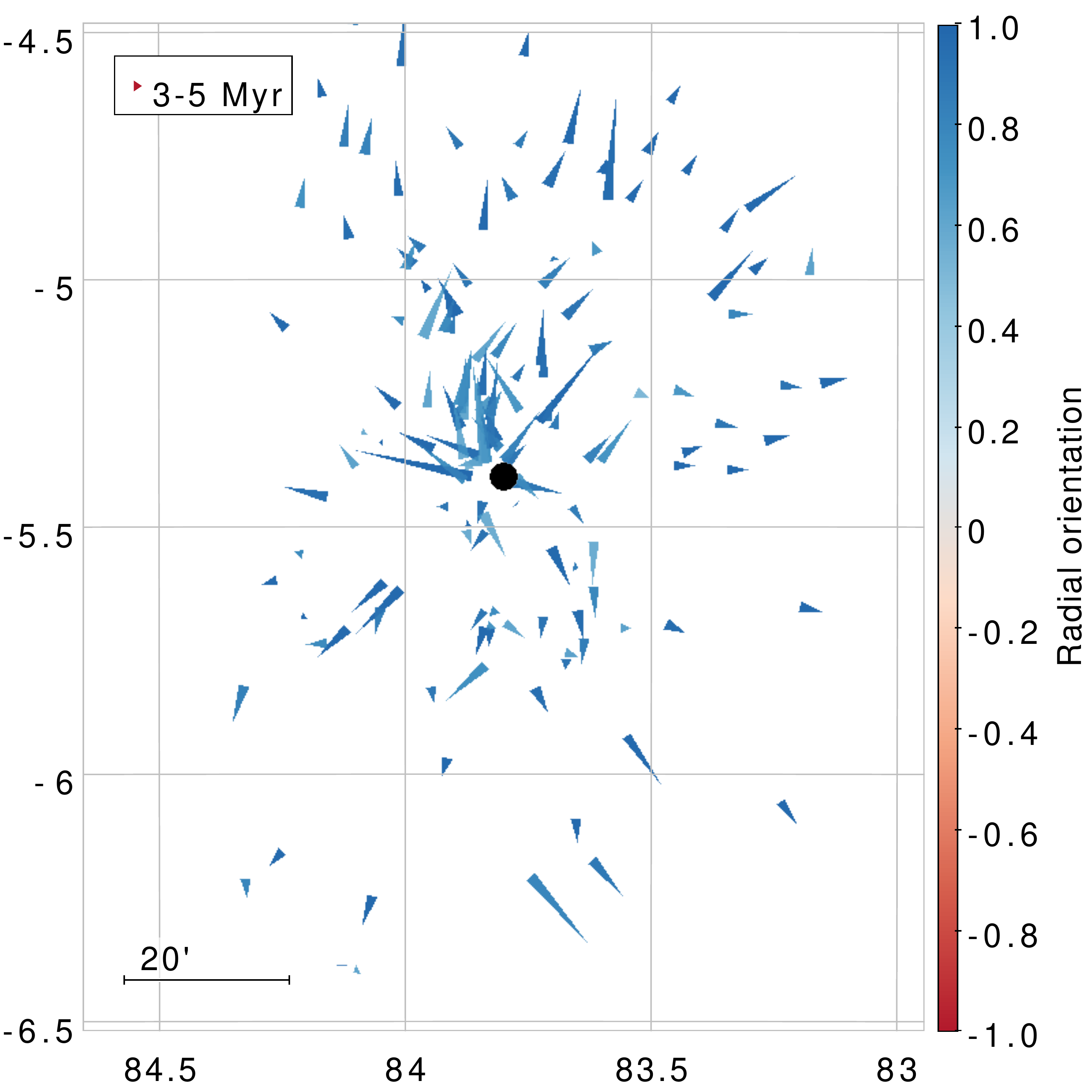}
\includegraphics[width={0.24\textwidth}]{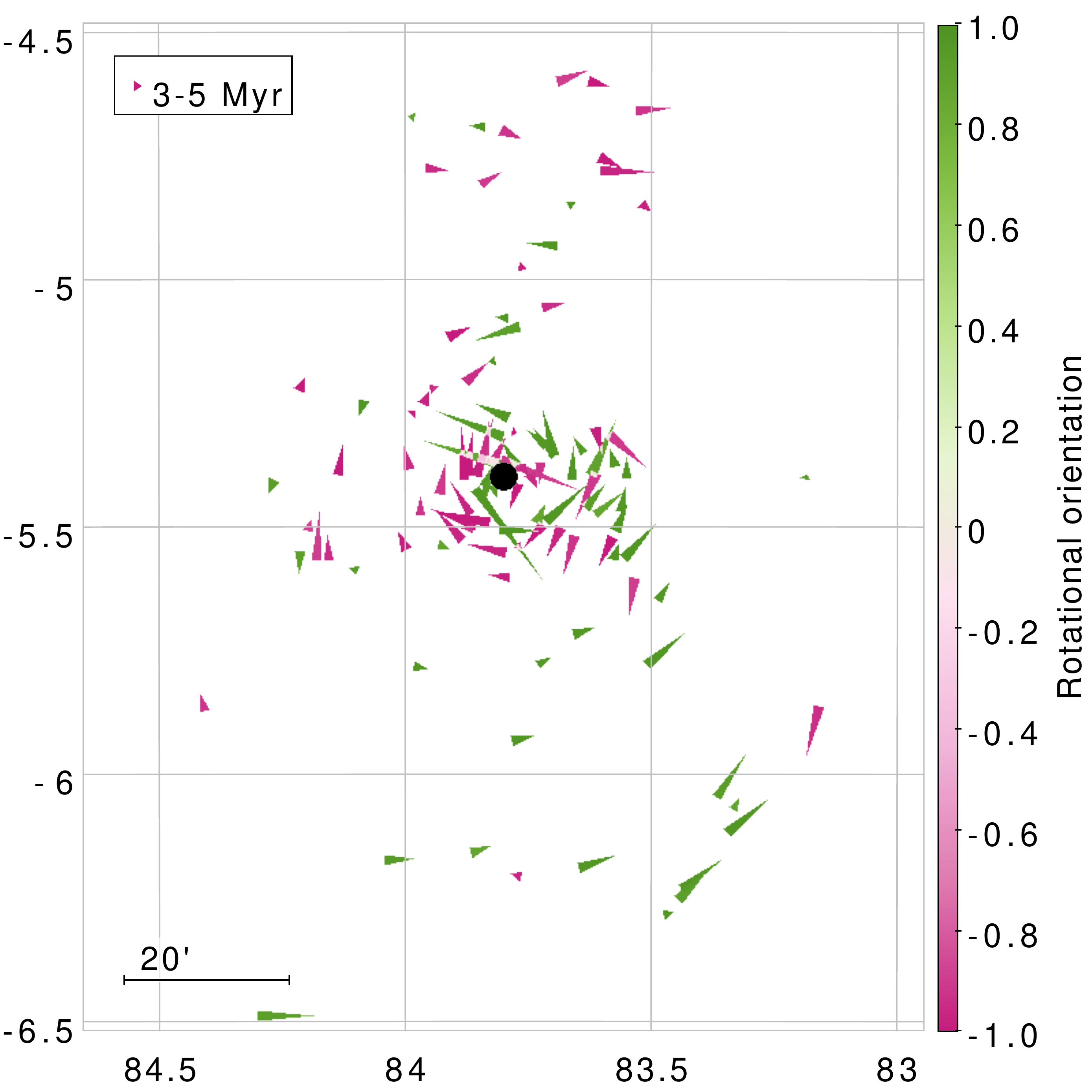}

\includegraphics[width={0.01\textwidth}]{dec.pdf}
\includegraphics[width={0.24\textwidth}]{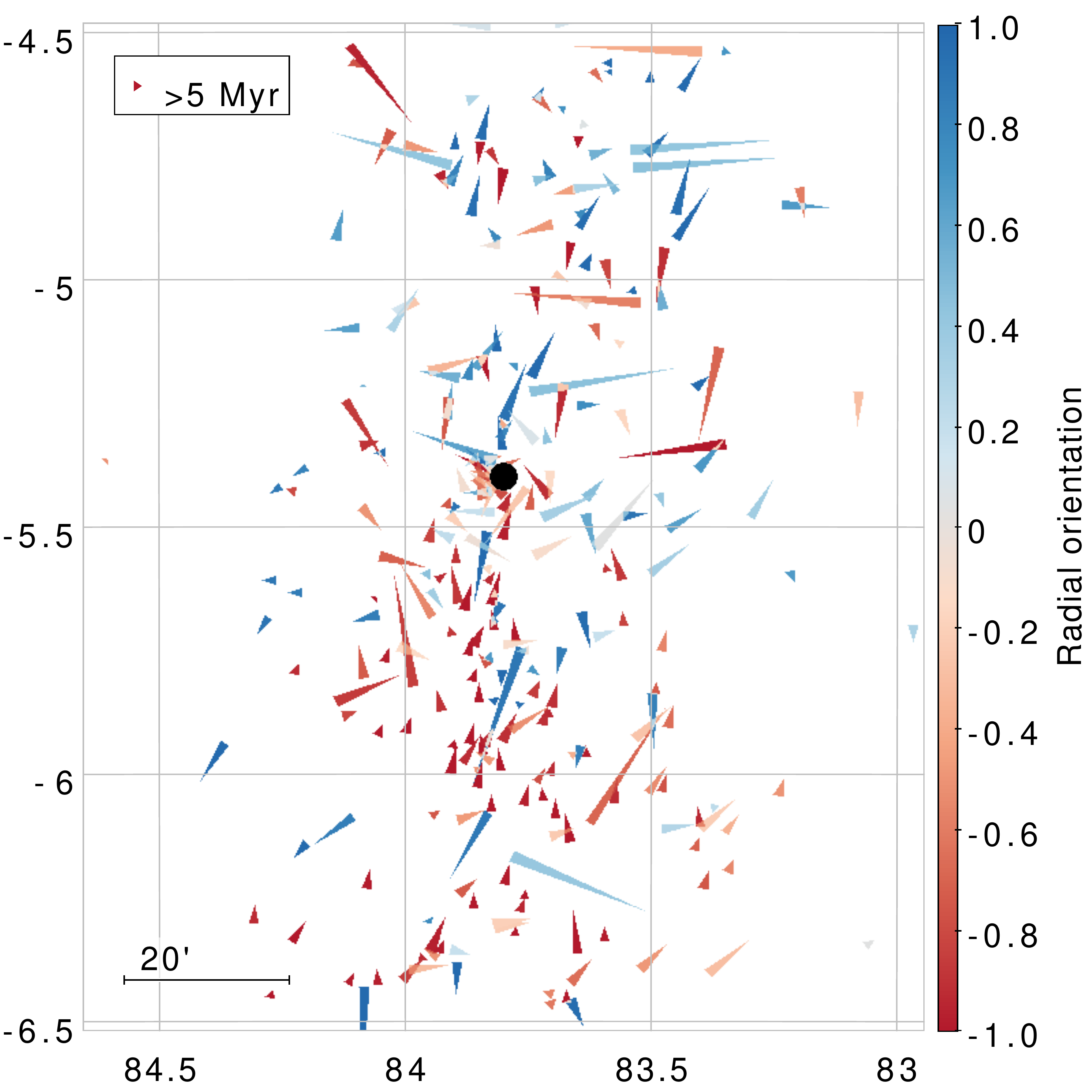}
\includegraphics[width={0.24\textwidth}]{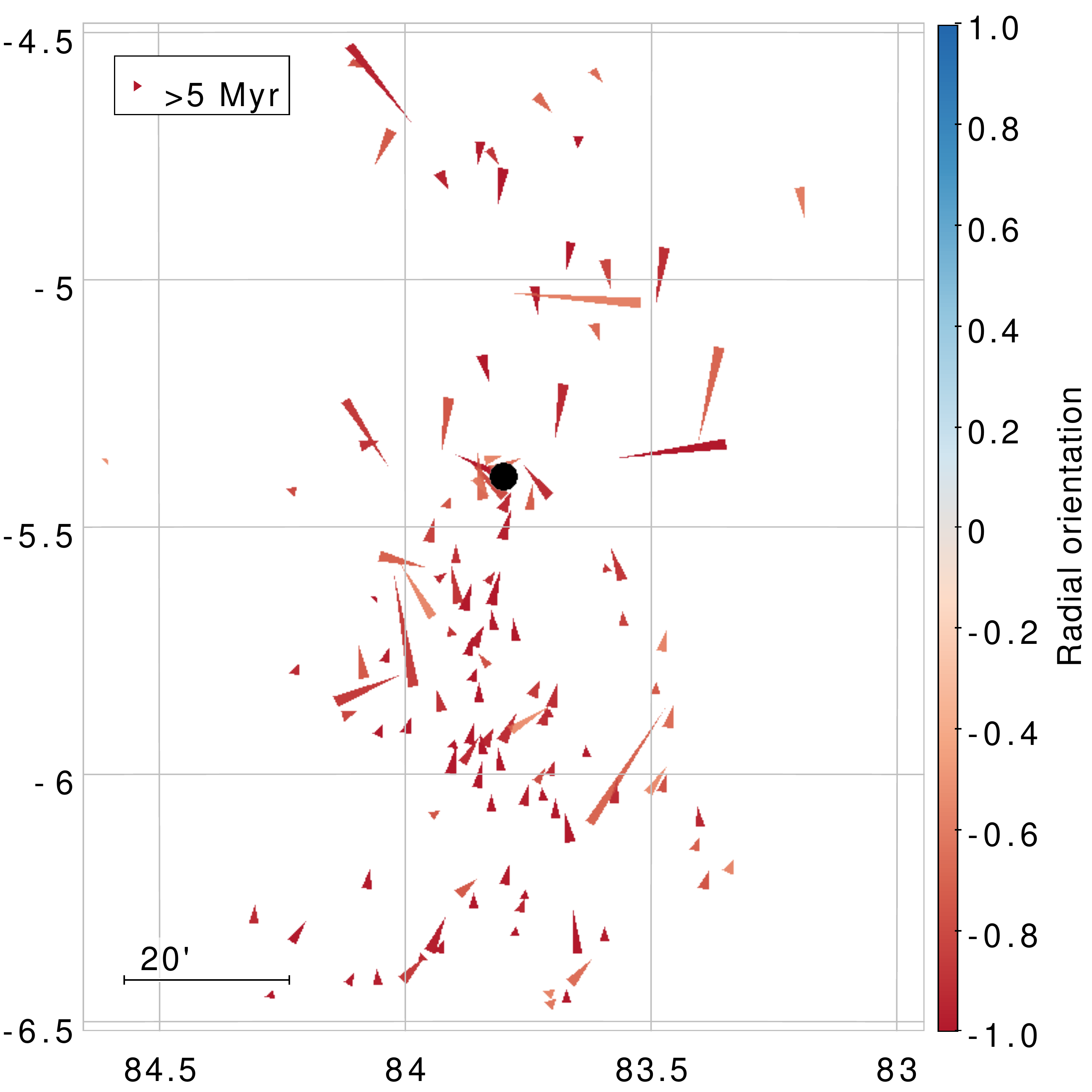}
\includegraphics[width={0.24\textwidth}]{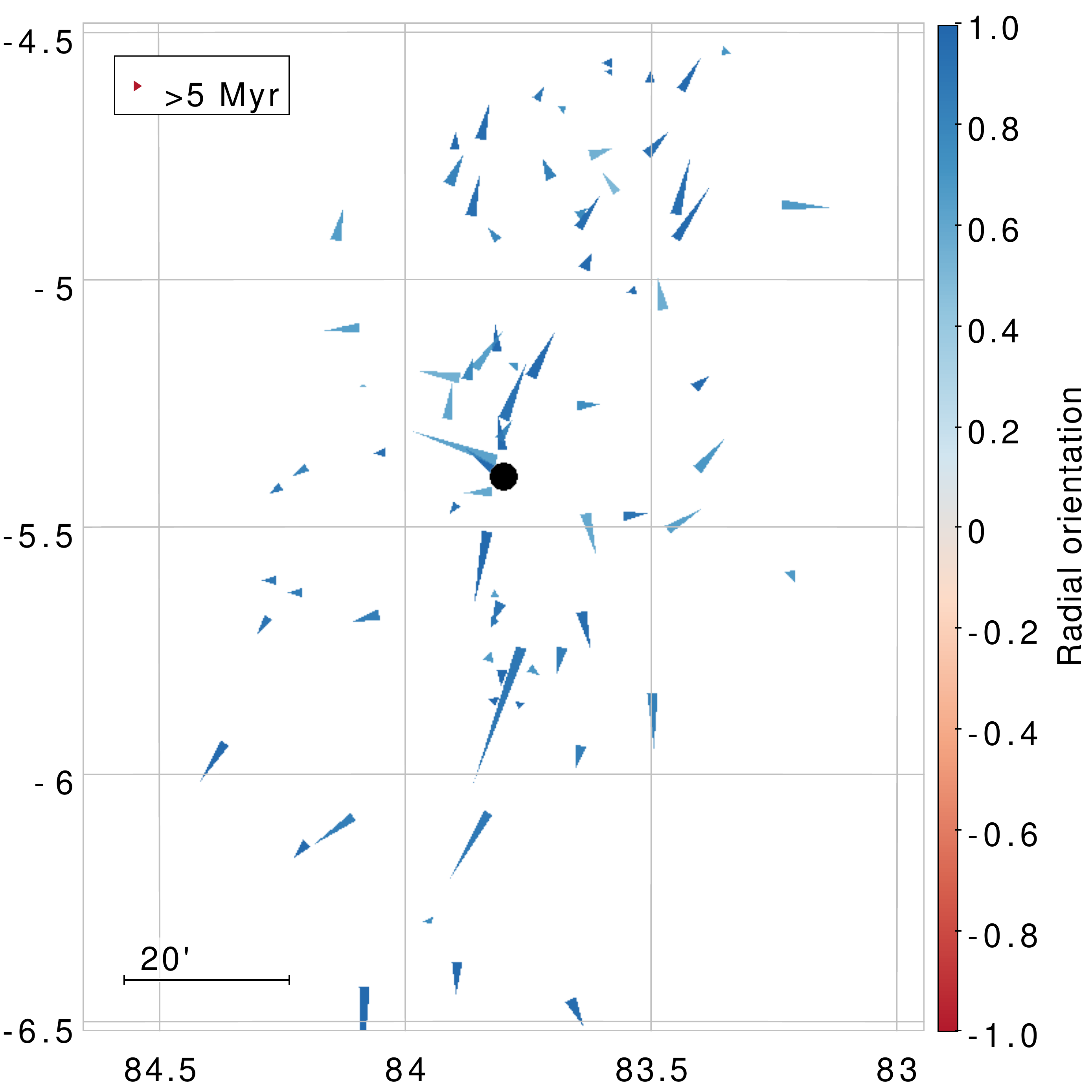}
\includegraphics[width={0.24\textwidth}]{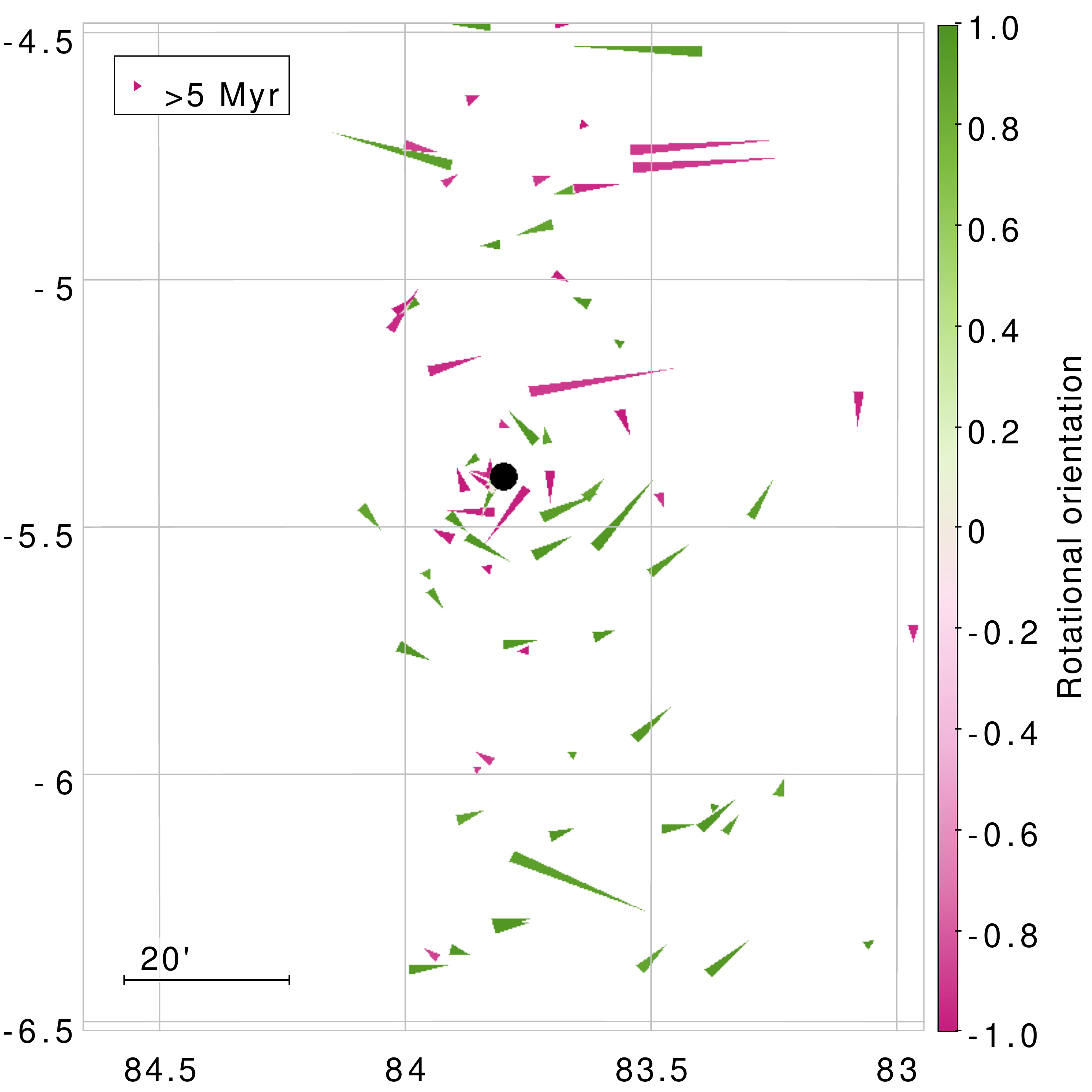}

\includegraphics[width={0.24\textwidth}]{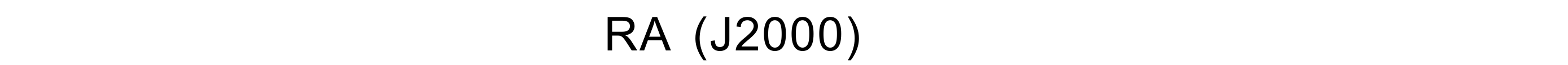}
\includegraphics[width={0.24\textwidth}]{ra.pdf}
\includegraphics[width={0.24\textwidth}]{ra.pdf}
\includegraphics[width={0.24\textwidth}]{ra.pdf}
\caption{Proper motions of stars towards ONC, separated into 5 age bins, shown in separate rows, based on their \logg, with the velocity vectors color-coded based on the orientation relative to the center: red for infall, blue for expansion, magenta for clockwise rotation, green for counter-clockwise rotation. The length of each vector corresponds to the motion of a star over the next 0.2 Myr. Typical velocities are 1.2 \masyr, or 2.2 \kms. The black dot shows the assumed center of the cluster. The first column shows the full sample, second column - only the sources that are falling towards the center, third column - only the sources outflowing from the center, fourth - sources that are preferentially moving tangentially around the center.
\label{fig:onc}}
\end{figure*}

\begin{figure*}
\includegraphics[width={0.01\textwidth}]{dec.pdf}
\includegraphics[width={0.24\textwidth}]{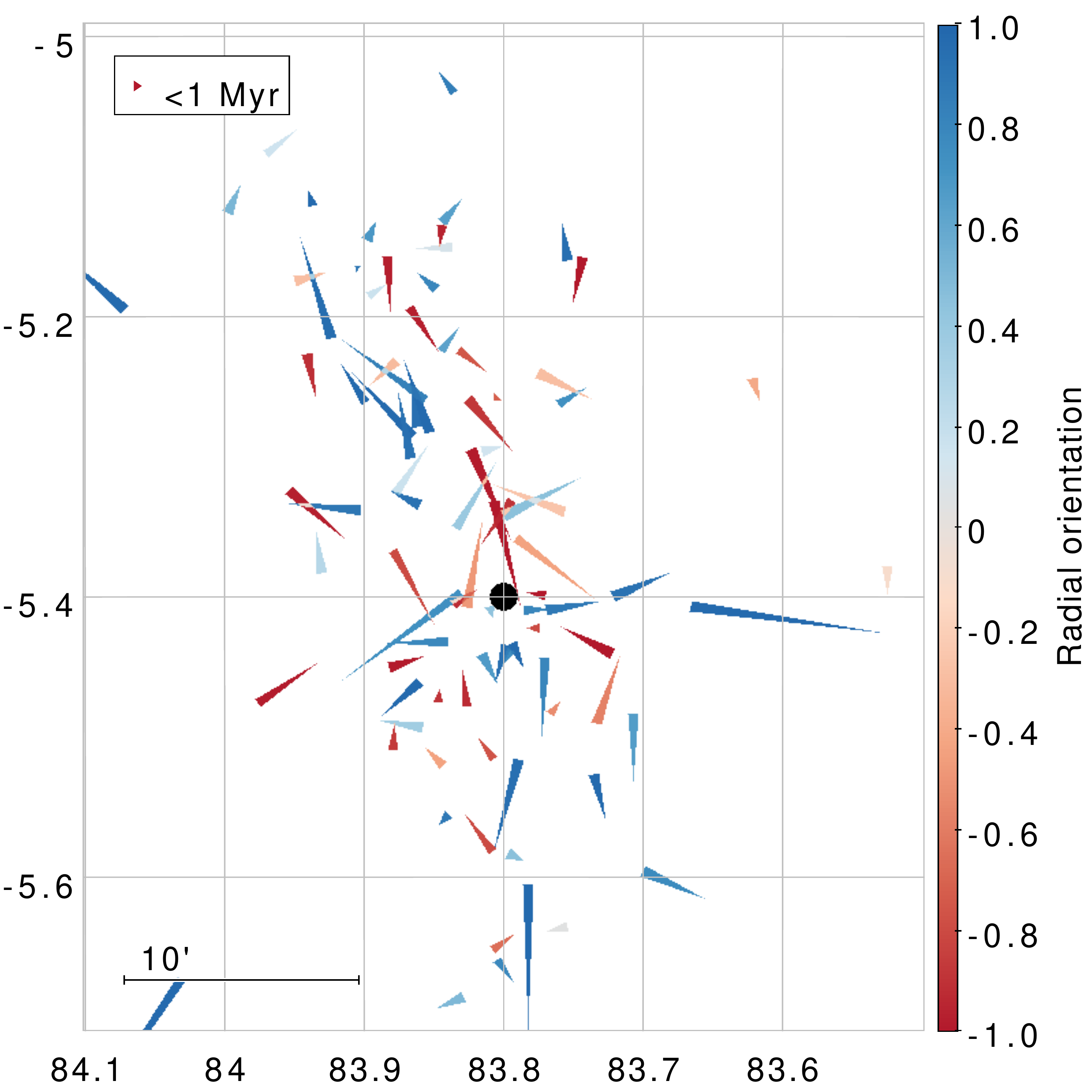}
\includegraphics[width={0.24\textwidth}]{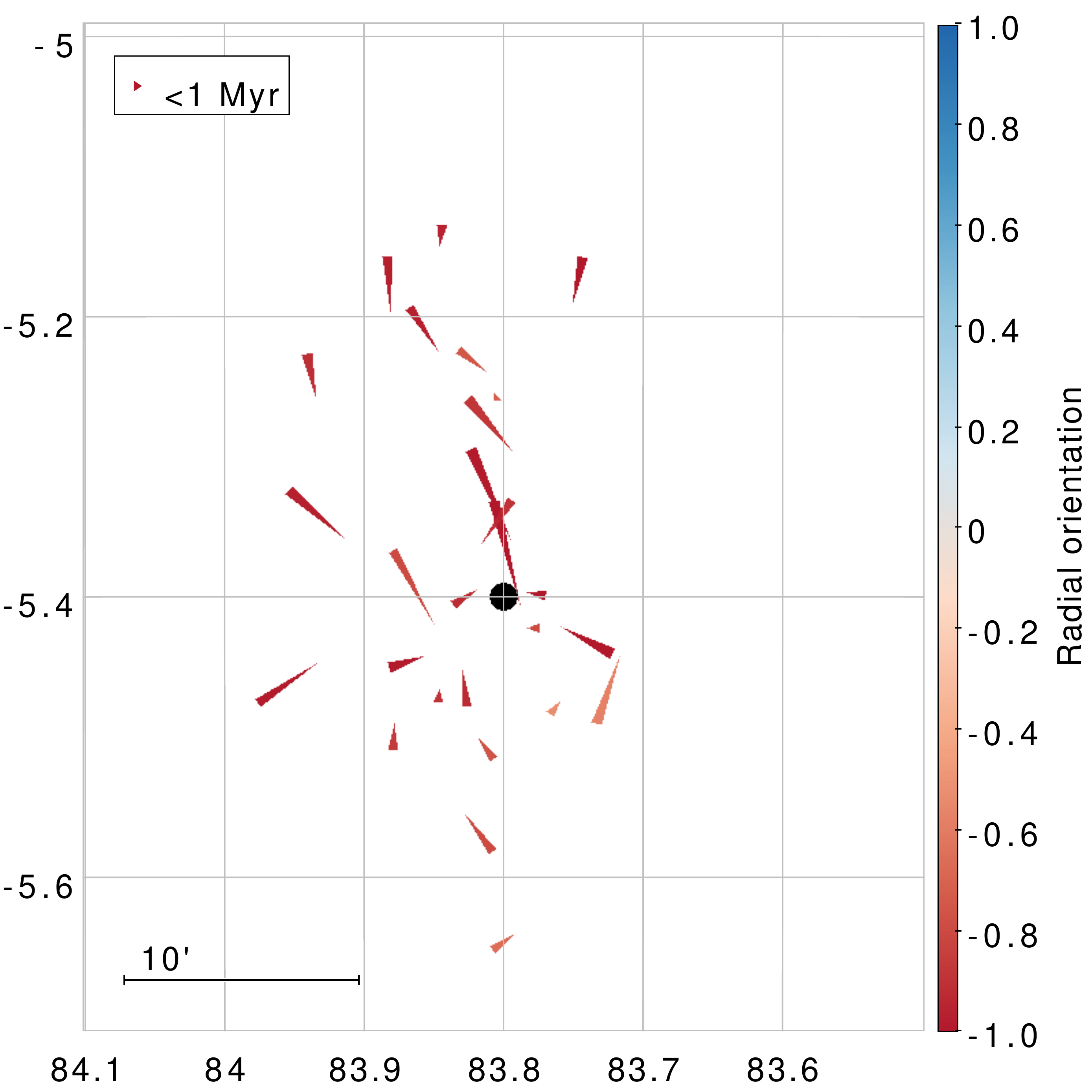}
\includegraphics[width={0.24\textwidth}]{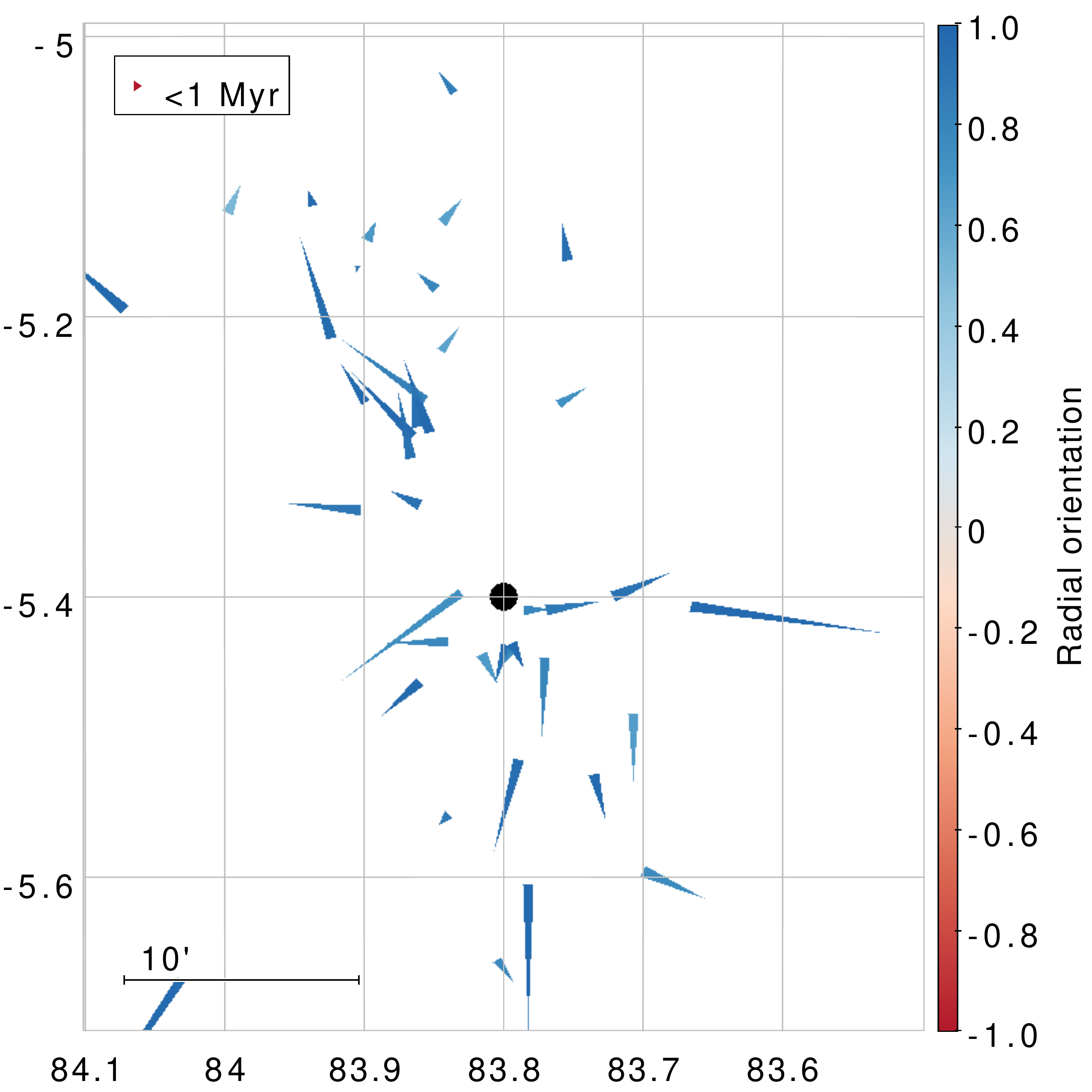}
\includegraphics[width={0.24\textwidth}]{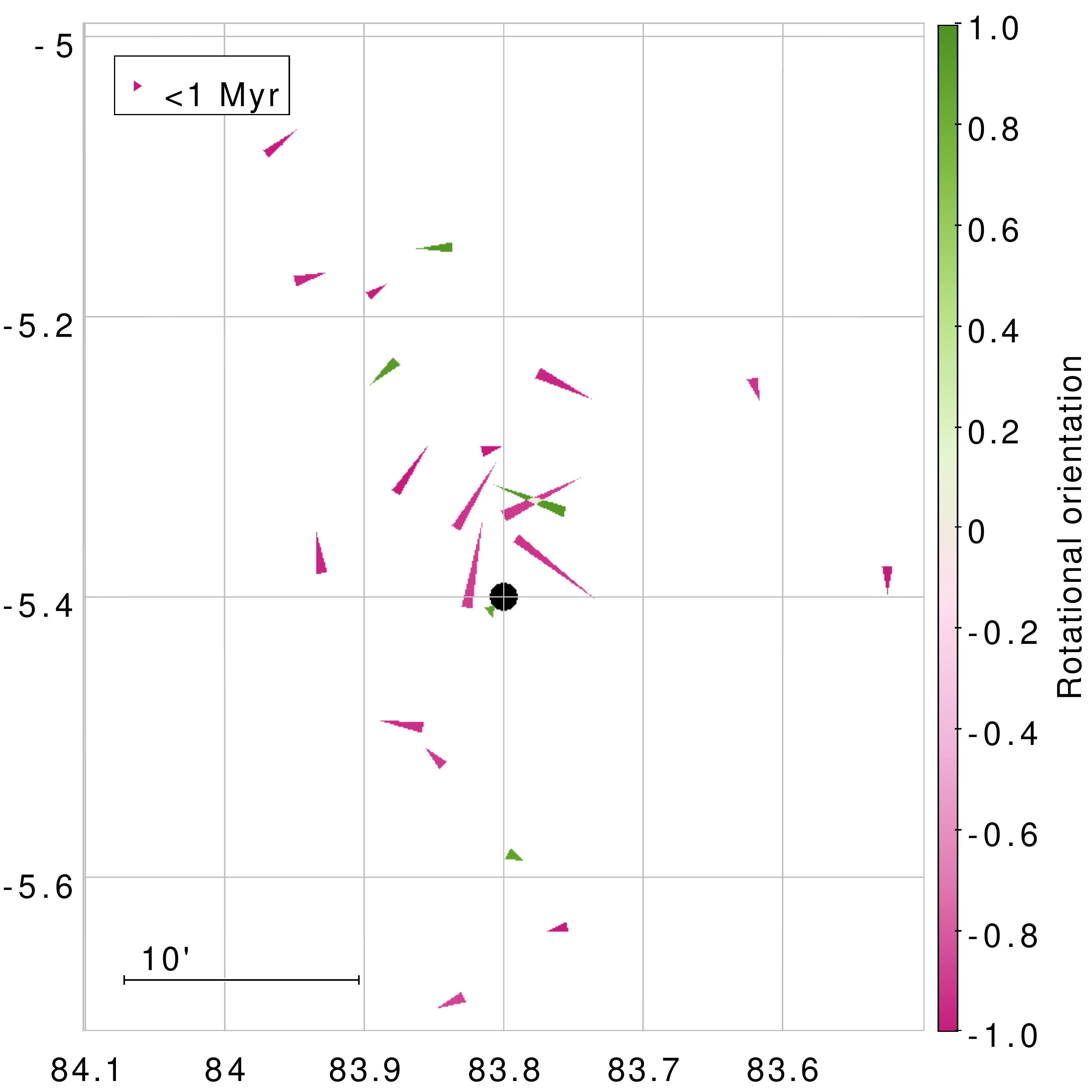}

\includegraphics[width={0.01\textwidth}]{dec.pdf}
\includegraphics[width={0.24\textwidth}]{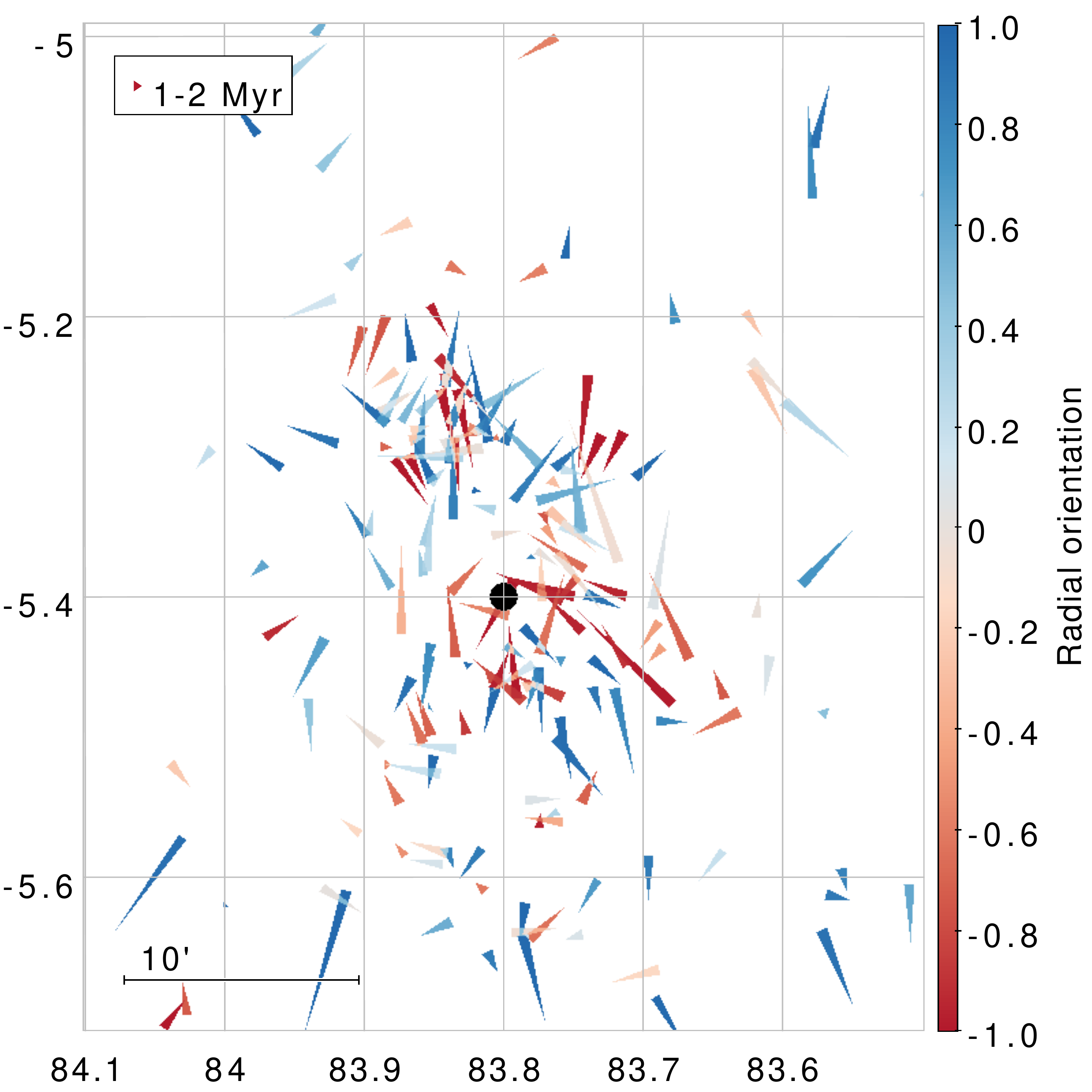}
\includegraphics[width={0.24\textwidth}]{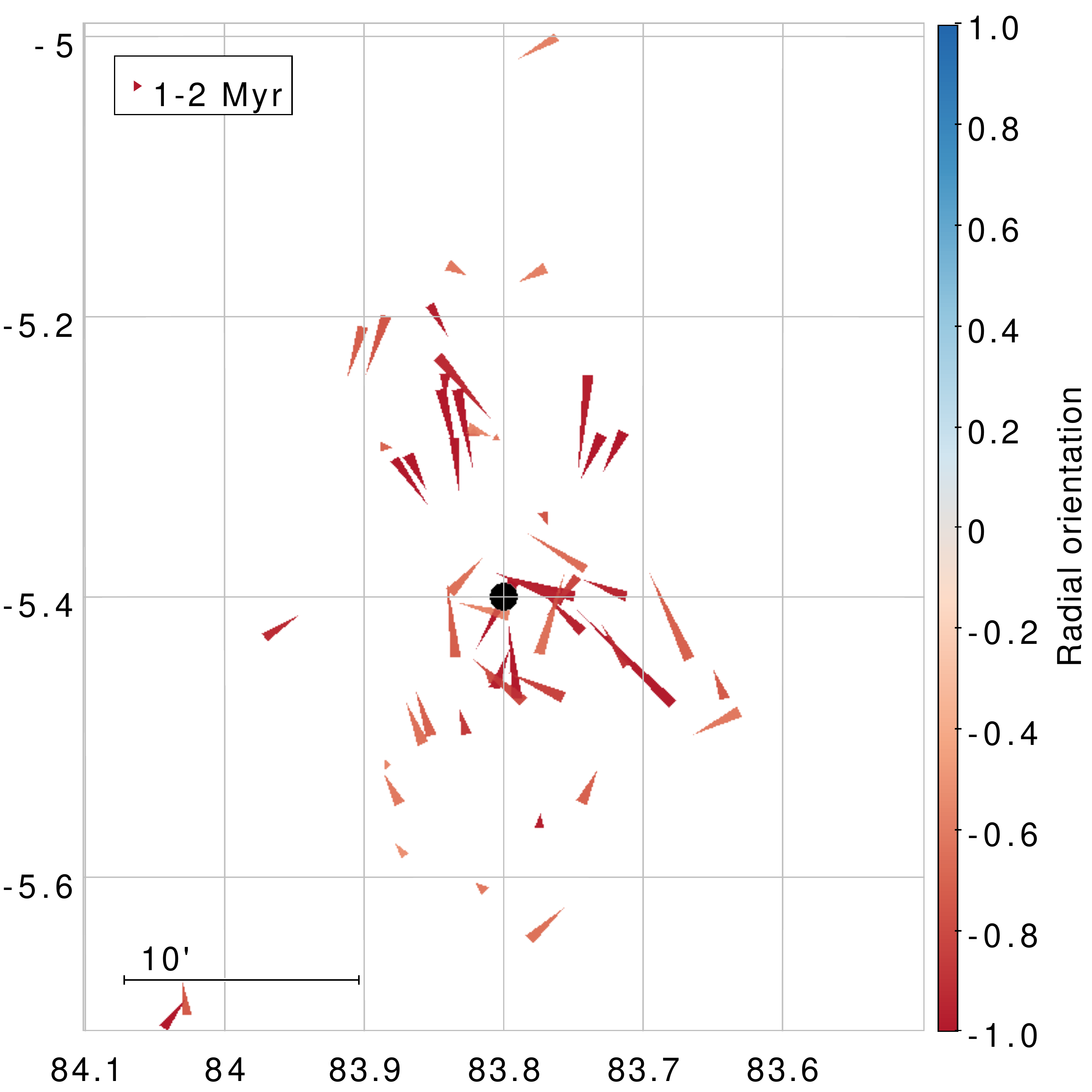}
\includegraphics[width={0.24\textwidth}]{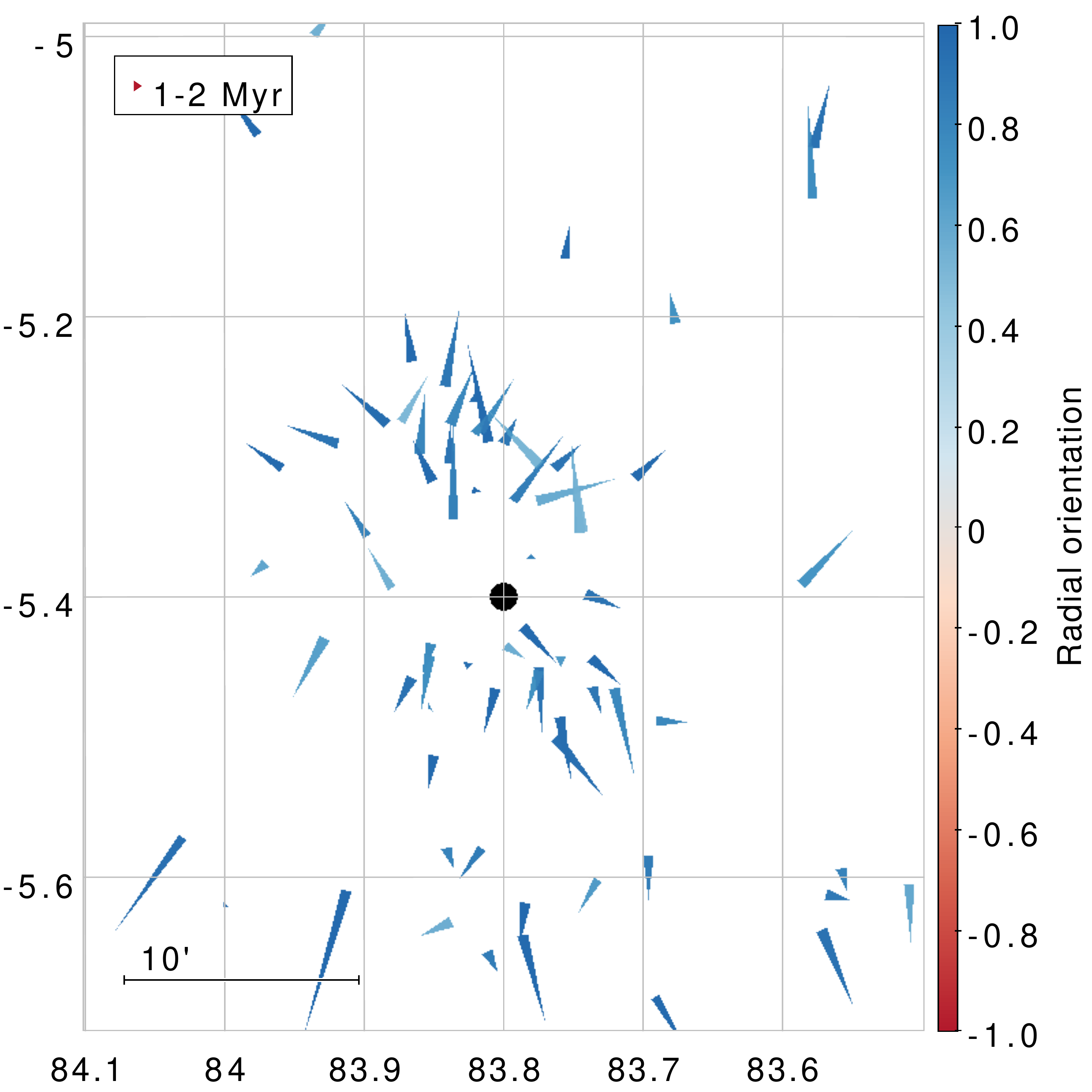}
\includegraphics[width={0.24\textwidth}]{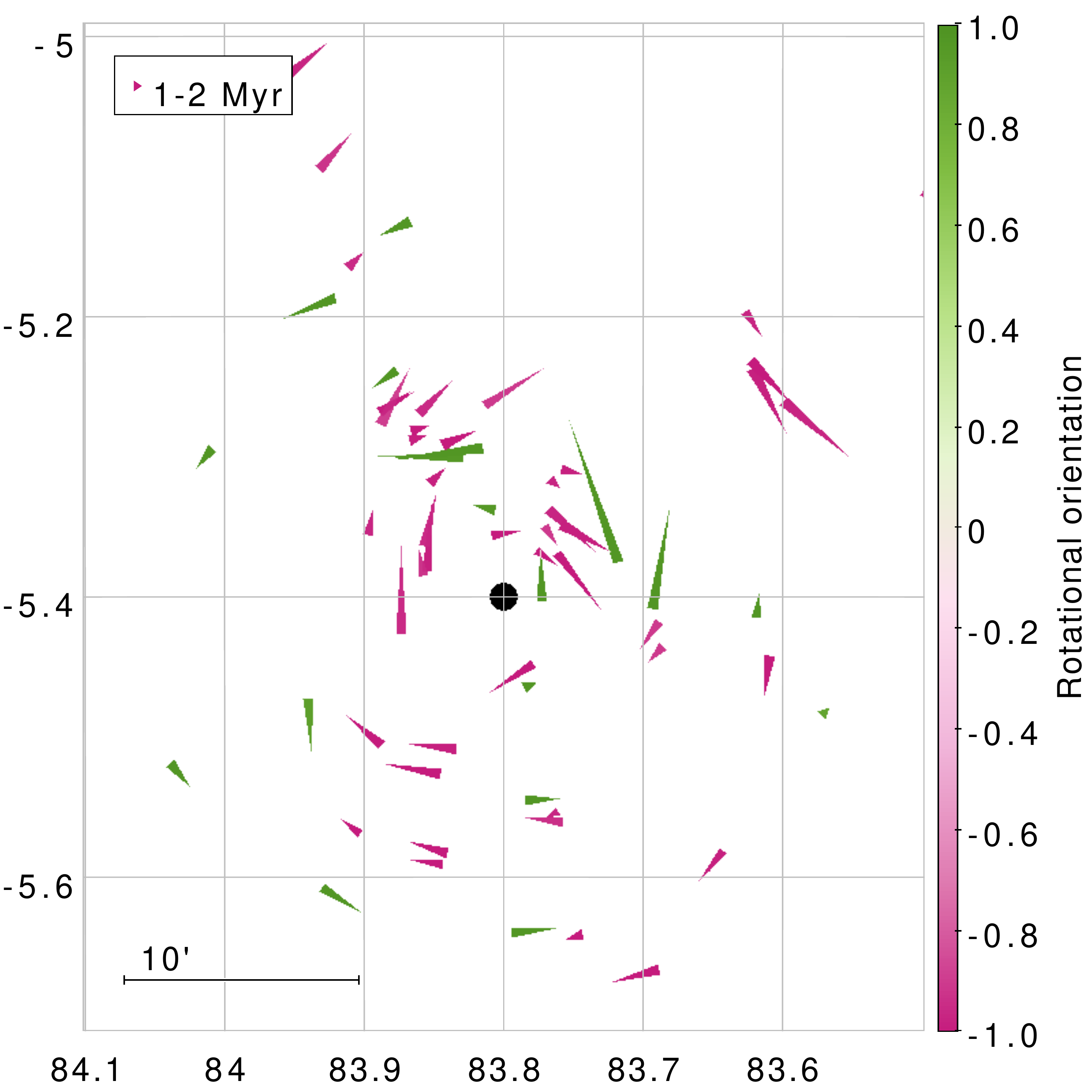}

\includegraphics[width={0.01\textwidth}]{dec.pdf}
\includegraphics[width={0.24\textwidth}]{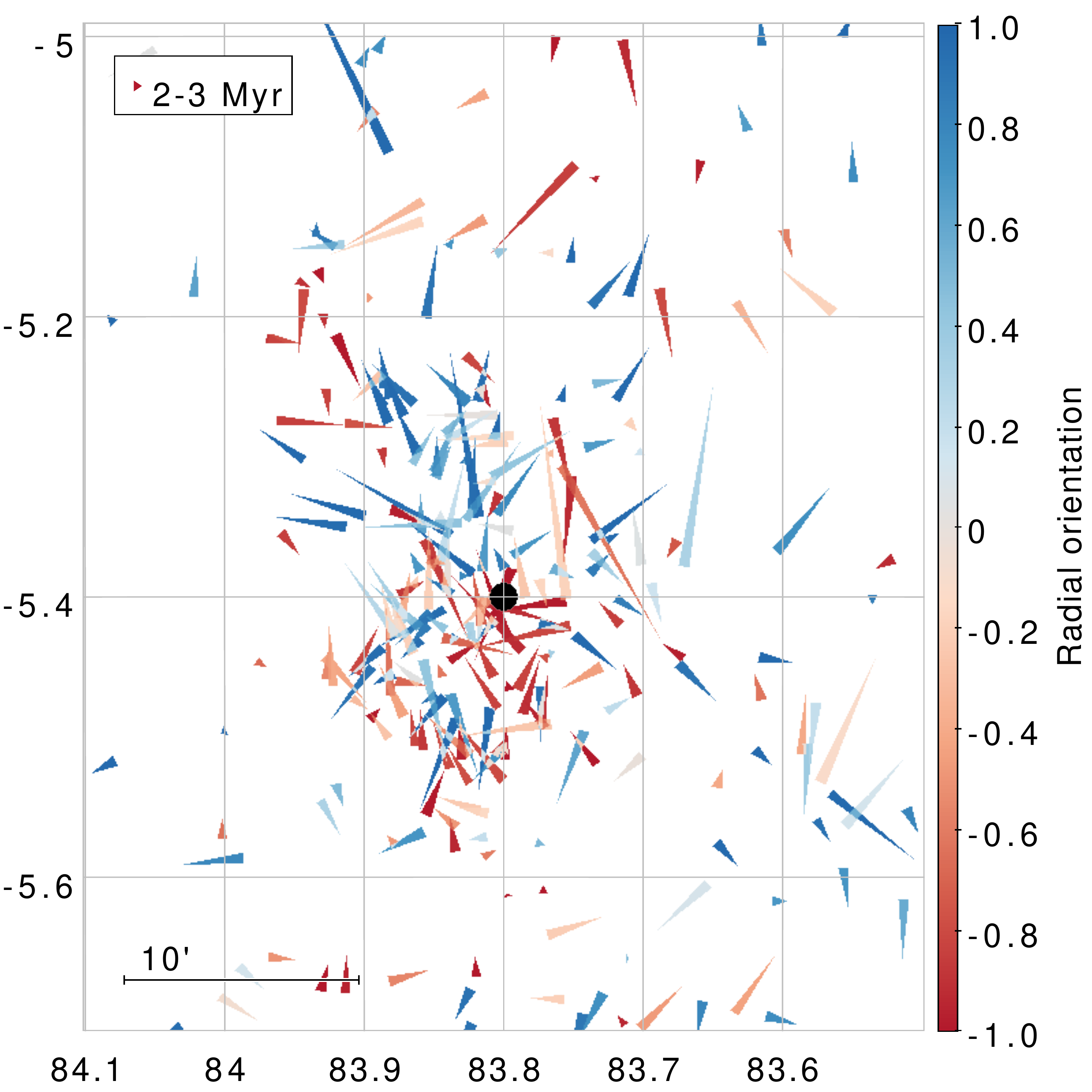}
\includegraphics[width={0.24\textwidth}]{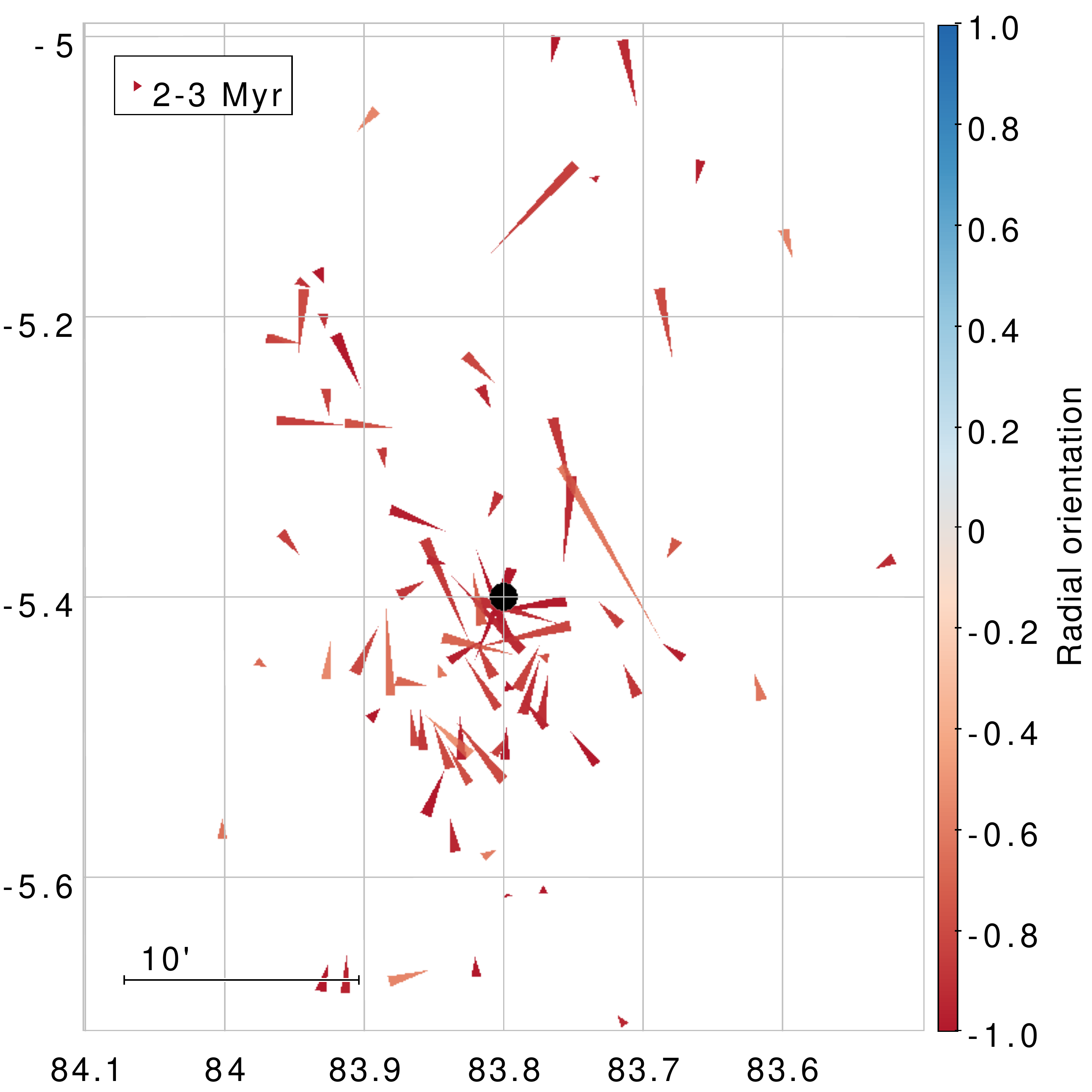}
\includegraphics[width={0.24\textwidth}]{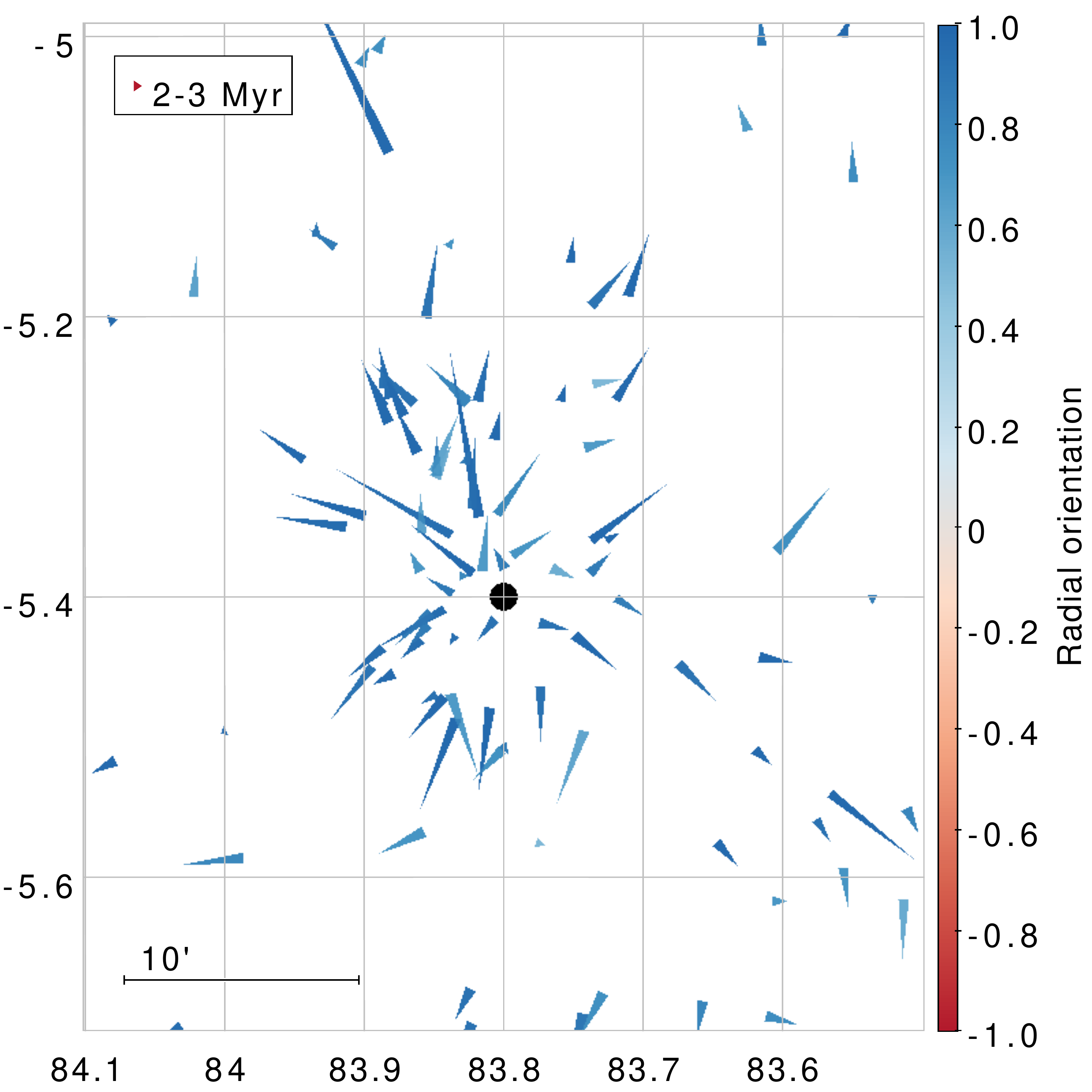}
\includegraphics[width={0.24\textwidth}]{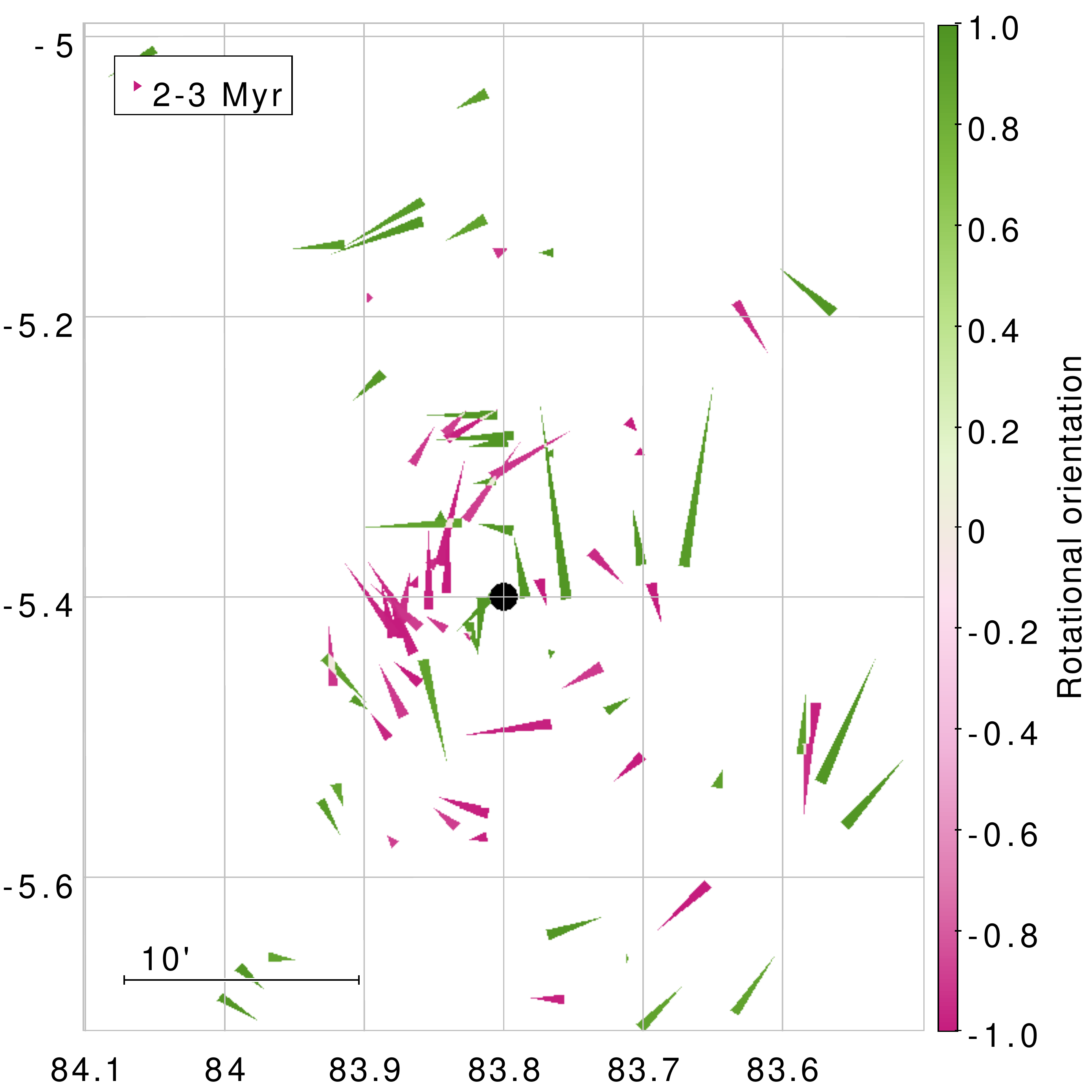}

\includegraphics[width={0.01\textwidth}]{dec.pdf}
\includegraphics[width={0.24\textwidth}]{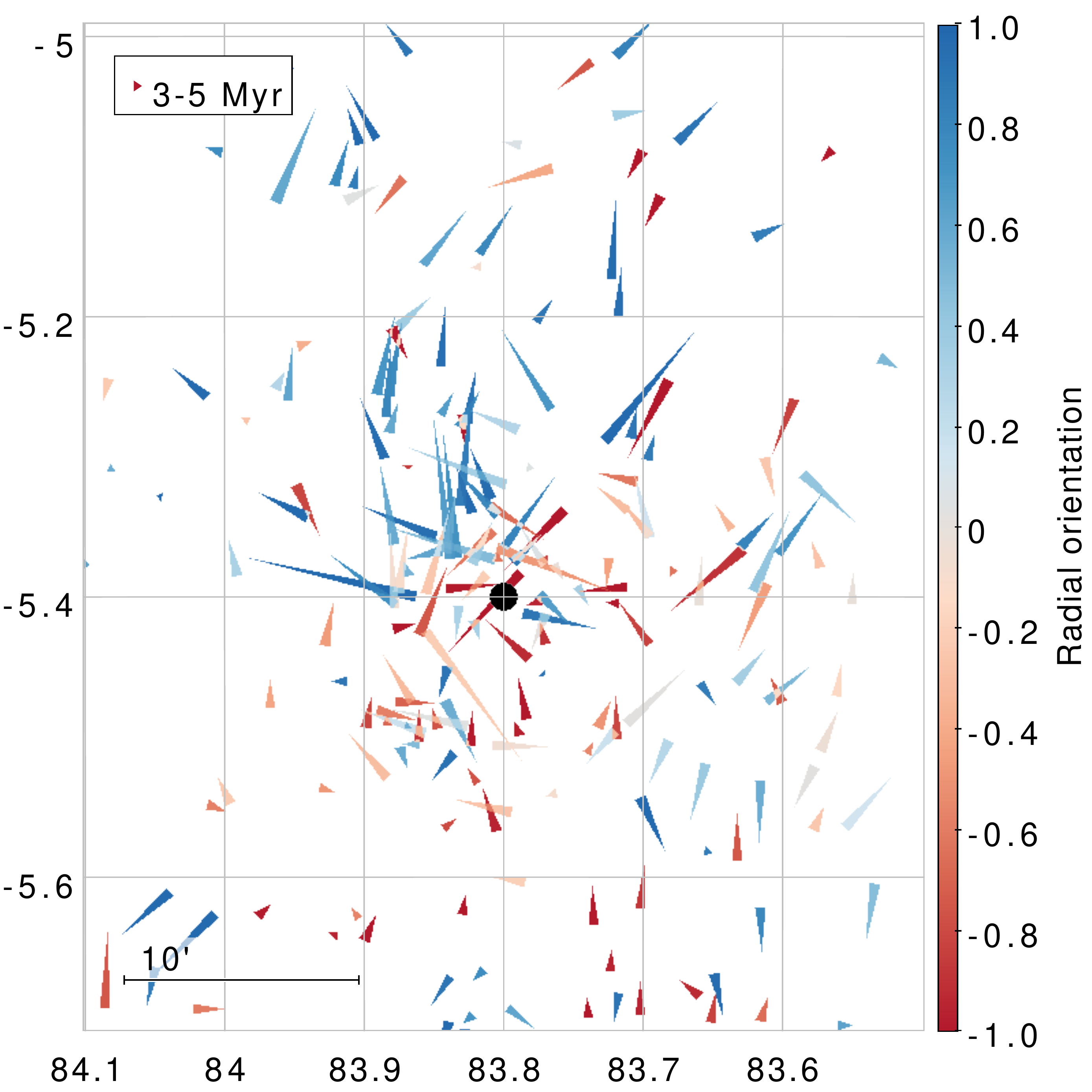}
\includegraphics[width={0.24\textwidth}]{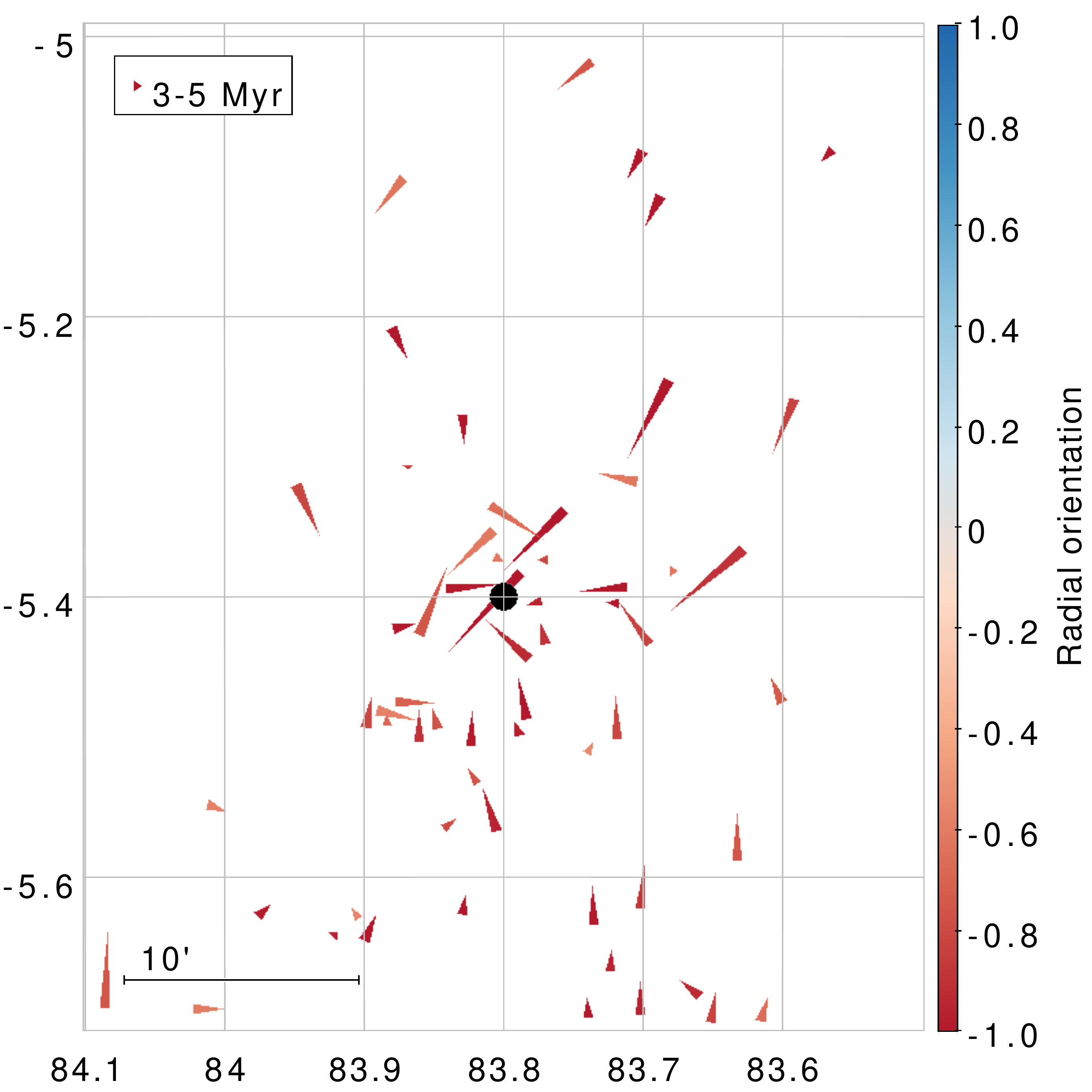}
\includegraphics[width={0.24\textwidth}]{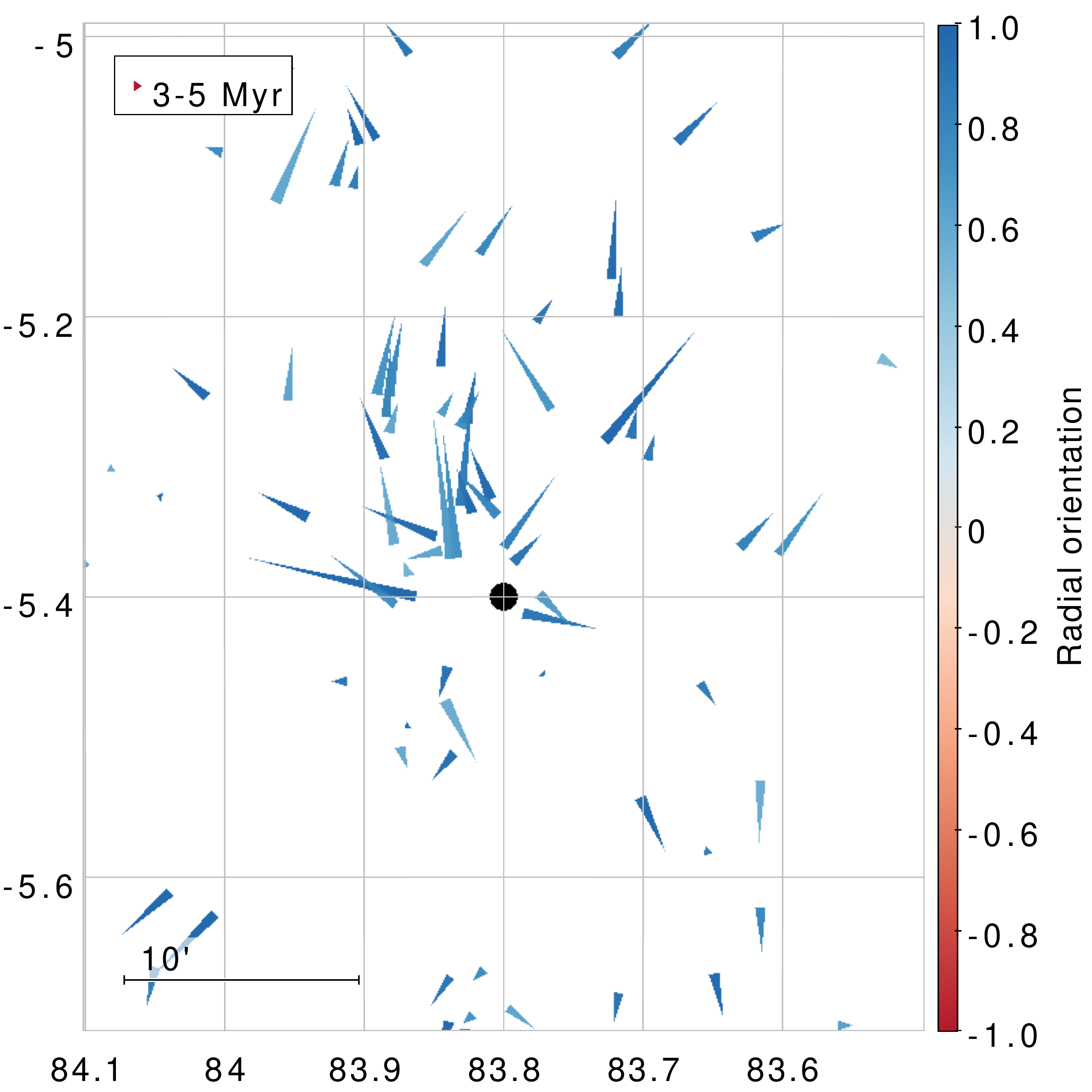}
\includegraphics[width={0.24\textwidth}]{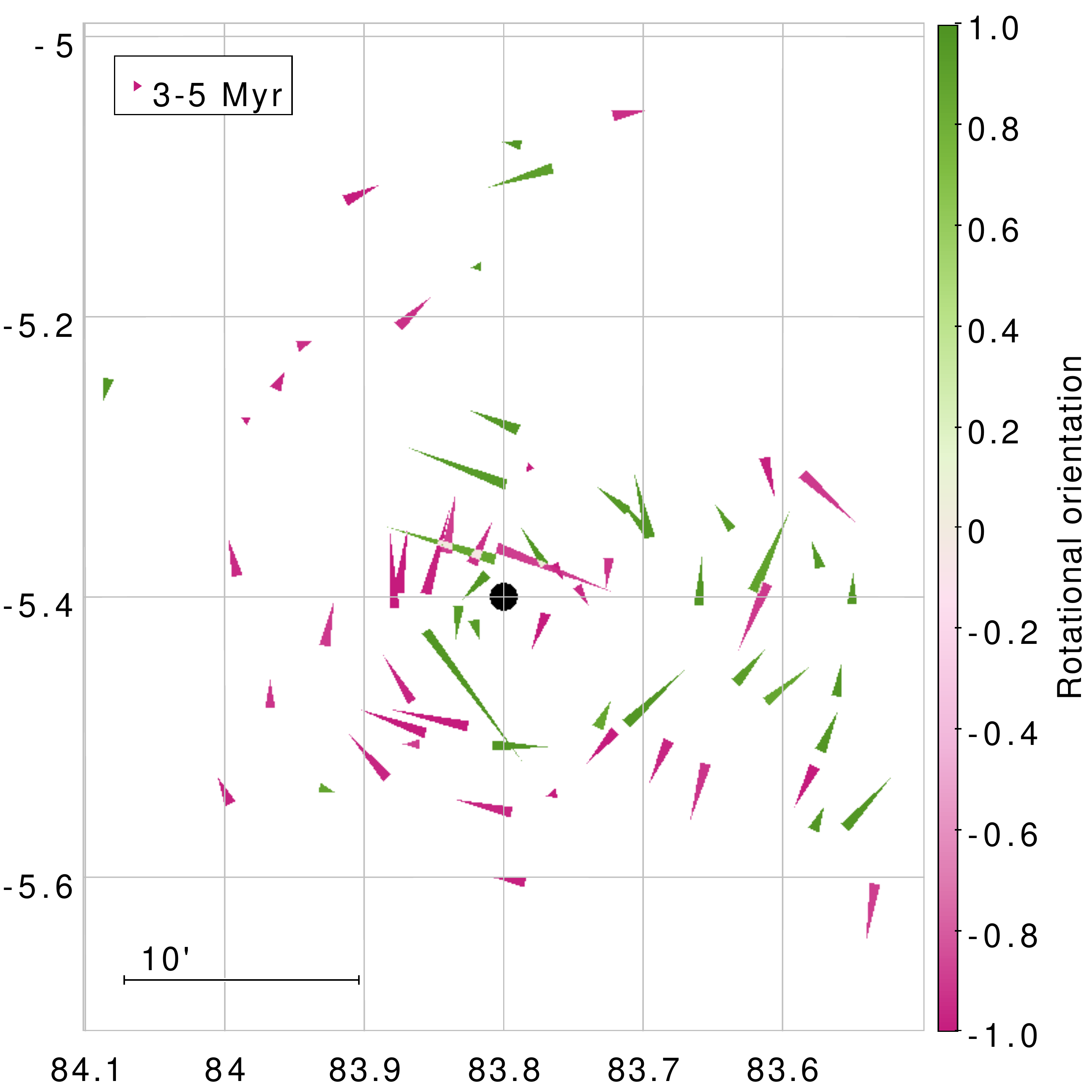}

\includegraphics[width={0.01\textwidth}]{dec.pdf}
\includegraphics[width={0.24\textwidth}]{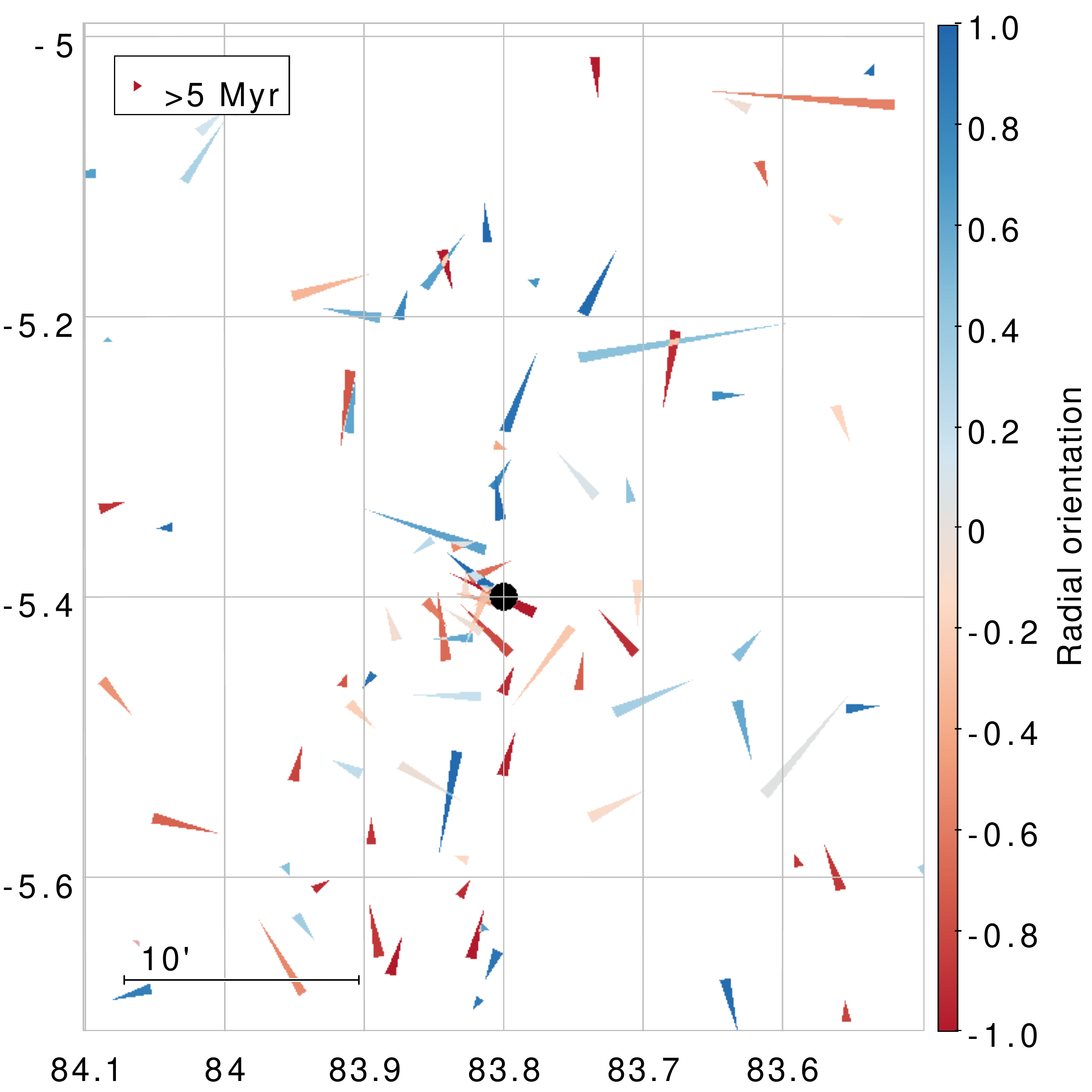}
\includegraphics[width={0.24\textwidth}]{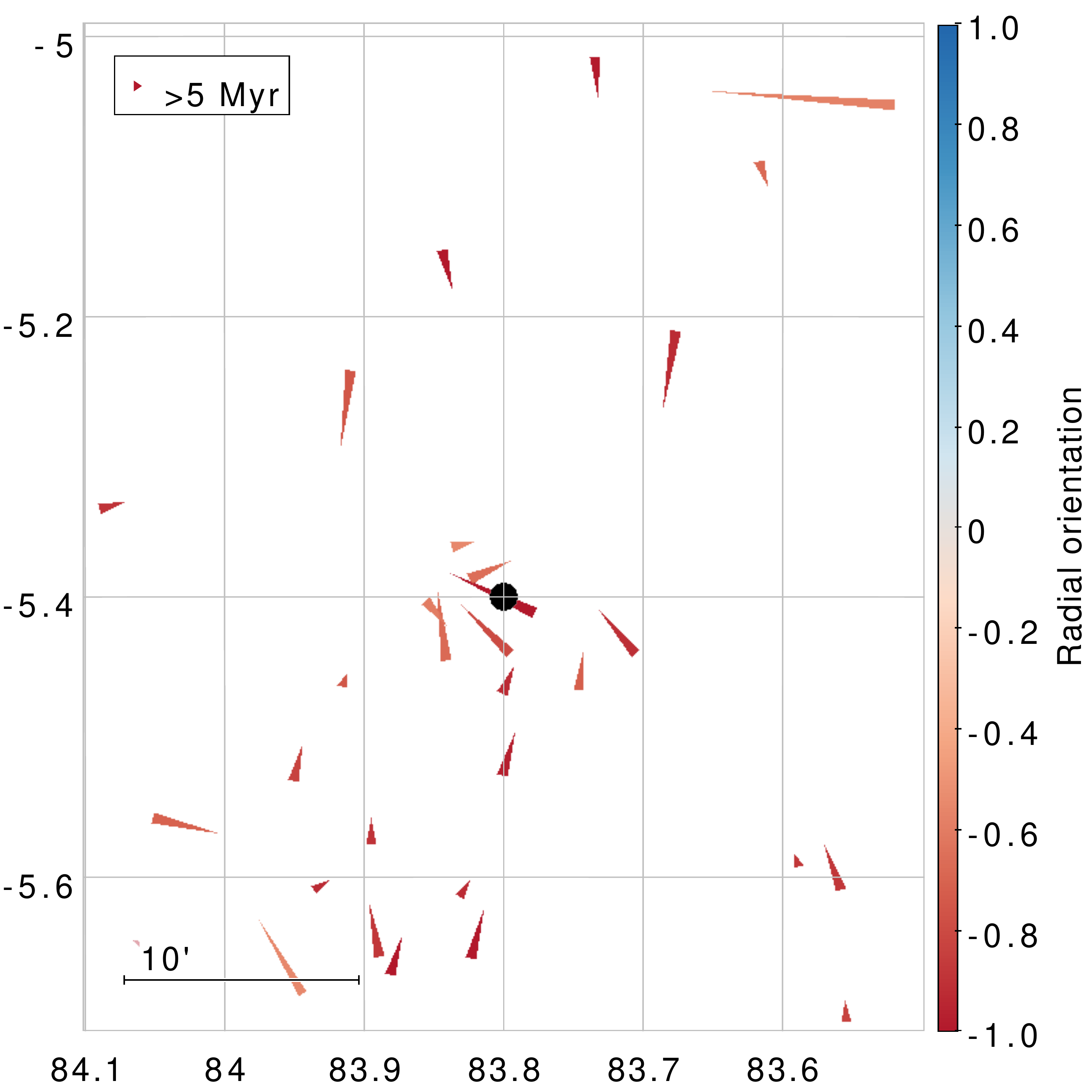}
\includegraphics[width={0.24\textwidth}]{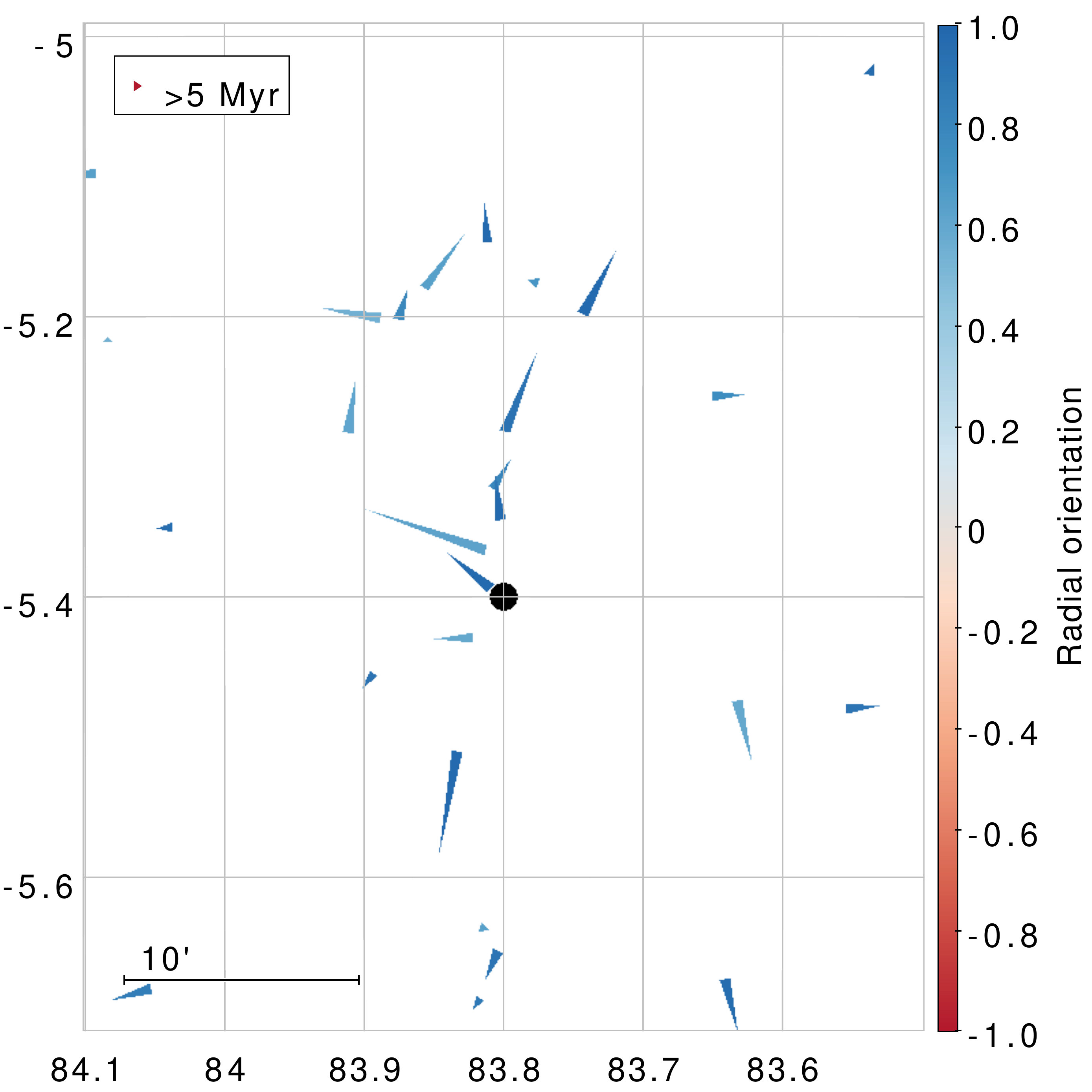}
\includegraphics[width={0.24\textwidth}]{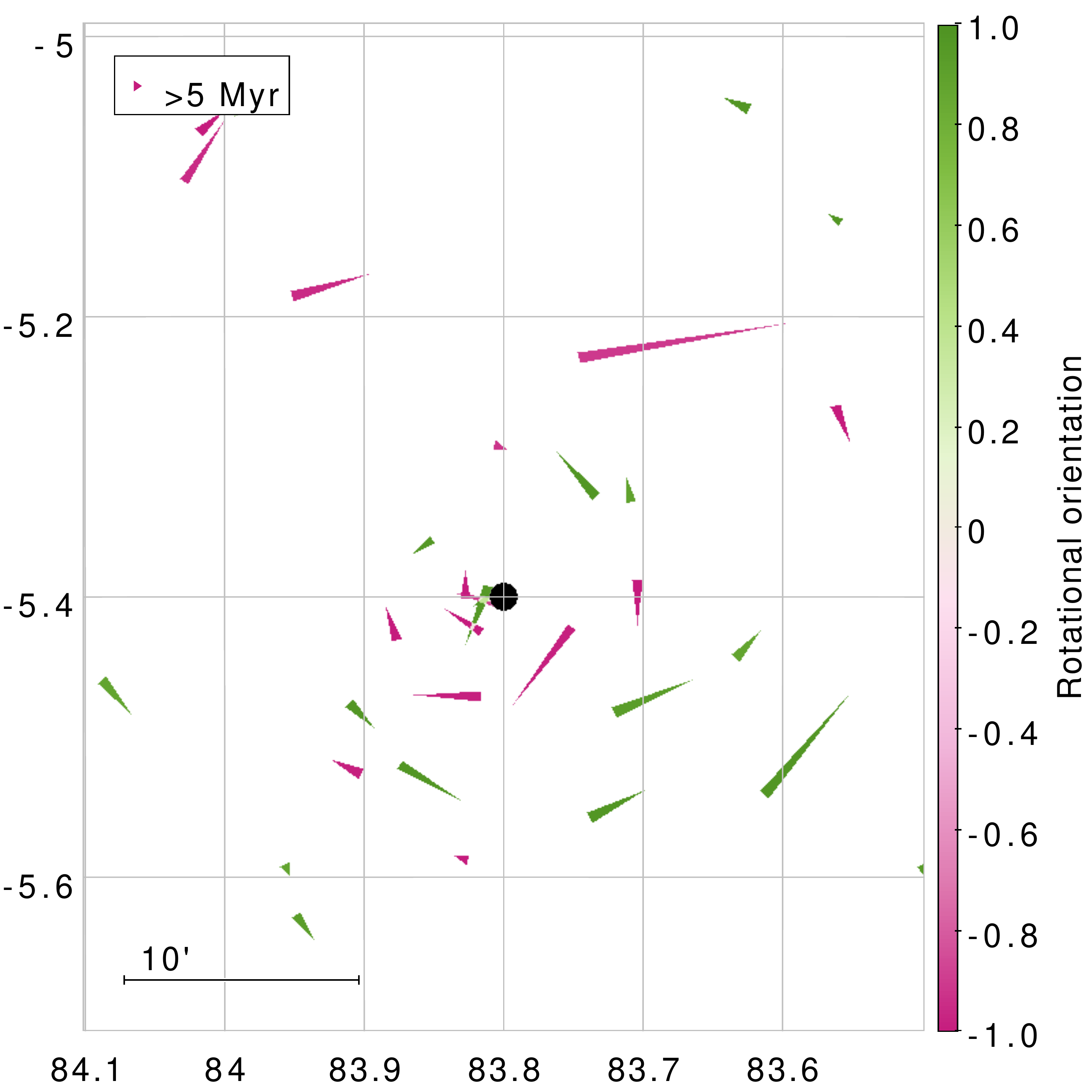}

\includegraphics[width={0.24\textwidth}]{ra.pdf}
\includegraphics[width={0.24\textwidth}]{ra.pdf}
\includegraphics[width={0.24\textwidth}]{ra.pdf}
\includegraphics[width={0.24\textwidth}]{ra.pdf}
\caption{Same as Figure \ref{fig:onc}, but zoomed in on Trapezium. The length of the vectors is decreased to the distance covered in 0.1 Myr.
\label{fig:trap}}
\end{figure*}

\subsection{Infall}

\begin{figure*}
\includegraphics[width={0.012\textwidth}]{dec.pdf}
\includegraphics[width={0.3\textwidth}]{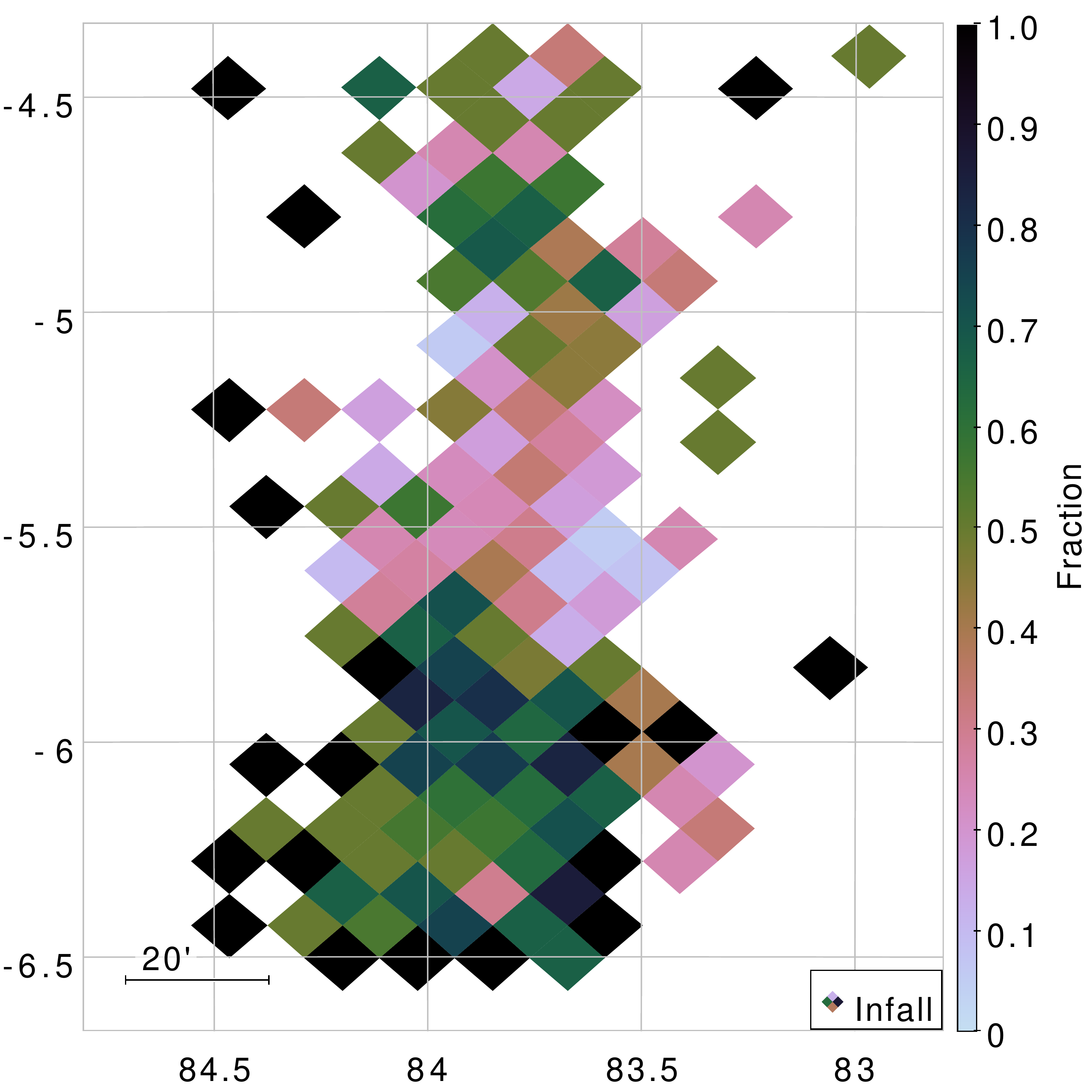}
\includegraphics[width={0.3\textwidth}]{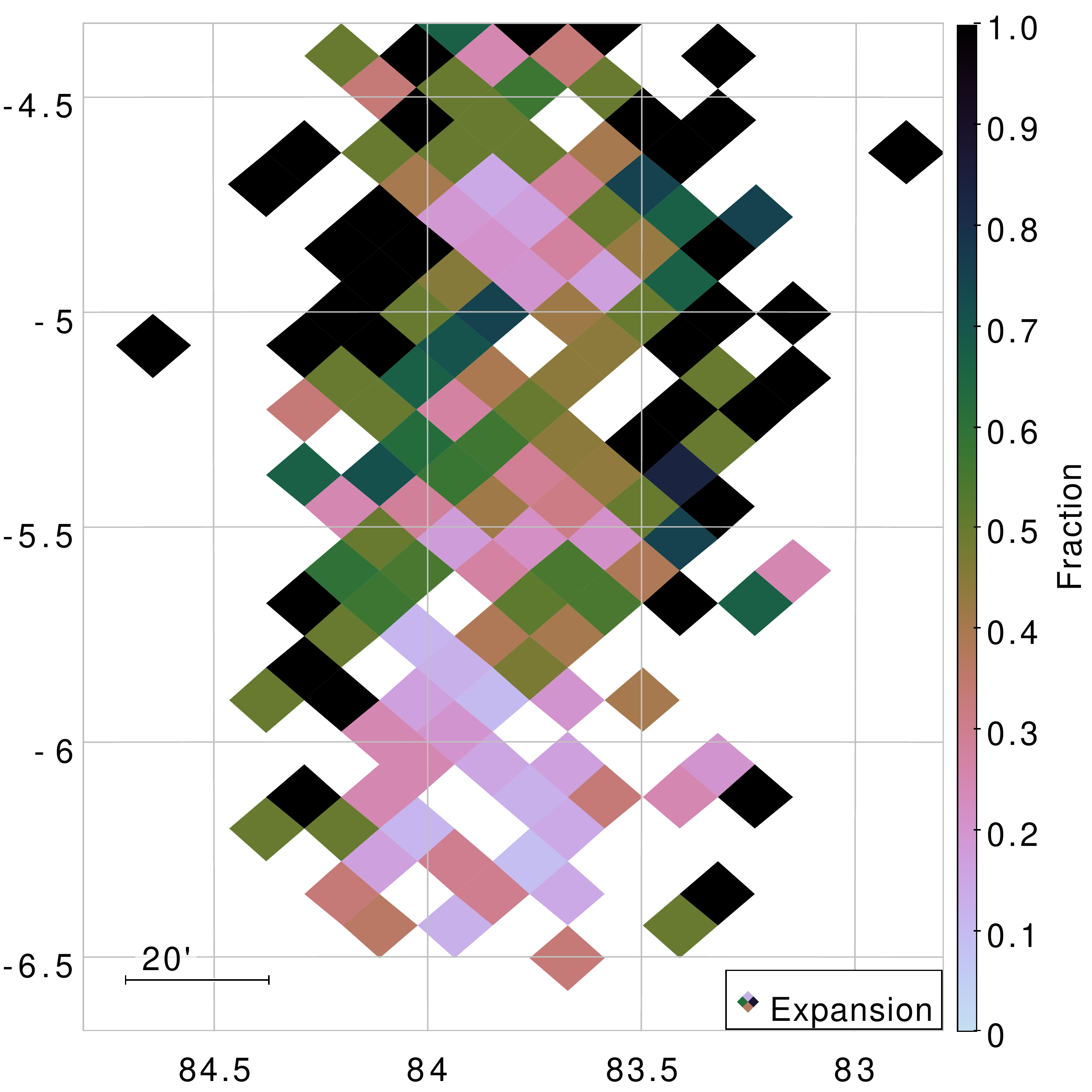}
\includegraphics[width={0.3\textwidth}]{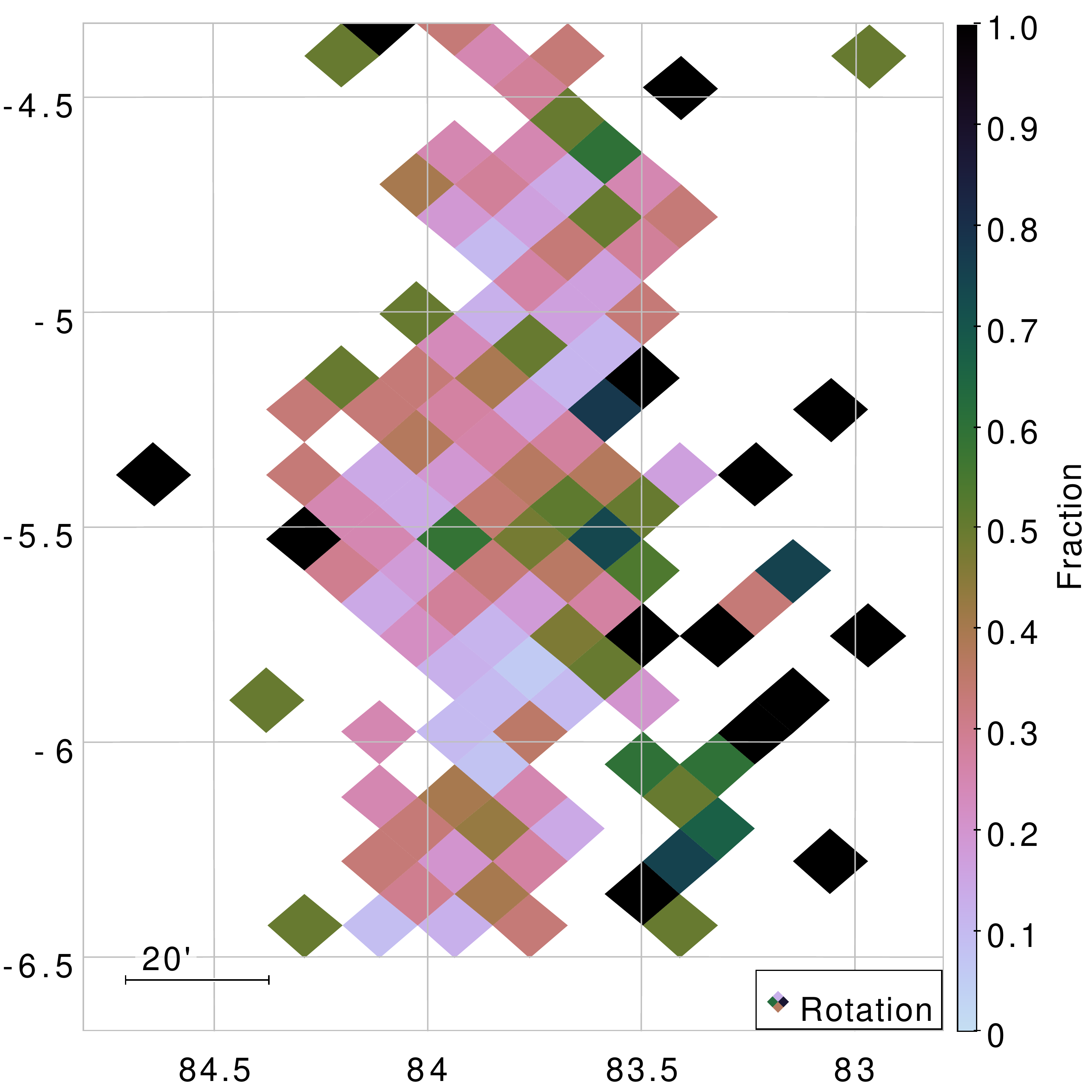}

\includegraphics[width={0.012\textwidth}]{dec.pdf}
\includegraphics[width={0.3\textwidth}]{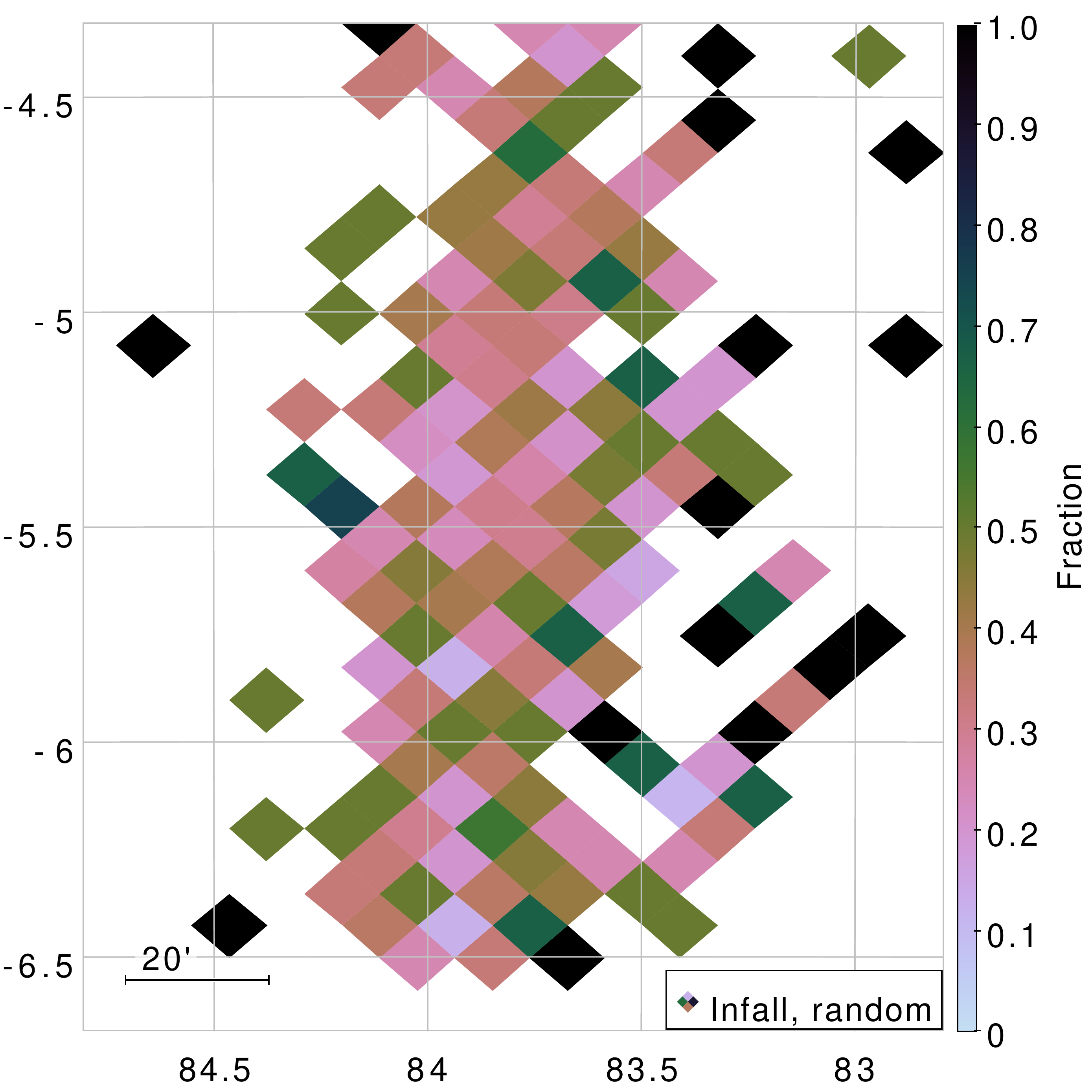}
\includegraphics[width={0.3\textwidth}]{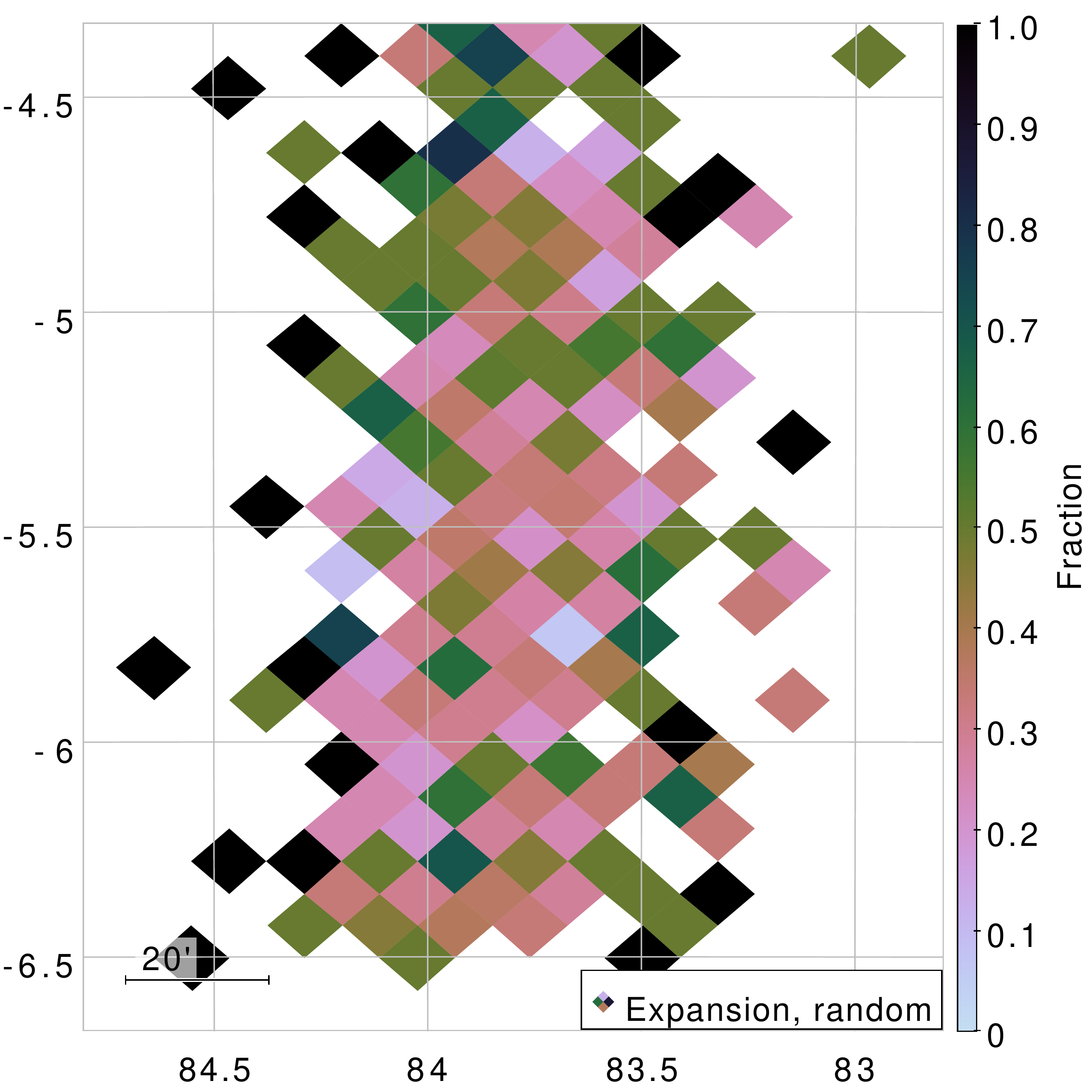}
\includegraphics[width={0.3\textwidth}]{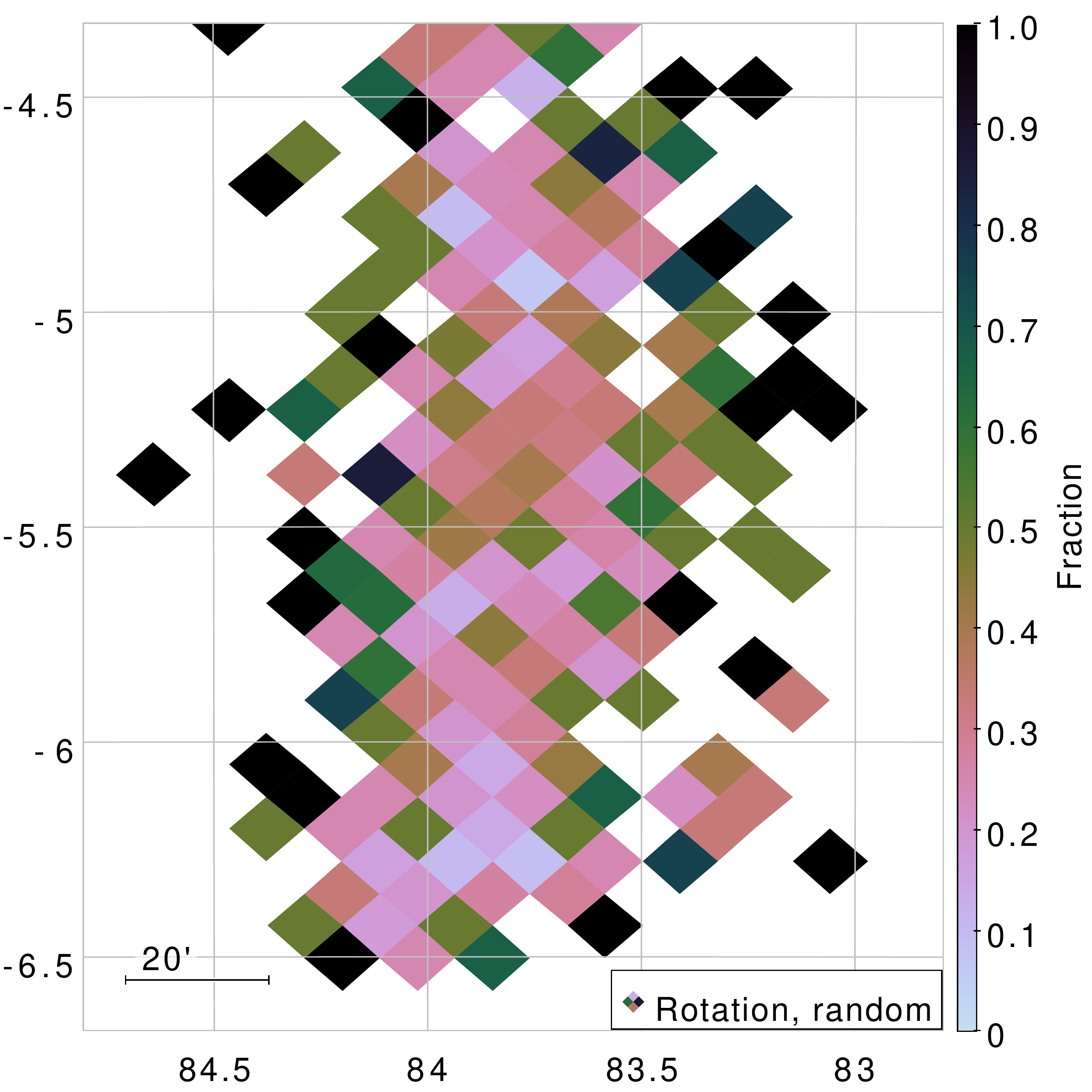}

\includegraphics[width={0.3\textwidth}]{ra.pdf}
\includegraphics[width={0.3\textwidth}]{ra.pdf}
\includegraphics[width={0.3\textwidth}]{ra.pdf}
\caption{Top: Map of the ONC showing fraction of sources in a given healpix that are preferentially falling into the center of the ONC, expanding outwards, and those that are preferentially moving tangentially around it. Bottom: same as above, but the velocities for each star have been generated randomly from a Gaussian distribution representing the cluster, to highlight the differences between the real data and the null hypothesis.
\label{fig:fraction}}
\end{figure*}

Figures \ref{fig:onc} and \ref{fig:trap} show the motions of infalling, expanding, and rotating sources in each of the five age bins. The stars in the oldest age bin ($\gtrsim$5 Myr) predate the formation of the central cluster as a whole. Although there are stars that fall into various cuts of $\cos\theta$, there is no organized expansion or rotation to speak of, they appear to primarily trace random motions of stars.

However, at the oldest ($\gtrsim$5 Myr) age bin, there does appear to be a significant infall of stars moving towards the Trapezium from south of the ONC, in the vicinity of NGC 1980 (we note that each panel shows the current proper motions of the stars of that age, not necessarily the motions at the time of their formation). This infall becomes even more prominent in the next age bin (3$\sim$5 Myr); furthermore, there does appear to be significant infall from north the northern part of the ONC, NGC 1977. This infall continues to persist in the younger age bins as well, until the density of sources in these regions deplete, as NGC 1980 and 1977 had stopped actively forming stars a few Myr ago. The infalling sources constitute the bulk of sources found north and south of the Trapezium, however, there are next to no infalling sources from either east or west of it (Figure \ref{fig:fraction}, top row). While we do not have proper motions for the molecular gas, given the similarity between radial velocity of the gas and stars that is often observed \citep[e.g.,][]{da-rio2017}, this suggests that the filament that has produced the ONC is contracting due to the gravitational potential of both the Trapezium and the entire cluster as a whole.

Given that the the central cluster in the ONC several times more massive and more than an order of magnitude denser than either NGC 1980 or 1977 \citep[e.g.,][]{megeath2016}, its potential provides the singularly most dominant force in the region. However, it is difficult to state definitively when the infall has begun: whether it has started when the cluster has accreted a large fraction of its mass, whether it was when it started forming its first stars, or whether it predates their formation and originated in the gas. However, given the dynamics of other stars in the Orion Complex \citep{kounkel2018a}, and the known sources of stellar feedback in the region (e.g., winds from $\theta^1$ Ori C), it is unlikely that anything other than gravity would be responsible for the observed infall in these regions.

In contrast, if we examine the sample with randomly generated proper motions, (Figure \ref{fig:fraction}, bottom), the distribution of sources that would have orientation of their proper motions suggestive of either infall, outflow, or rotation are expected to be more homogeneous and uniform across the cluster. Comparing the distribution of fractions of stars that are infalling per healpix (i.e., distribution shown in Figure \ref{fig:fraction}, considering a set of values across all healpix within a given range of $\delta$) in the real data vs the random sample via a two-sided KS test shows that the two populations are distinct at $>4\sigma$ level. The difference becomes more pronounced when segregating the ONC into 3 portions: the top ($\delta\gtrsim -5^\circ$), middle ($-5.8^\circ\lesssim\delta\lesssim-5^\circ$), and bottom ($\delta\lesssim-5.8^\circ$). The resulting KS-test statistics are shown in Table \ref{tab:ks}.

We do note that ONC is not an isolated system, and that it is all can also be affected by the gravity of the larger Orion A molecular cloud, as well as gravity of the Orion Complex as a whole, and these forces could also influence the dynamics of the ONC. However, on the scales of a few pc (which is the scale over which we observe the infall), self gravity of the cluster would dominate over the gravity of the extended structure.

\begin{table}
	\centering
	\caption{KS test statistics comparing the fraction of sources that are expanding, infalling, or rotating, in the ONC compared to the random sample. Asterisks highlight the instances that appear to be most significantly discrepant at $>3\sigma$.
\label{tab:ks}}
	\begin{tabular}{cccc} 
		\hline
		 & Infall & Expansion & Rotation\\
		\hline
Full & 1.31e-05 (4.3$\sigma^*$) & 0.030 (2.1$\sigma$) & 0.110 (1.6$\sigma$) \\
Top & 0.00600 (2.7$\sigma$) & 0.180 (1.3$\sigma$) & 0.0870 (1.7$\sigma$)\\
Middle & 0.0775 (1.8$\sigma$) & 0.00255 (3.0$\sigma^*$) & 0.361 (0.9$\sigma$)\\
Bottom & 7.01e-07 (5.0$\sigma^*$) & 0.0362 (2.0$\sigma$) & 0.514 (0.6$\sigma$)\\
		\hline
	\end{tabular}
\end{table}

\subsection{Expansion}

Stars with age $<$5 Myr begin to trace the central overdensity associated with the Trapezium, and the younger they are, the more centrally concentrated they become. Around the cluster core (including along the OMC 2/3 filament), there is little to no ``organized'' infall of stars towards the center. However, as soon as the central cluster begins to develop, there is a significant signature of both rotation and expansion, overshadowing the infalling sources that are there. This is consistent with the simulations, where infall is easier to definitively observe at wider separations from a massive cluster than in its immediate proximity \citep{kuznetsova2015}.

The bulk of the expanding sources can be traced back directly to the cluster center and they tend to be distributed spherically around it. However, the expanding sources dominate the sample around the outer edges on the east and the west side of the cluster due to this spherical distribution, as there is no overlap with the infalling sources that are only found along the filament (Figure \ref{fig:fraction}). That preference for the outer edges of the cluster to be dominated by expansion has previously been noted by \citet{kounkel2018a}, \citet{kuhn2019}, and \citet{getman2019}.

We do note that at the ages of 3--5 Myr, the sources that appear to move away from the cluster center have their motion preferentially due north, with very few sources moving due south, in contrast to the younger age bins where the expansion does appear to be more spherically symmetric, or in contrast to the sources moving towards the cluster center or those moving tangentially around it in the similar 3--5 age bin which also appear to be more uniformly distributed. It is unclear why such an asymmetry may have arisen, as there is no obvious bias in targeting of such sources in the underlying survey. Alternatively -- the mean cluster position and velocity were carefully selected using within the cluster as a whole, but there may have been a slight evolution in their precise placement at earlier epochs.

\subsection{Rotation}

\begin{figure}
\includegraphics[width=\columnwidth]{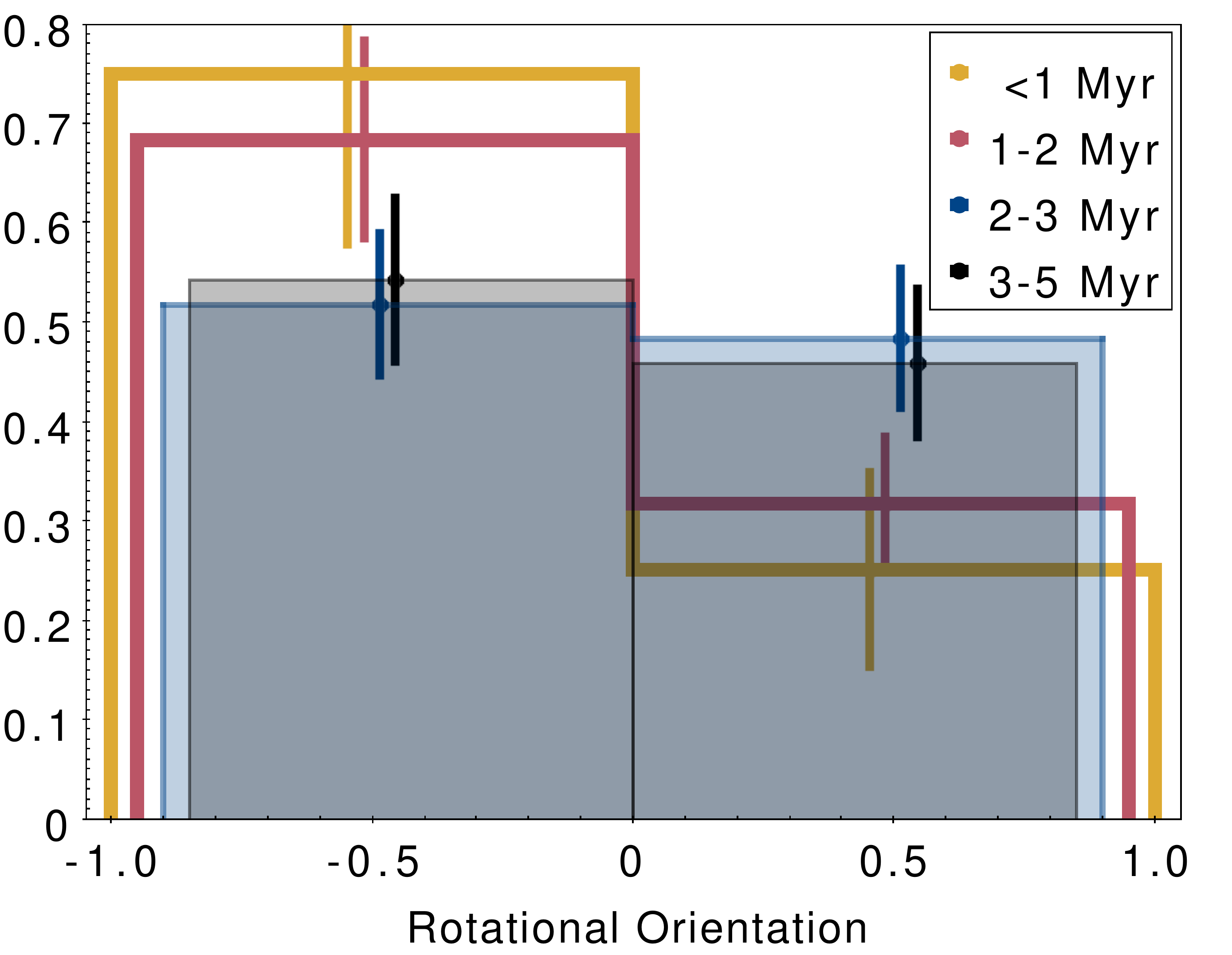}
\caption{Distribution of direction of motion of sources within 0.4$^\circ$ around the cluster center that preferentially exhibiting tangential motion. Orientation $<0$ shows the sources moving clockwise, $>0$ are those moving counterclockwise. Sources are separated into 4 age bins.
\label{fig:trapeziumrot}}
\end{figure}

Examining the rotation around the Trapezium, within 0.4$^\circ$ of the cluster center, in sources that are preferentially moving tangentially, there is little to no preference in the direction among the stars that are older than 2 Myr. However, there is a strong preference for clockwise motion in younger stars, by a factor of $\sim$2 among 1--2 Myr old stars, and by a factor of $\sim$3 among $<1$ Myr old stars (Figure \ref{fig:trapeziumrot}). This preference for clockwise motion in youngest stars does not depend on the precise point adopted as the cluster center around which rotation is evaluated. However, the excess is strongest when evaluated in the center of the Trapezium, and it weakens the further it is displaced.

It is difficult to say, however, if the older stars used to have a preferred orientation that has since been washed out over their lifetime through the dynamical evolution of the cluster, with the younger stars being the only ones with some memory of it, or if the angular momentum of the cluster has evolved over time as the cluster grew in such a manner as to develop a semi-coherent rotation only for the currently forming stars.

Rotation of the molecular gas in ONC has been previously predicted in the toy model of the formation of the cluster from \citet{hartmann2007}, and it is expected that the stars would inherit the angular momentum from the gas. A weak signature of rotation of the central cluster has also been reported by \citet{theissen2022}.

\begin{figure}
\includegraphics[width=\columnwidth]{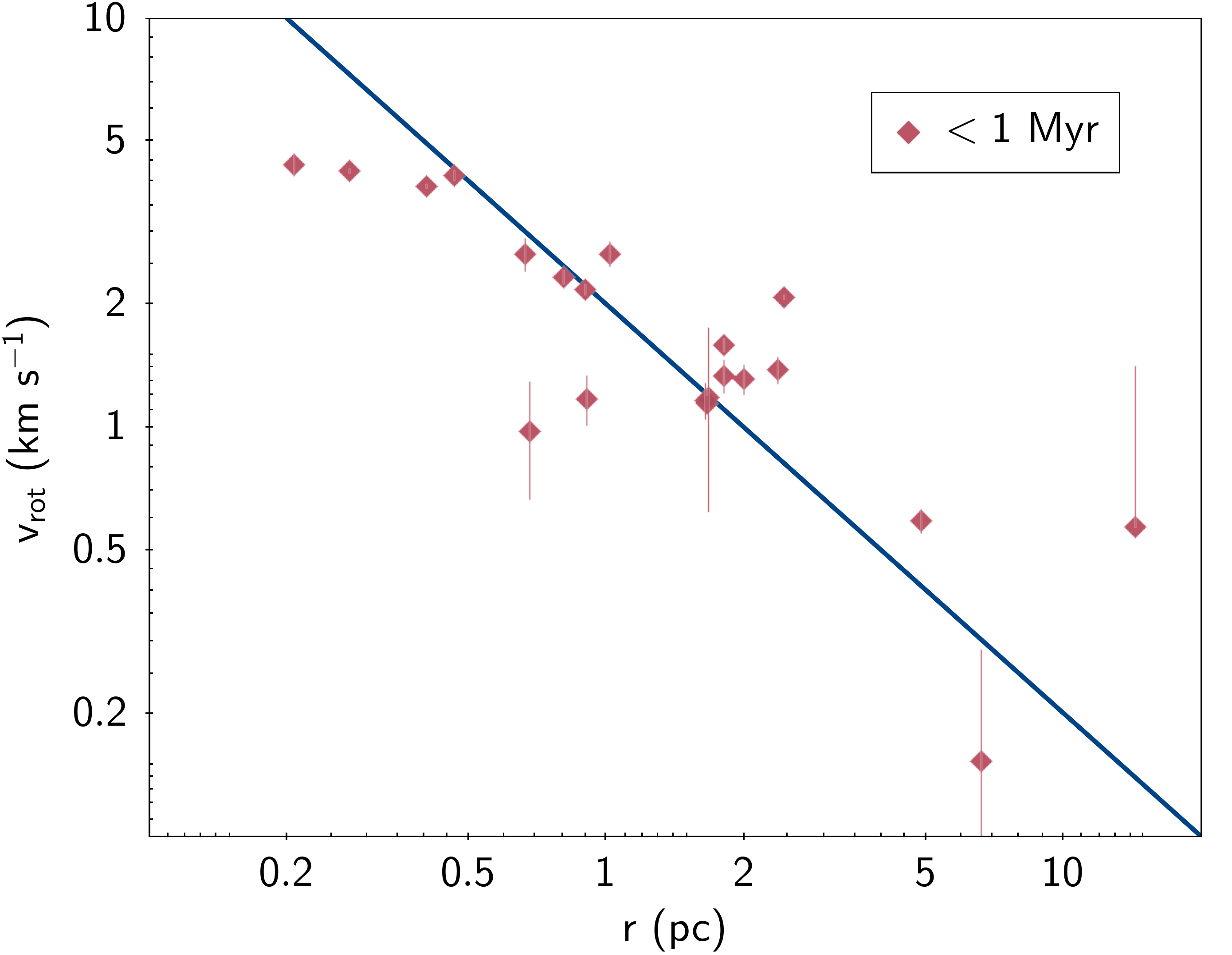}
\caption{Tangential velocity as a function of radial distance of the stars with age $<$1 Myr that are preferentialy rotating around the cluster center. The fitted blue line follows the relation of $v=2/r$. The uncertainties in $v_{rot}$ include propagated errors in proper motions, parallax, and angle; most are smaller than the symbol size.
\label{fig:momentum}}
\end{figure}

Among the youngest ($<$1 Myr) sources that are rotating around the cluster center, the specific angular momentum can be most easily seen to be conserved, with tangential velocity inversely proportional to the radius at which these stars are found (Figure \ref{fig:momentum}). From this constant $r\times v{rot}$, it is possible to estimate the specific angular momentum of $\sim7.4\times10^{20}$ cm$^2$ s$^{-1}$, with both $r$ and $v$ being 2d projections in the plane of the sky of these youngest sources rotating around the cluster. This specific angular momentum is a factor of $\sim$1.8 higher than the specific angular momentum of the Orion B molecular cloud \citep{hsieh2021}. However, in the older stars (including 1--2 Myr), the coherence between $v_{rot}$ and $r$ is no longer apparent, that is to say, $v$ has a significant scatter relative to $r$ which washes away any trends.

\subsection{Radial velocity component}

\begin{figure*}
\includegraphics[width={0.245\textwidth}]{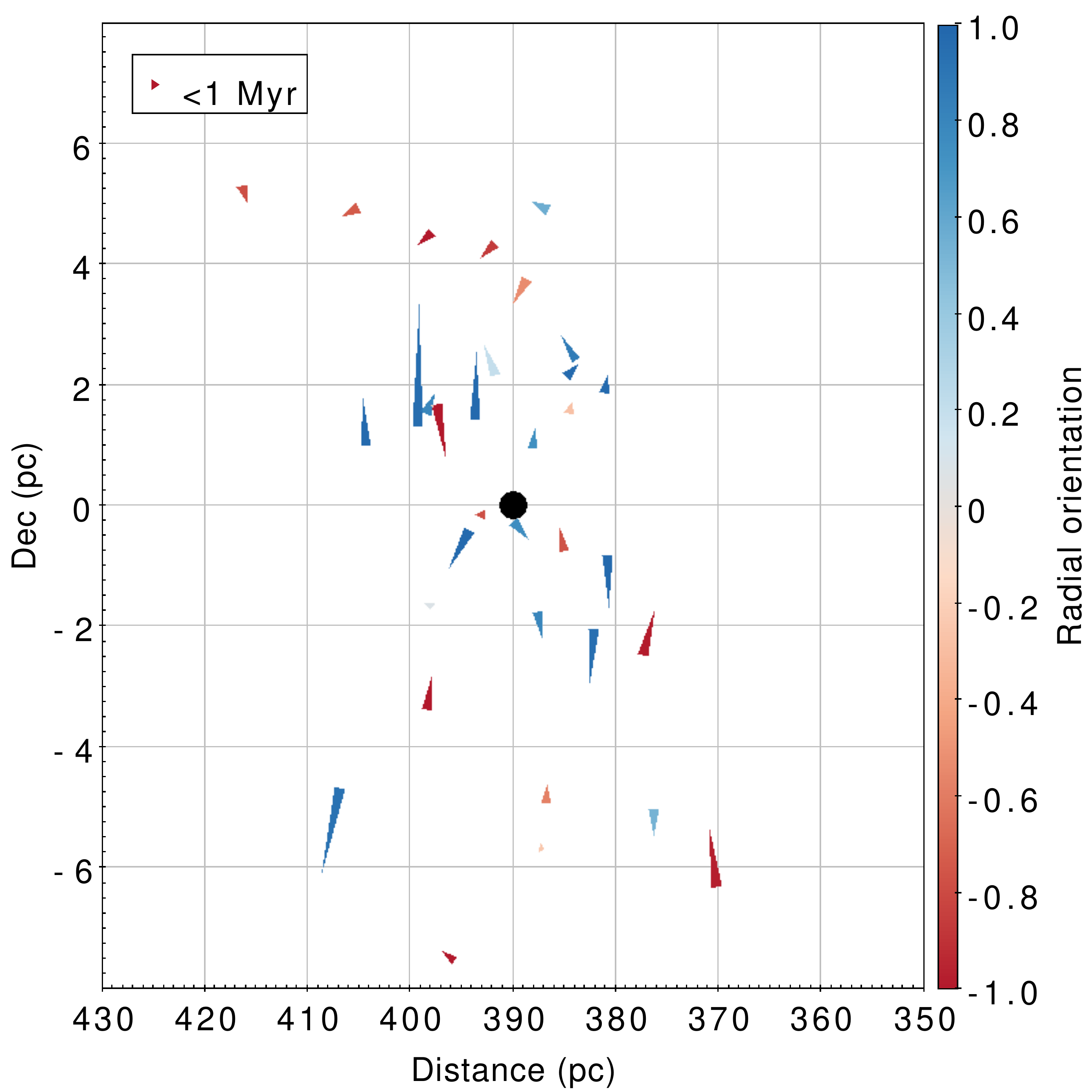}
\includegraphics[width={0.245\textwidth}]{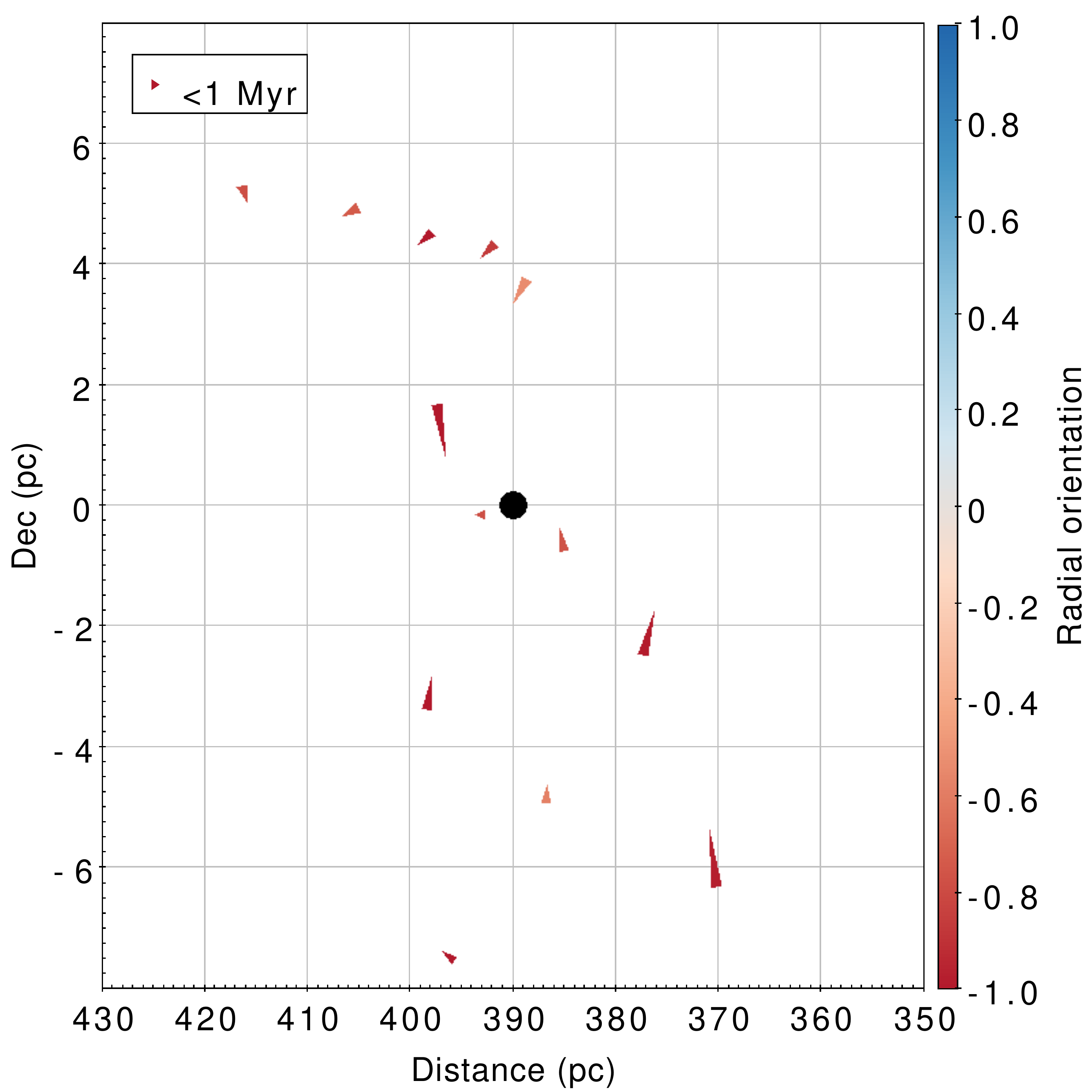}
\includegraphics[width={0.245\textwidth}]{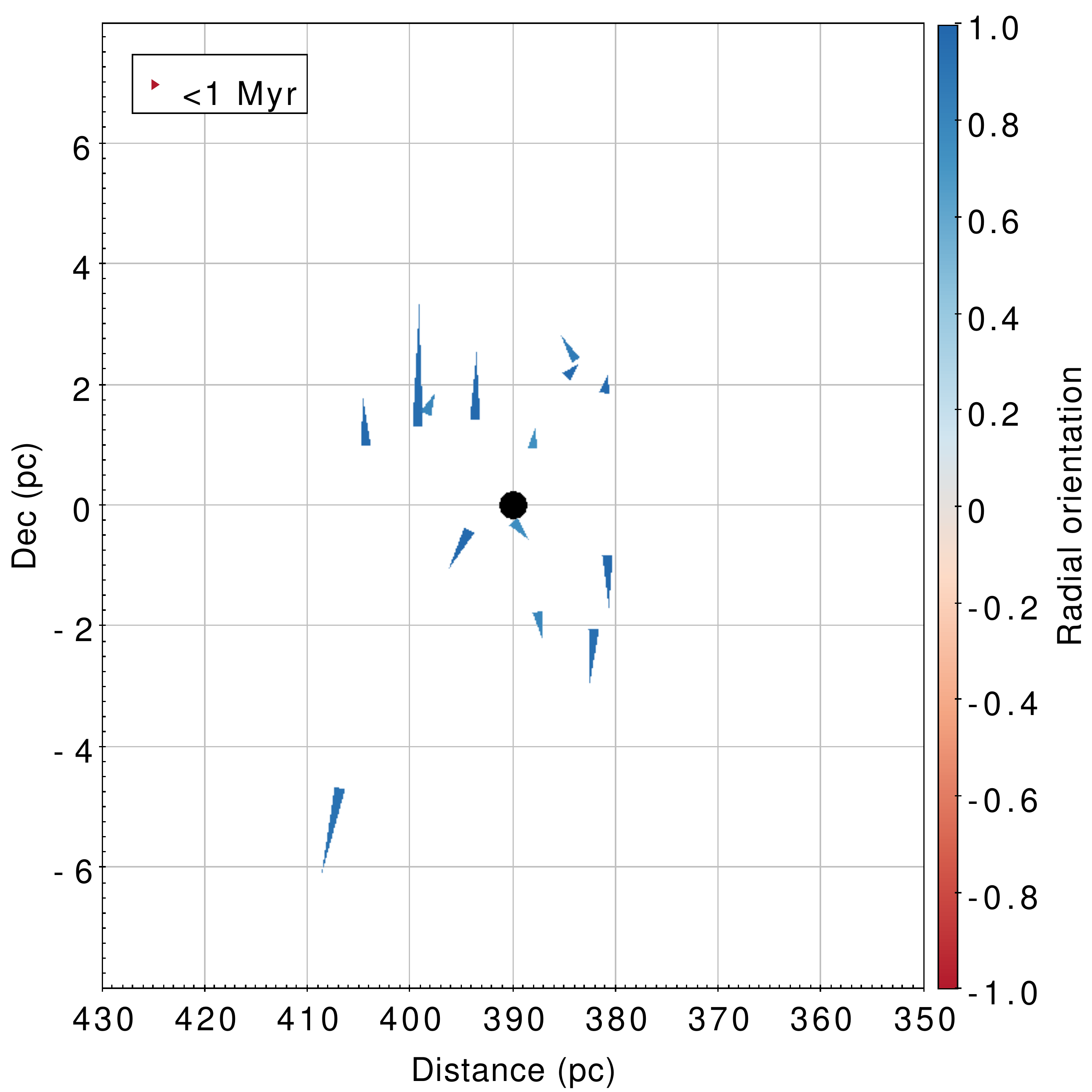}
\includegraphics[width={0.245\textwidth}]{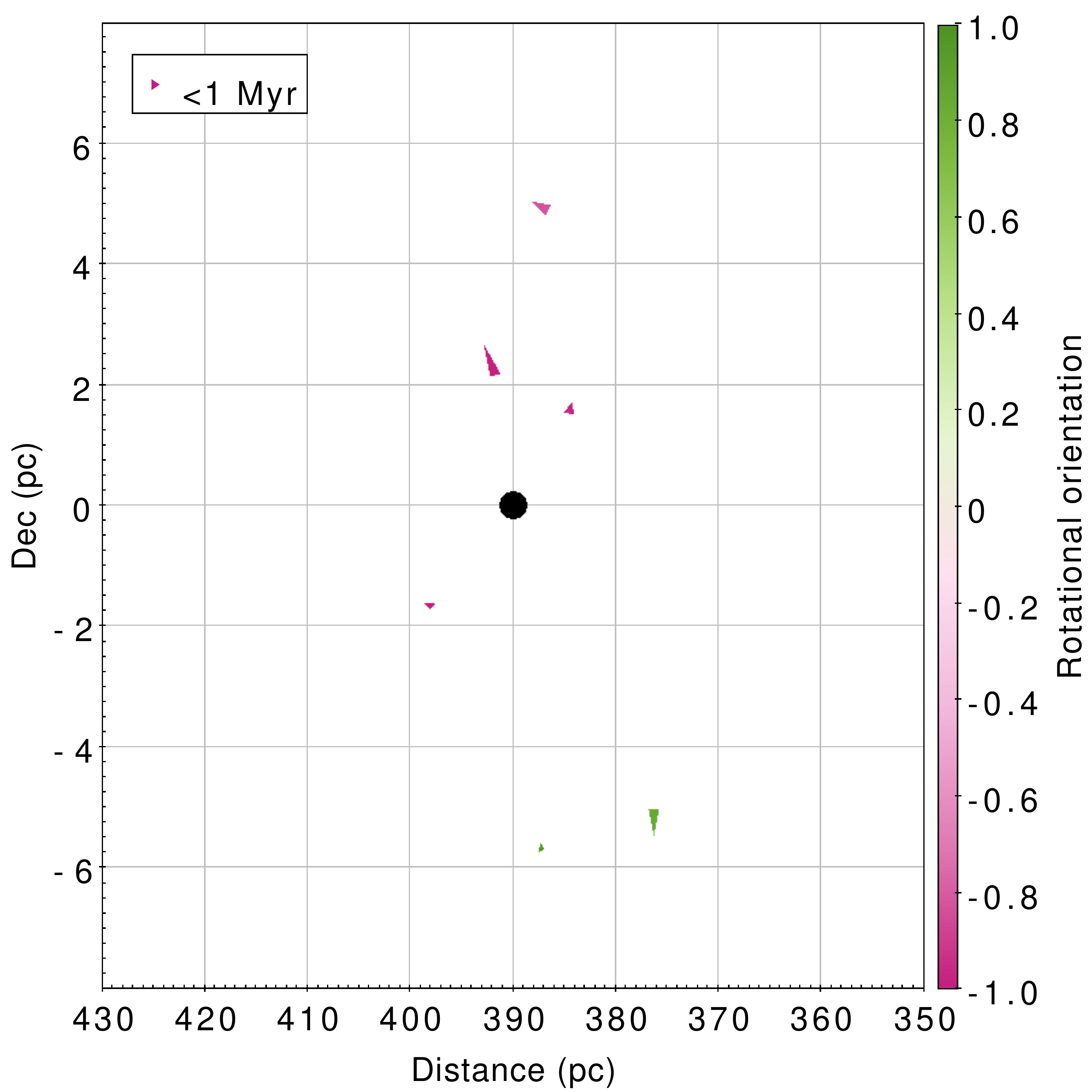}

\includegraphics[width={0.245\textwidth}]{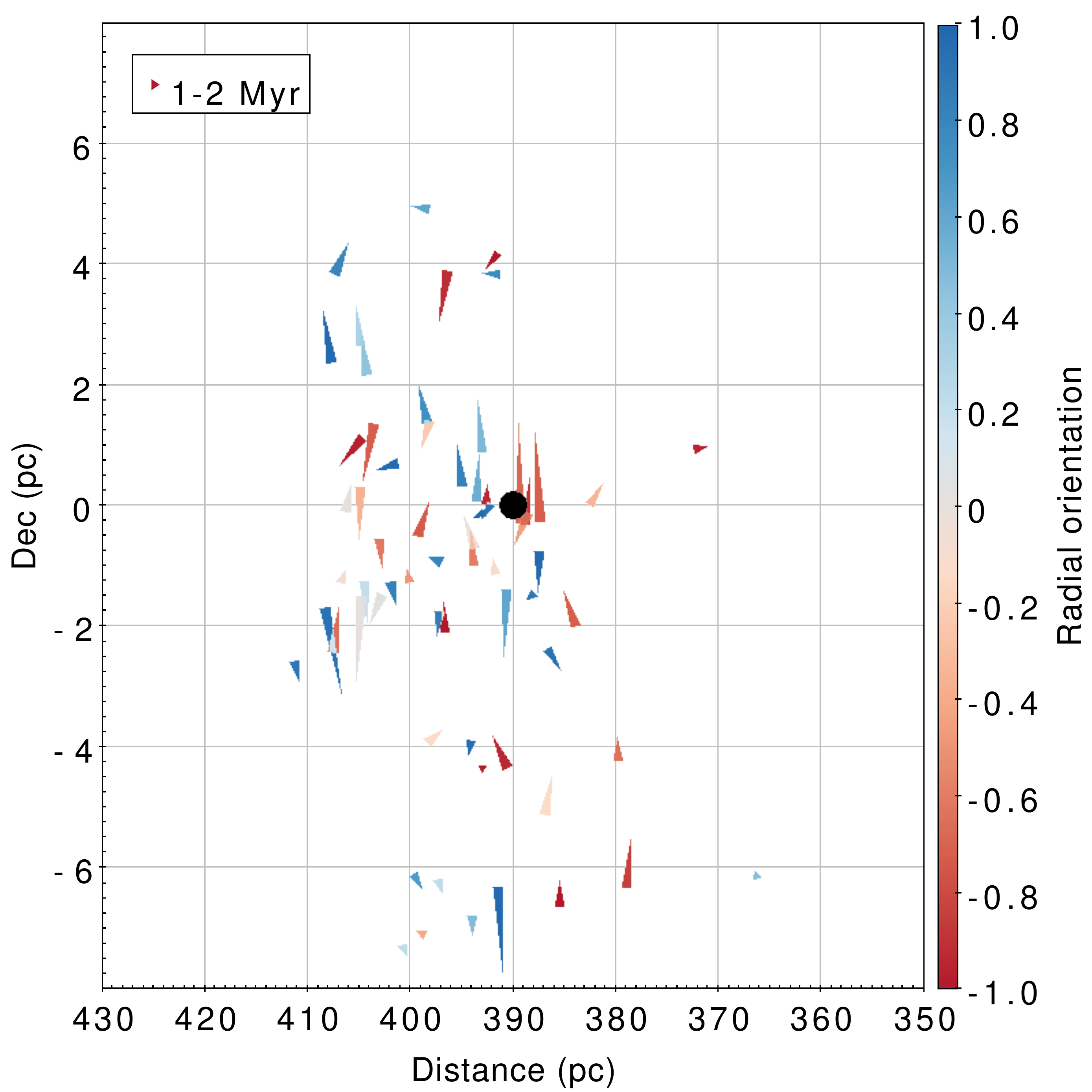}
\includegraphics[width={0.245\textwidth}]{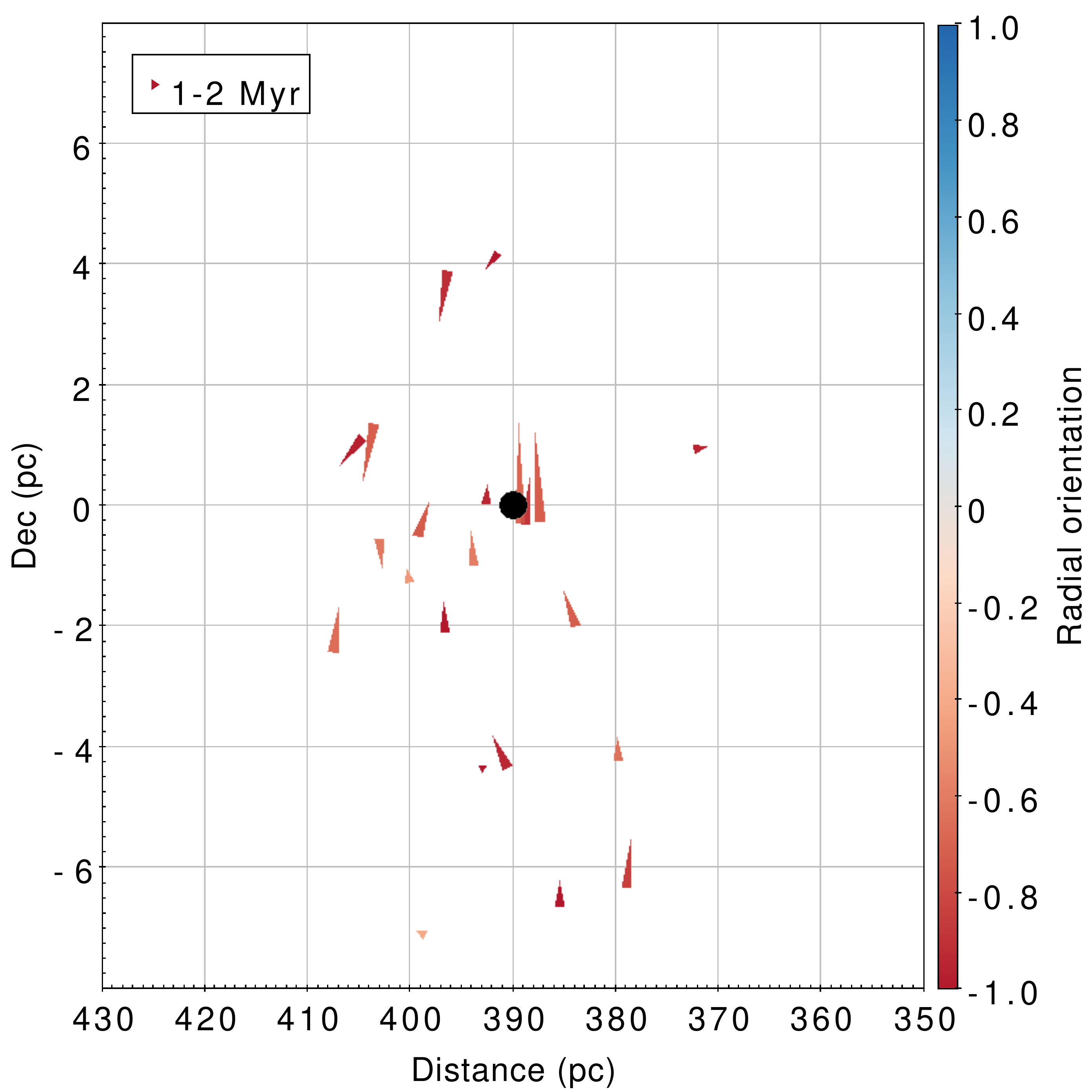}
\includegraphics[width={0.245\textwidth}]{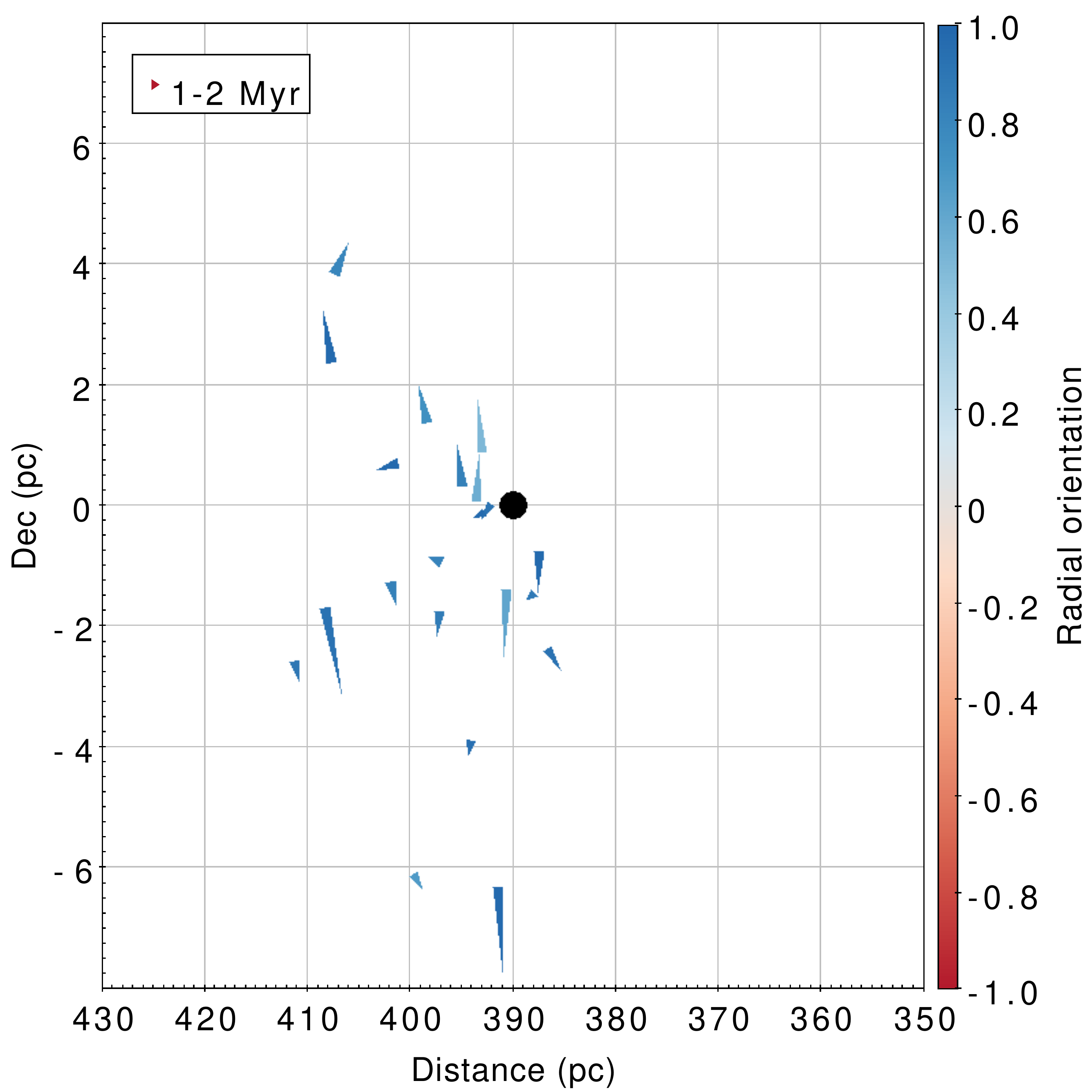}
\includegraphics[width={0.245\textwidth}]{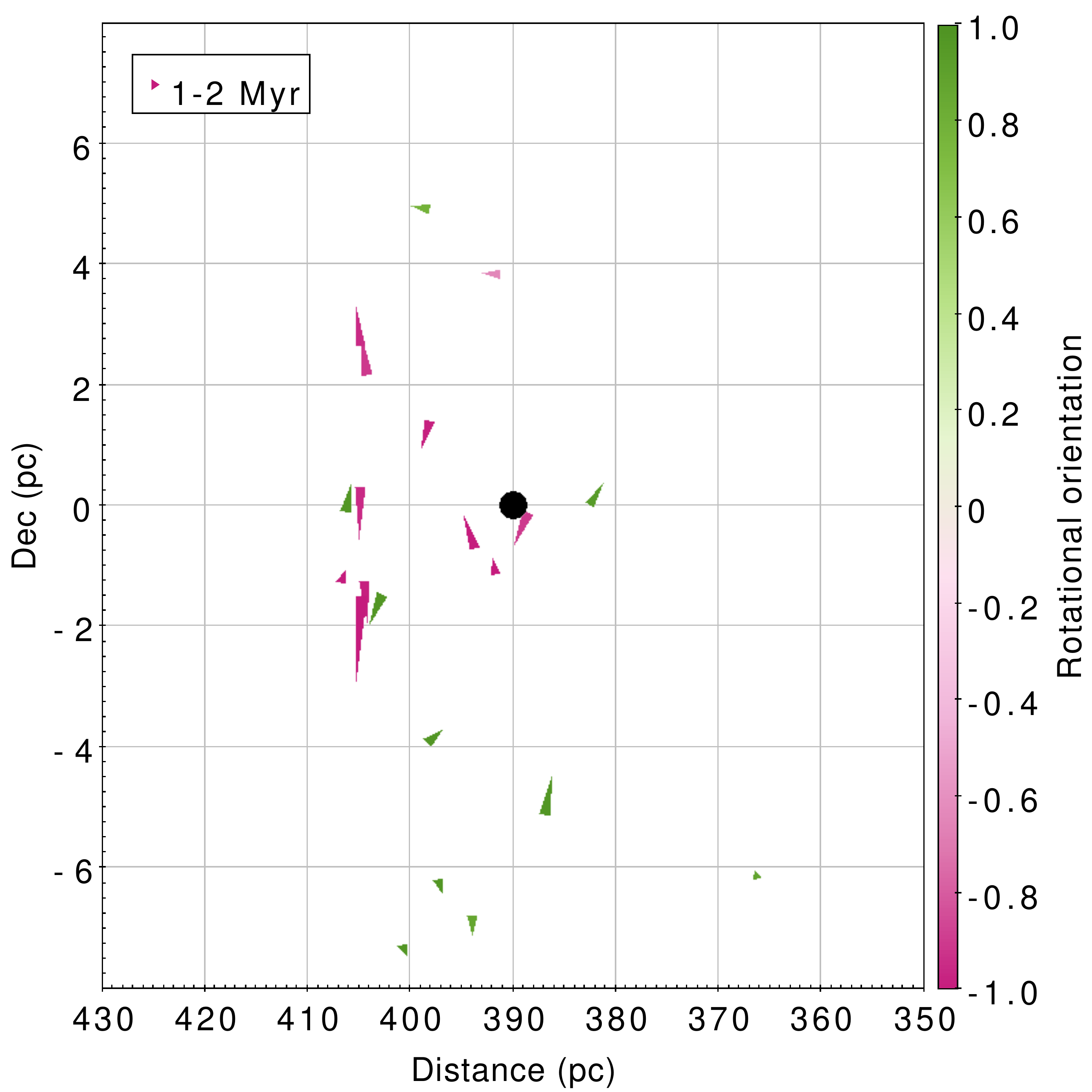}

\includegraphics[width={0.245\textwidth}]{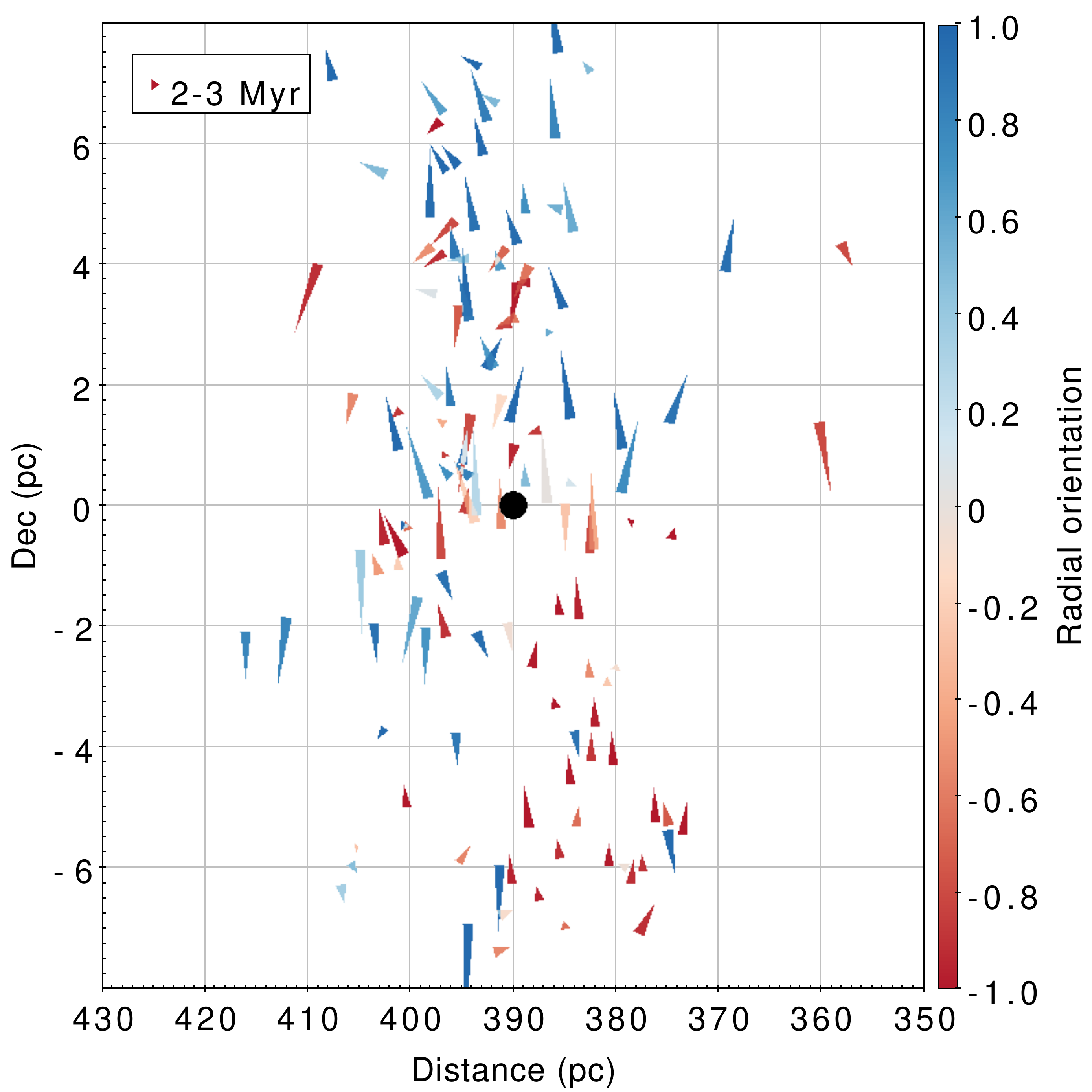}
\includegraphics[width={0.245\textwidth}]{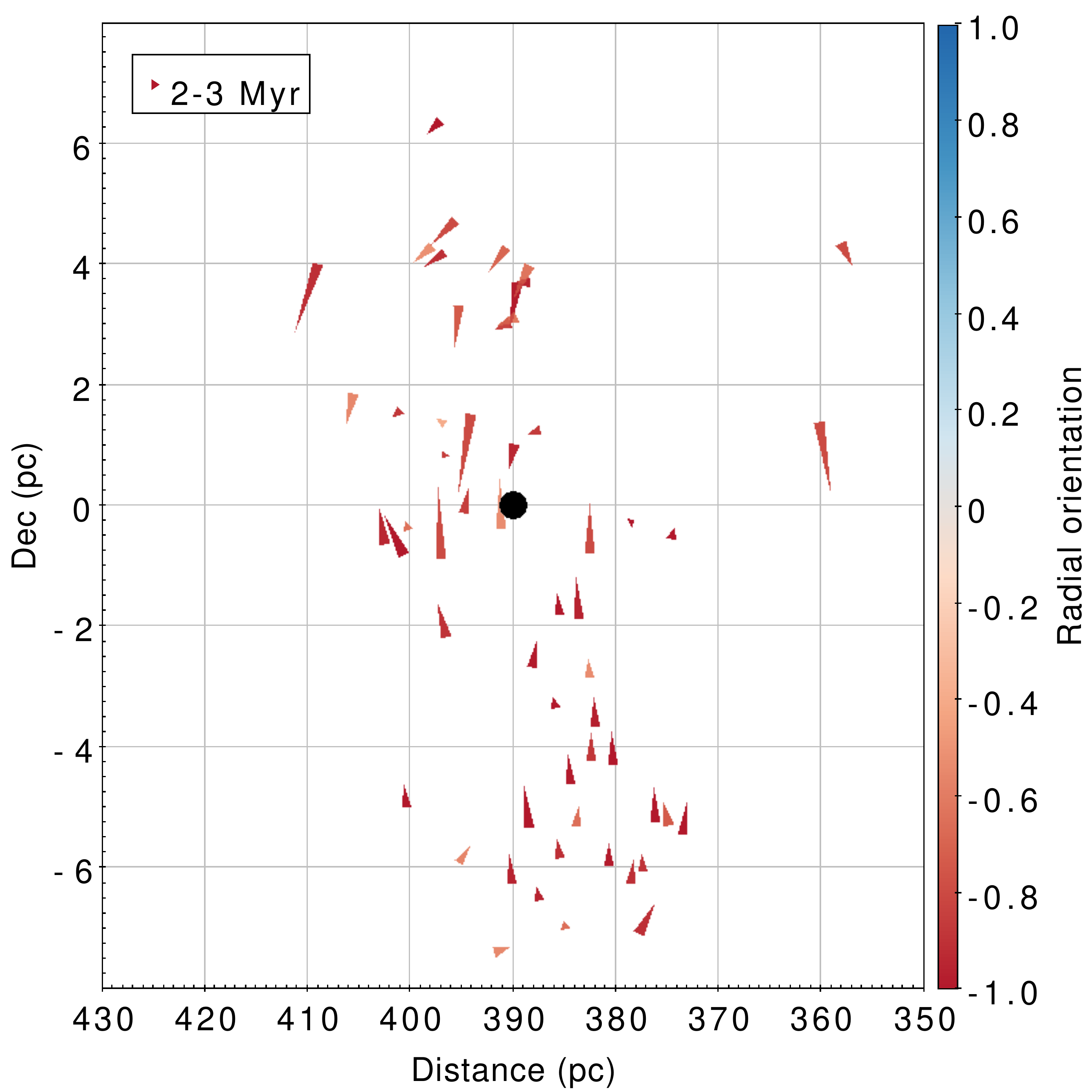}
\includegraphics[width={0.245\textwidth}]{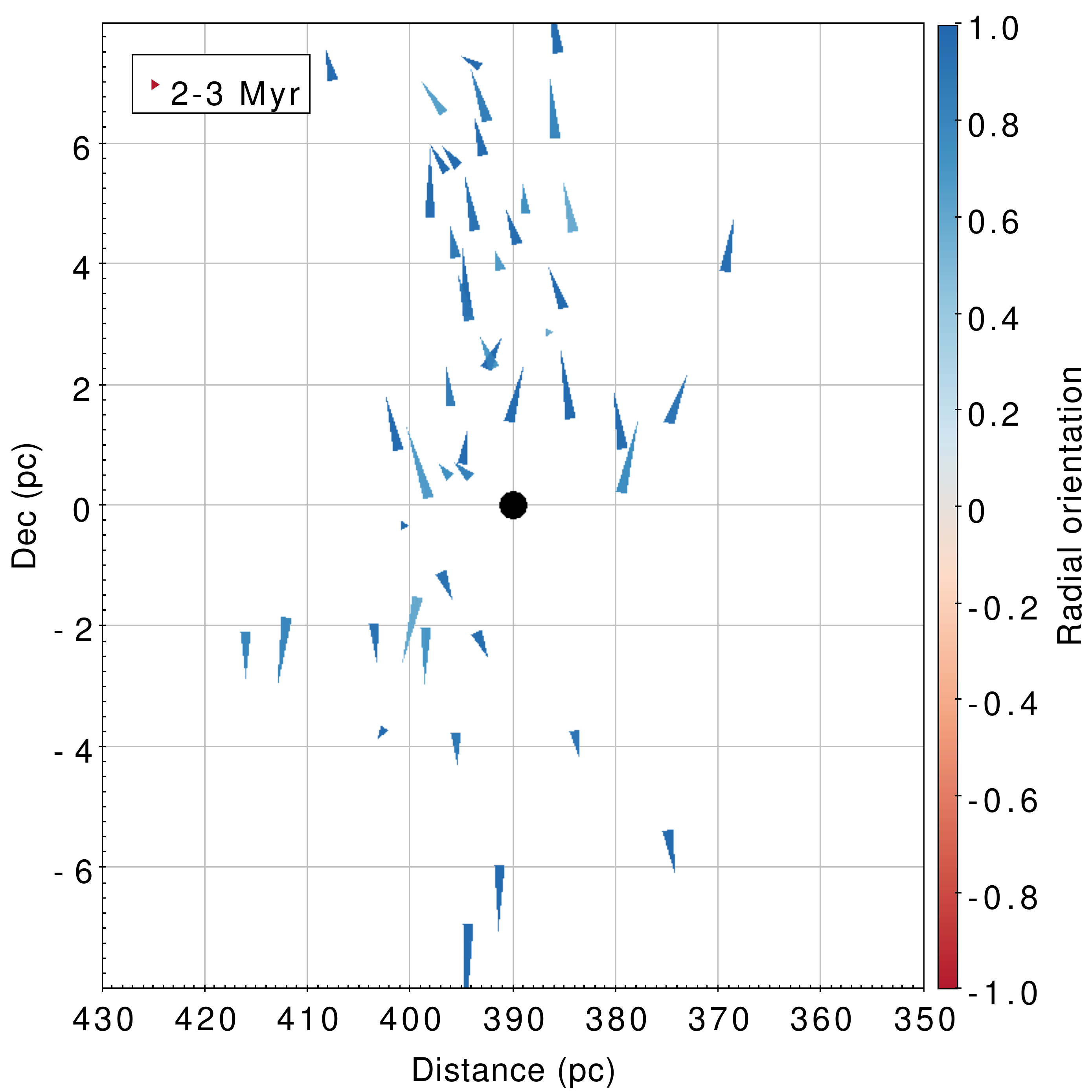}
\includegraphics[width={0.245\textwidth}]{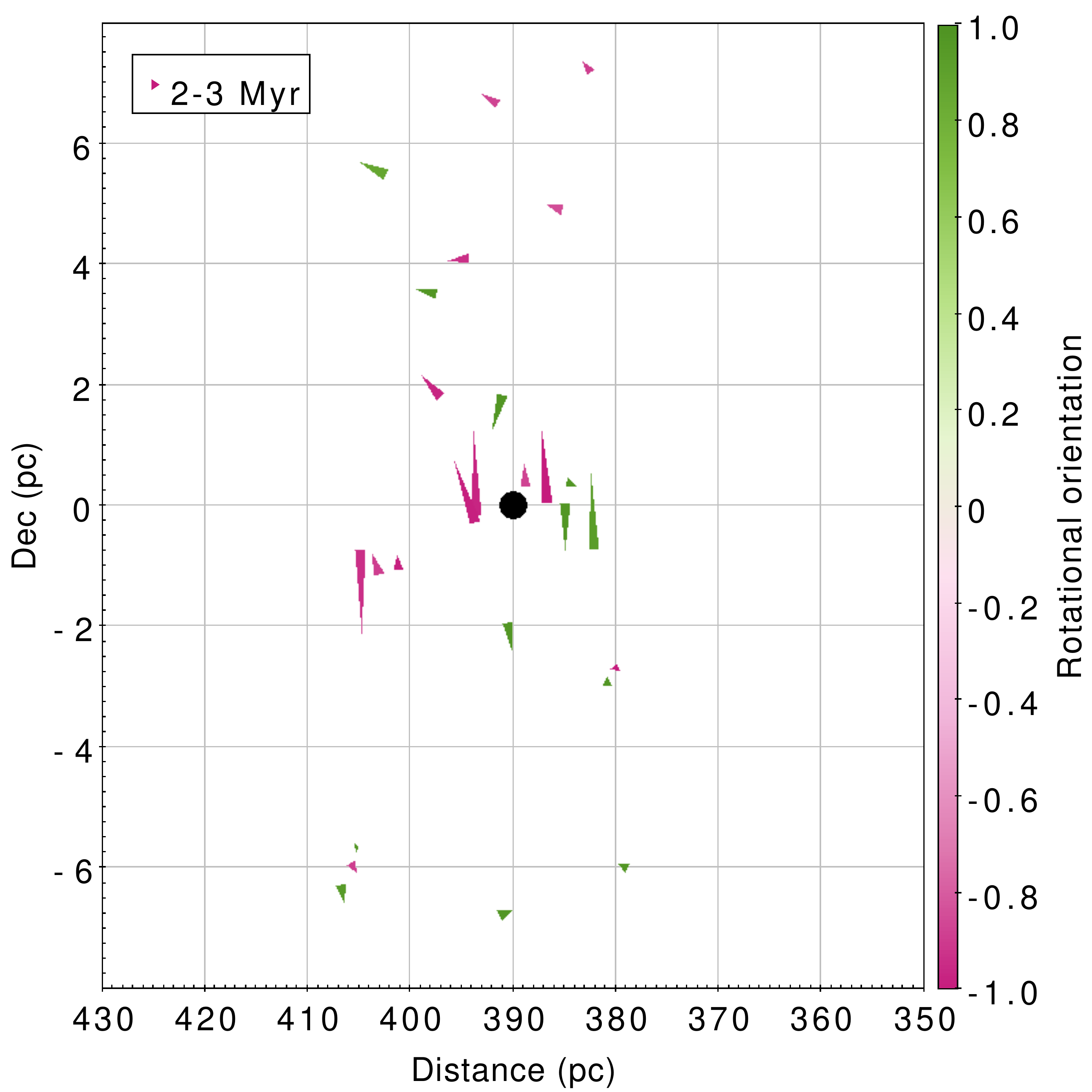}

\includegraphics[width={0.245\textwidth}]{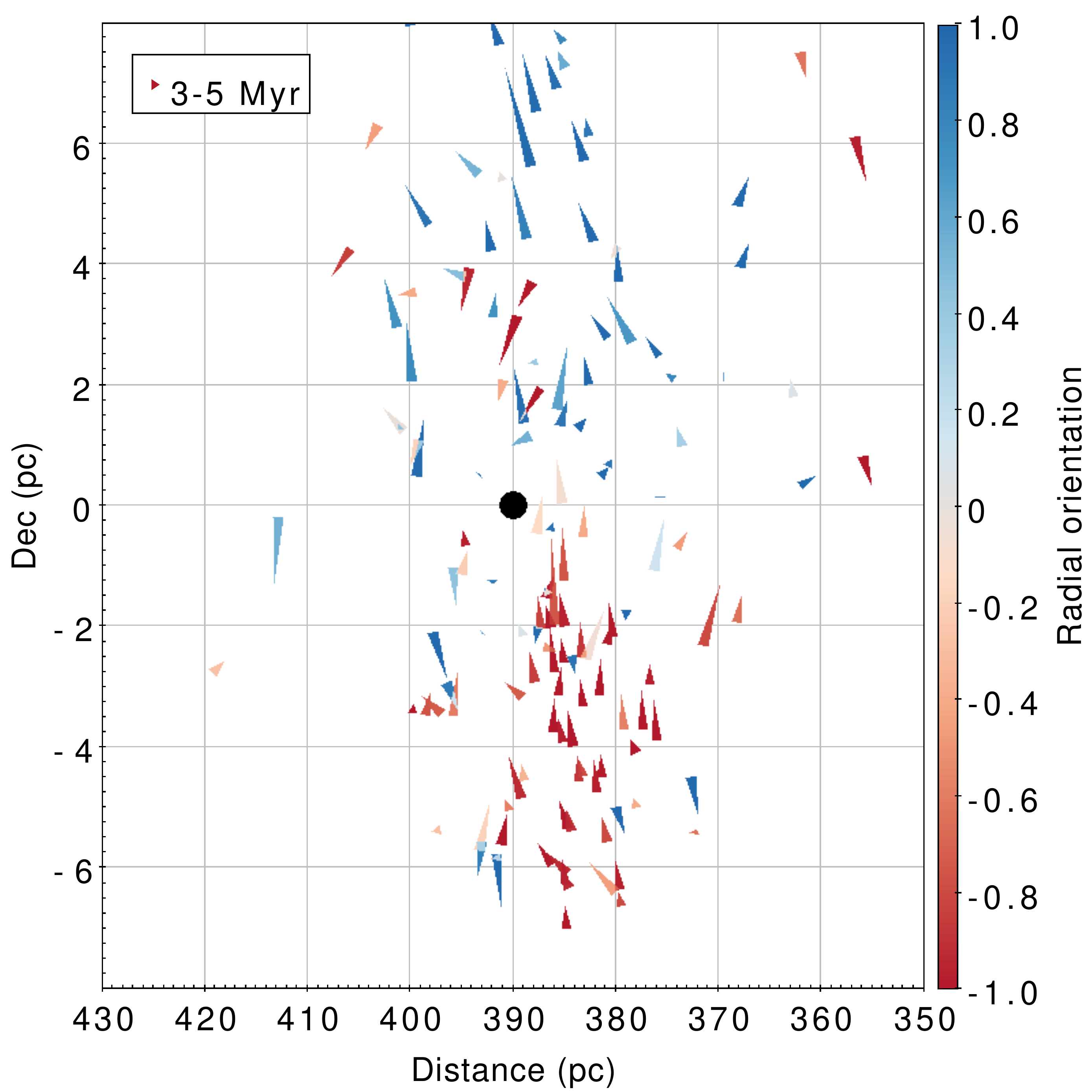}
\includegraphics[width={0.245\textwidth}]{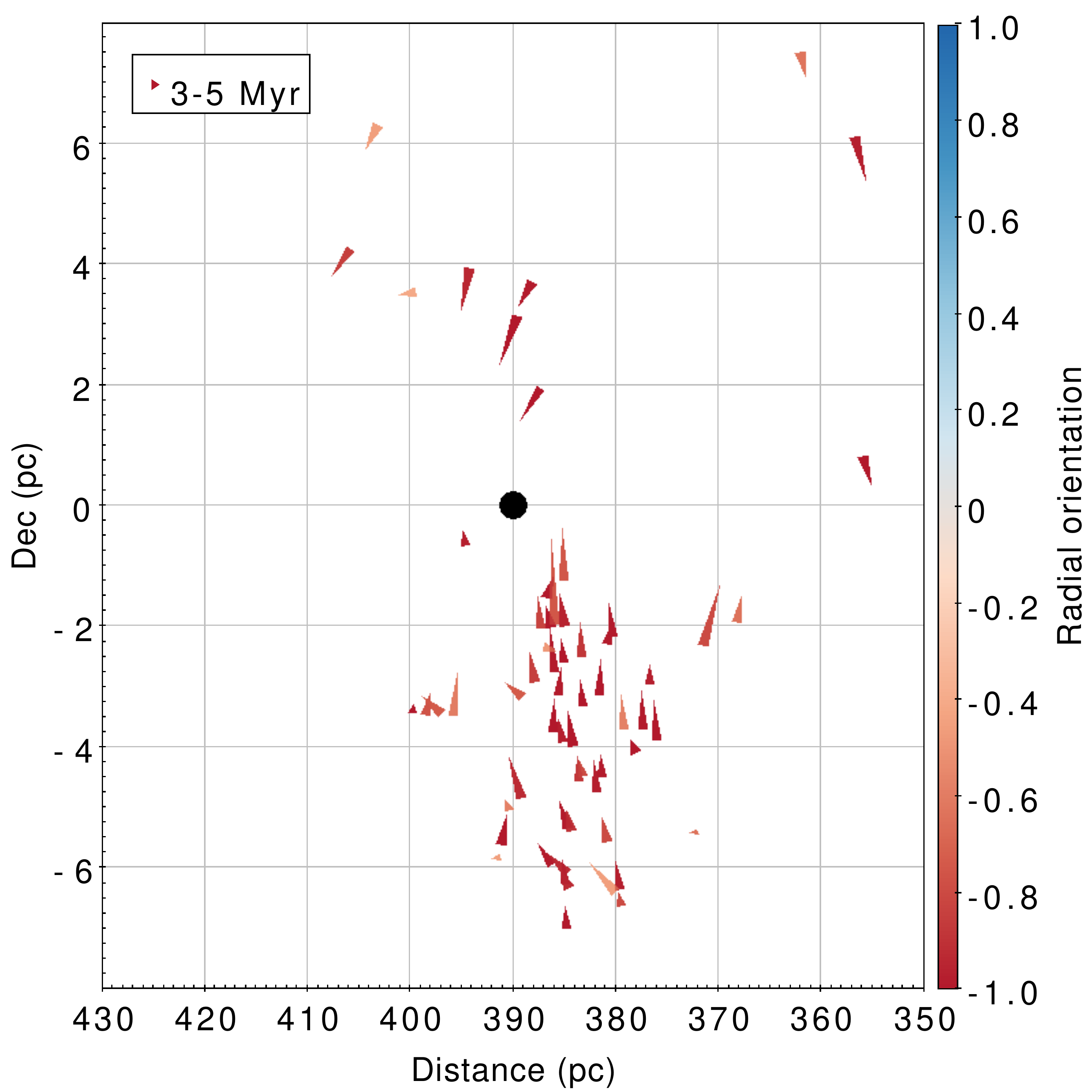}
\includegraphics[width={0.245\textwidth}]{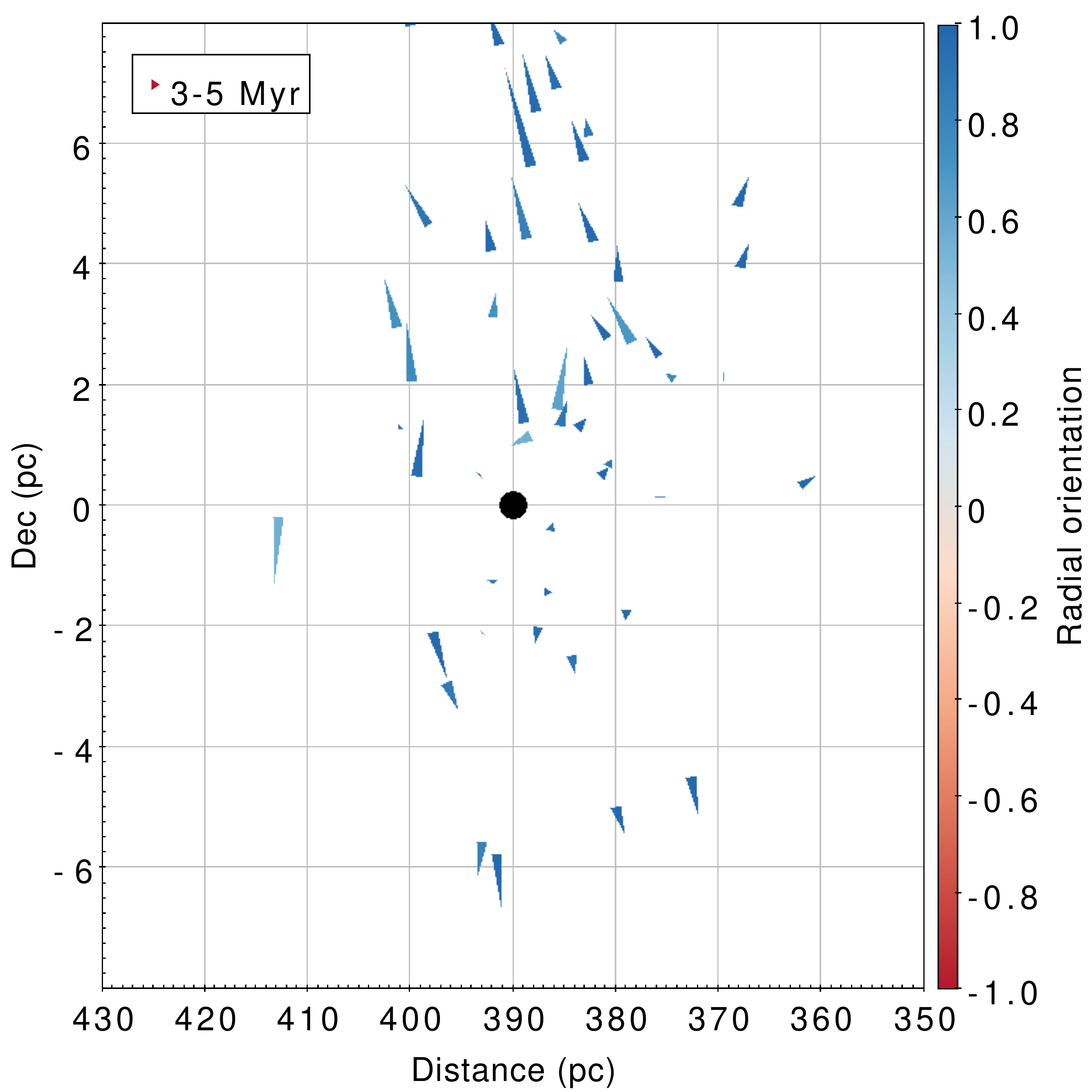}
\includegraphics[width={0.245\textwidth}]{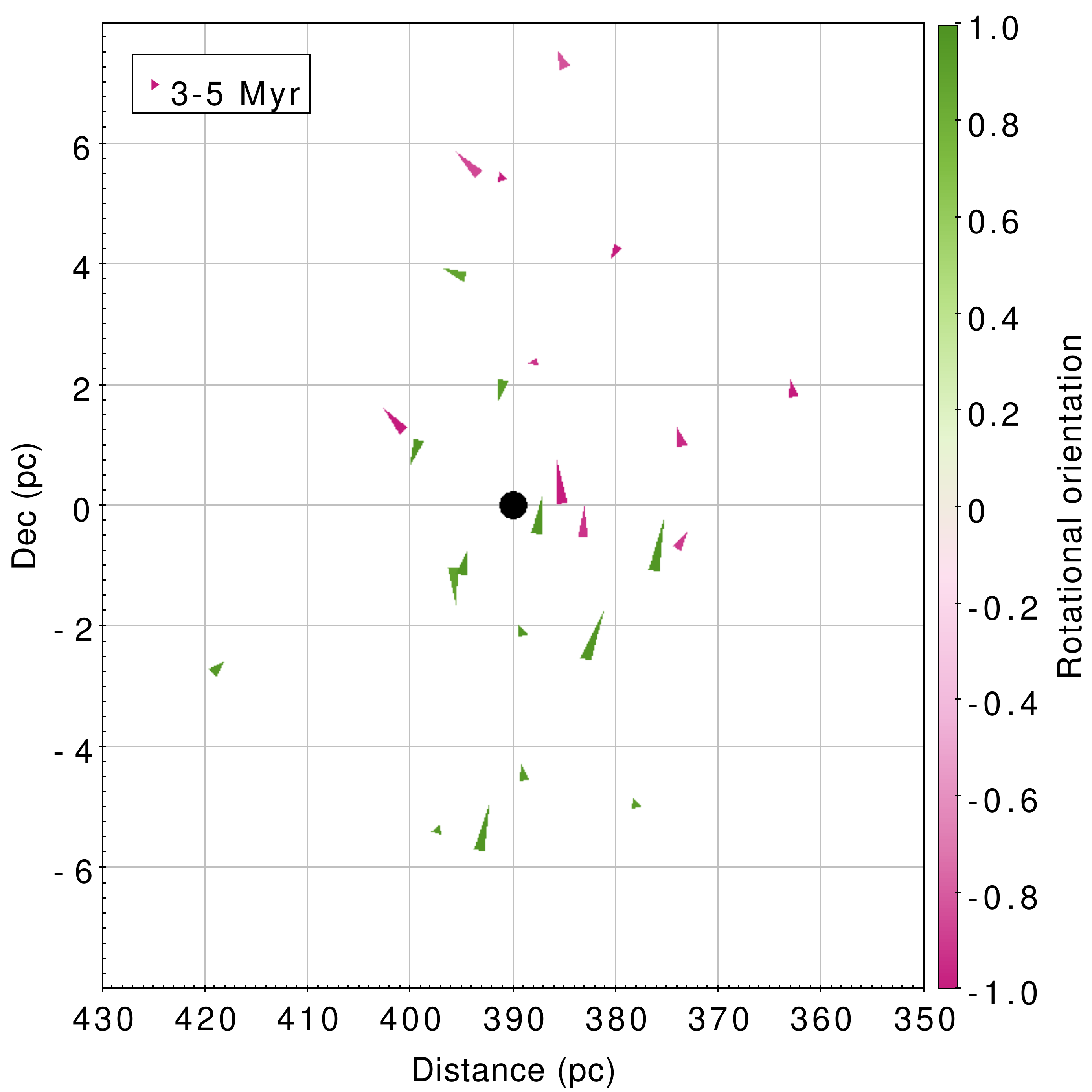}

\includegraphics[width={0.245\textwidth}]{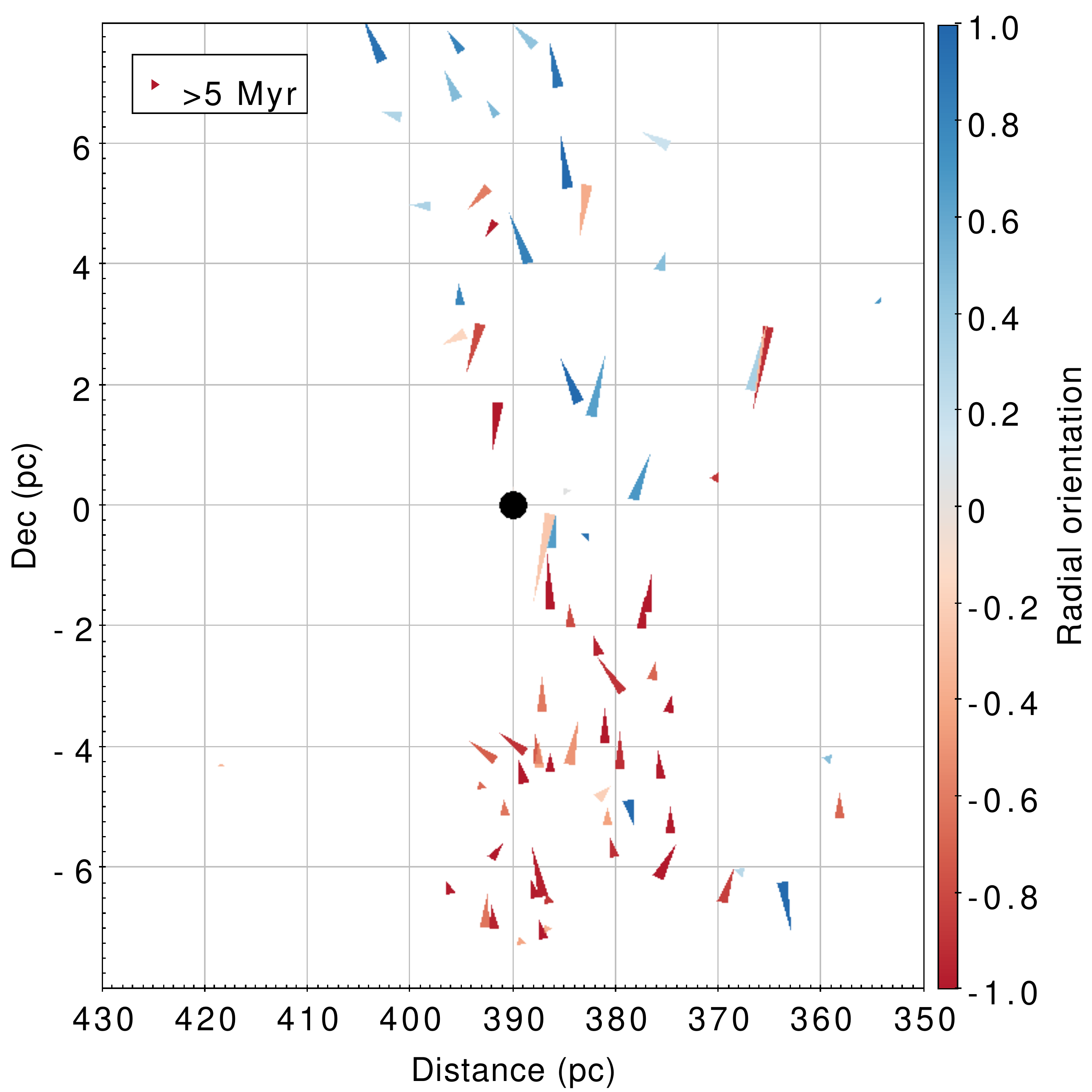}
\includegraphics[width={0.245\textwidth}]{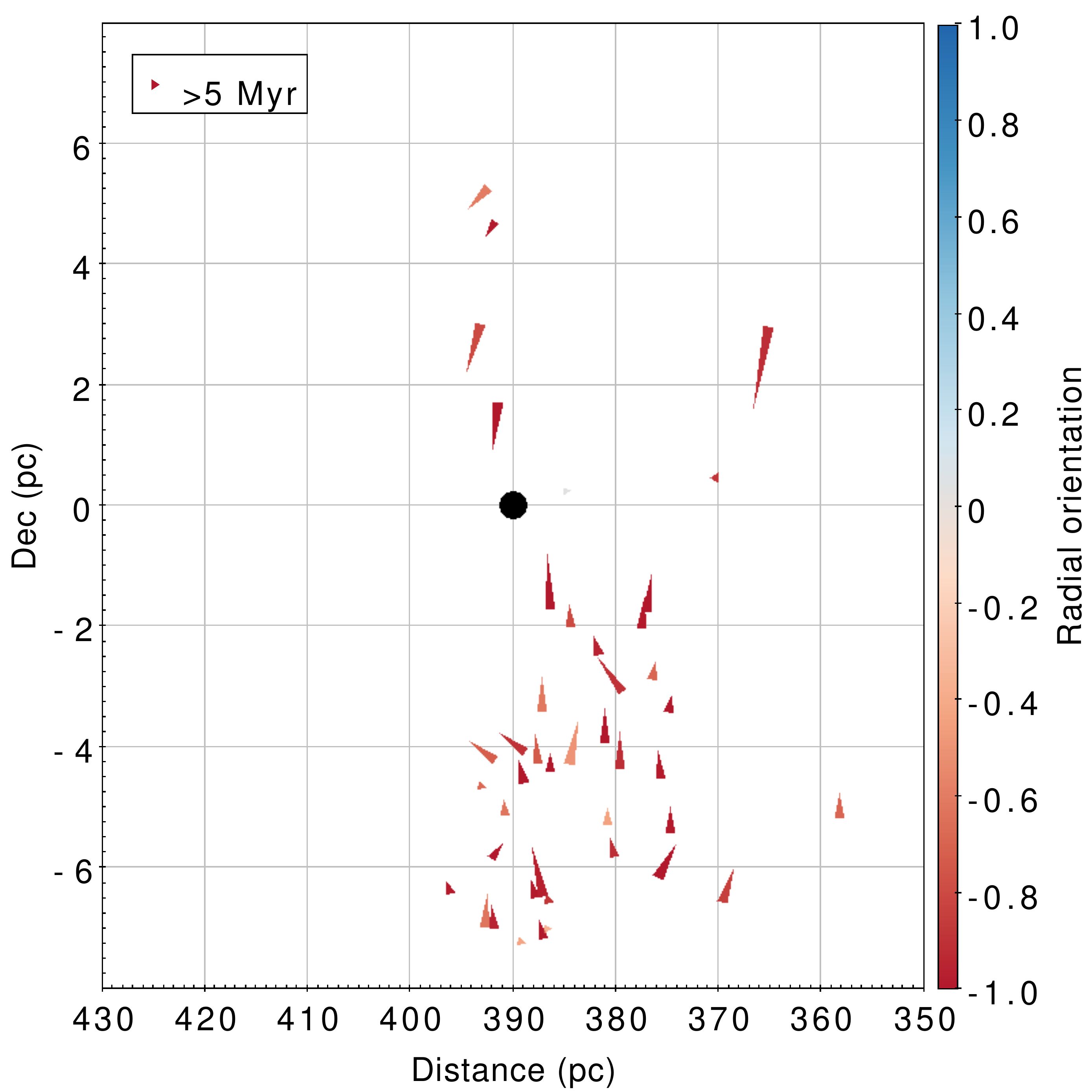}
\includegraphics[width={0.245\textwidth}]{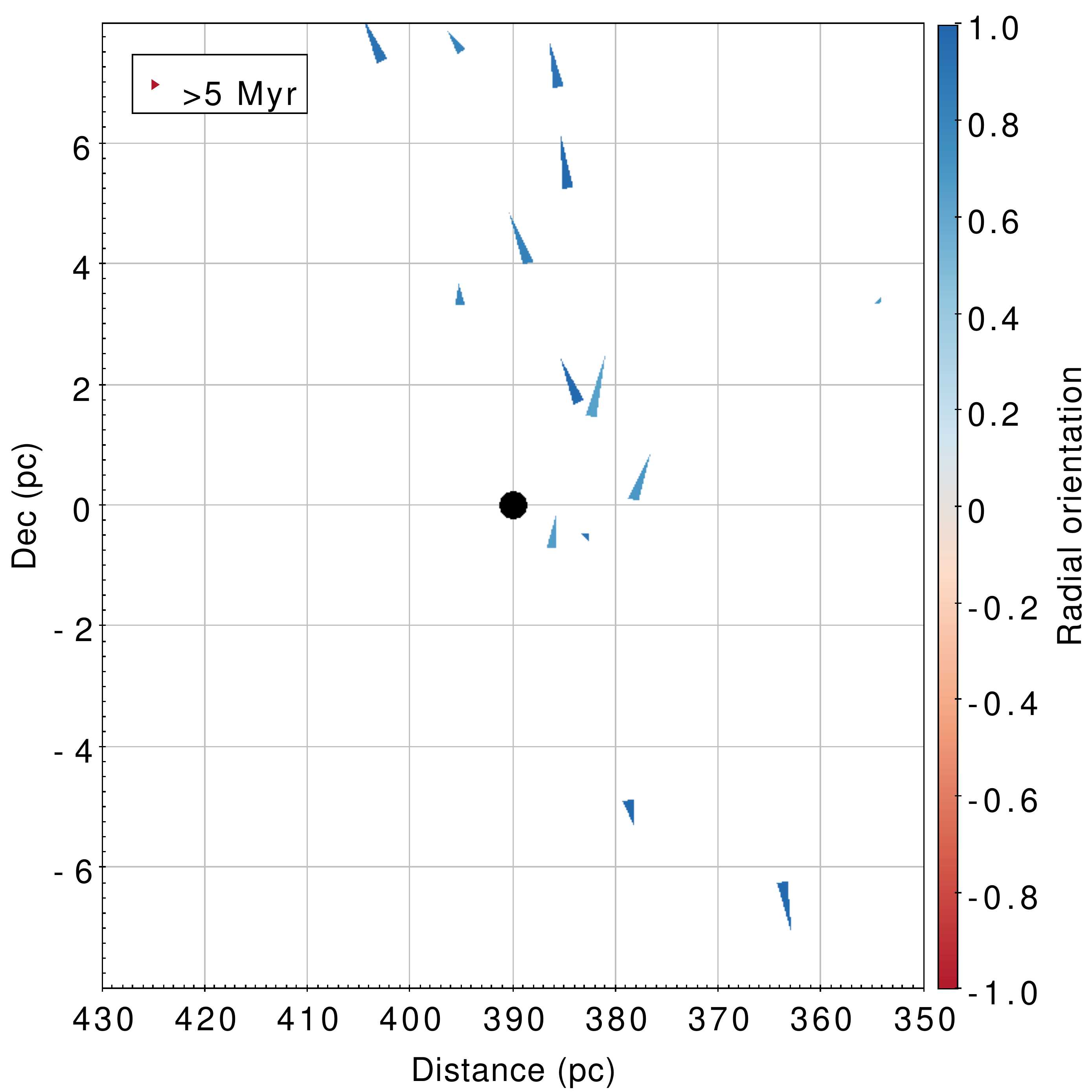}
\includegraphics[width={0.245\textwidth}]{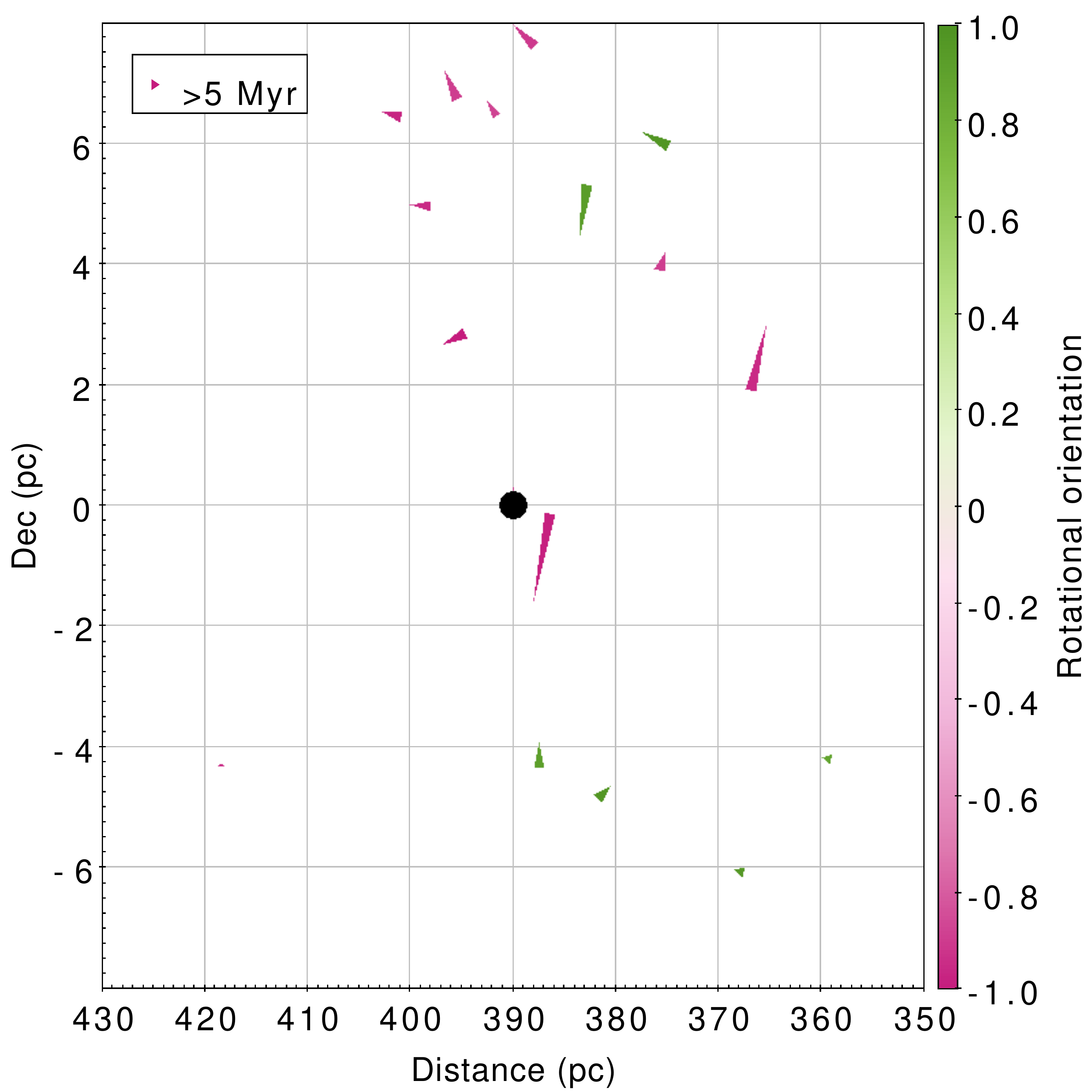}
\caption{Same as Figure \ref{fig:onc}, but showing distance from the Sun versus the position along the filament in $\delta$ converted to physical units, with vectors converted from the $\mu_\delta$ and radial velocities. The length of the vectors corresponds to the distance covered in 0.4 Myr. Typical uncertainties in distance are $\sim$6 pc.
\label{fig:rv}}
\end{figure*}

\begin{figure*}
\includegraphics[width=\columnwidth]{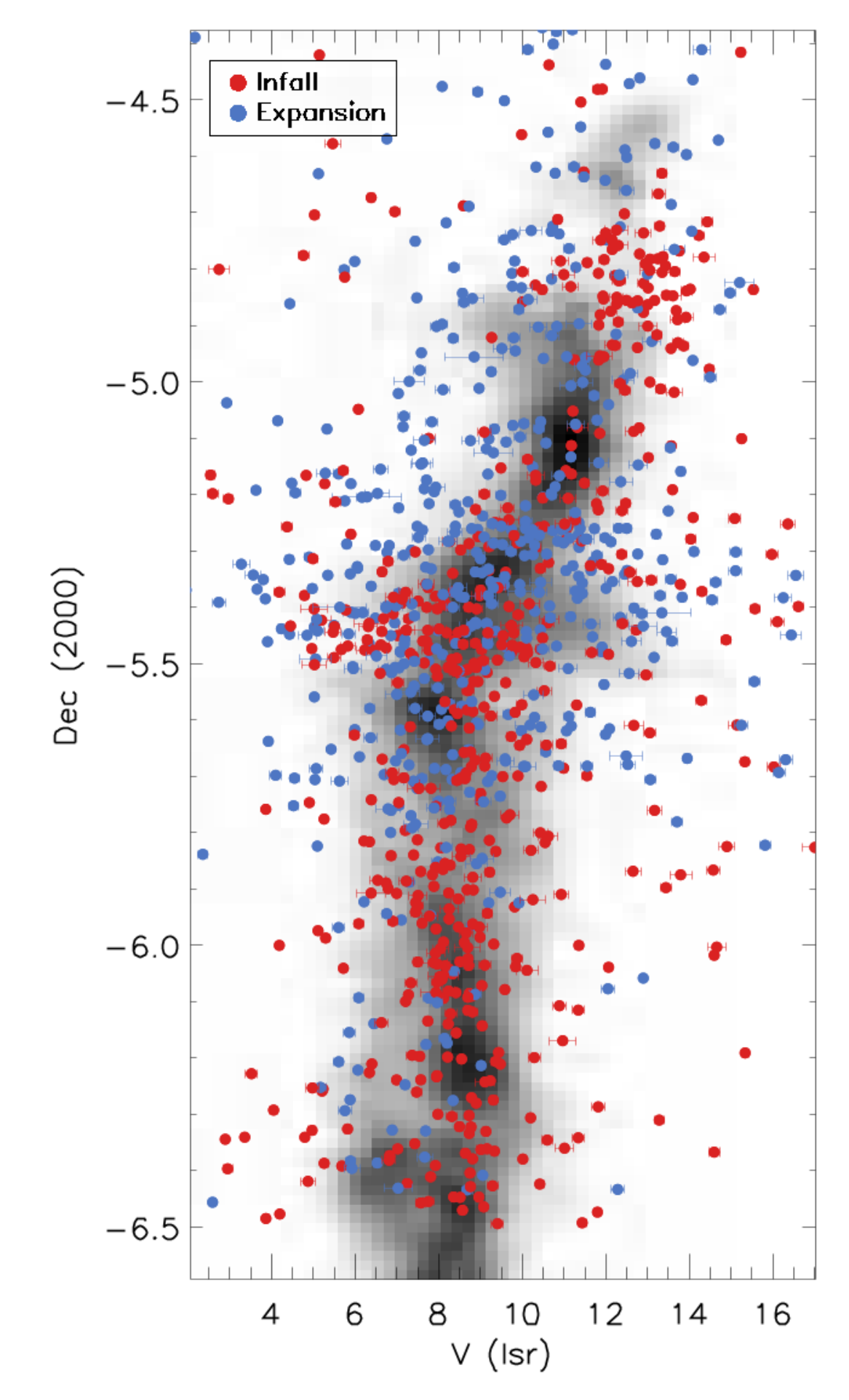}
\includegraphics[width=\columnwidth]{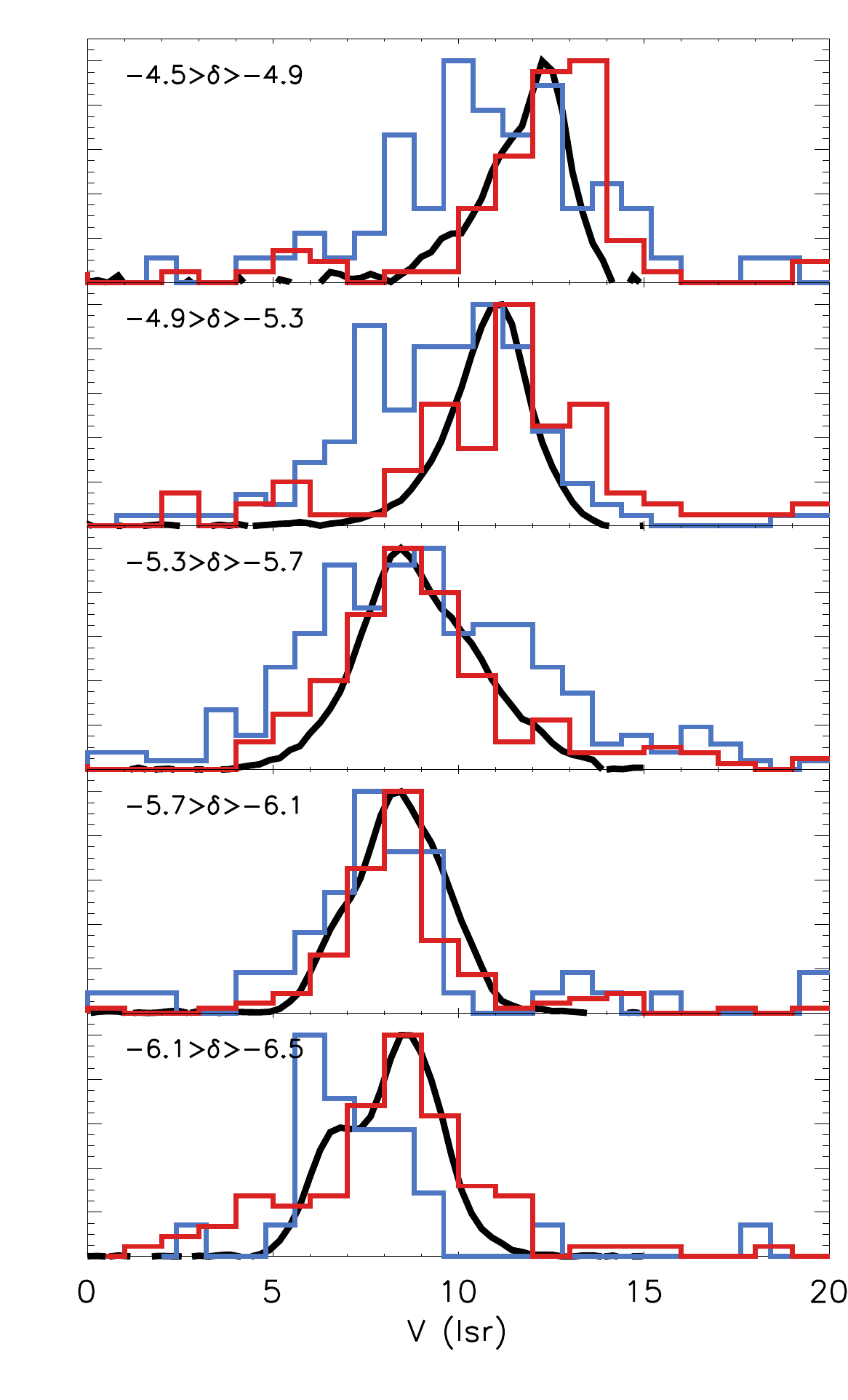}
\caption{Radial velocity of the $^{13}$CO molecular gas \citep{bally1987}, and that of the stars in the sample, in the local standard of rest reference frame. Left panel shows the full view of the cluster, right panel is binned in 5 discrete slices along $\delta$. The velocity distribution of the gas is shown in greyscale in the background of the left panel, and as a black line in each tier of the panel on the right. Infalling and outflowing stars are shown in red and blue respectively. On the right each distribution is scaled relative to its peak. Typical uncertainty in RV is 0.1 \kms.
\label{fig:rv1}}
\end{figure*}

\begin{table*}
	\centering
	\caption{Radial velocity statistics 
\label{tab:stats}}
\begin{threeparttable}
	\begin{tabular}{c|cccc|cccc|cccc} 
		\hline
		 \multicolumn{1}{c|}{} & \multicolumn{4}{c|}{Infall} & \multicolumn{4}{c|}{Expansion}  & \multicolumn{4}{c}{Rotation}\\
		 
		 \multicolumn{1}{c|}{} &$v_{med}$\tnote{a} &
		 $\sigma_{v}$\tnote{a} & $N_*$ &
		 \multicolumn{1}{c|}{\logg\tnote{b}} & $v_{med}$ &
		 $\sigma_{v}$ & $N_*$ & \multicolumn{1}{c|}{\logg} & $v_{r}$ & 
		 $\sigma_{v}$ & $N_*$ & \logg\\
		\hline
$-4.9<\delta<-4.5^\circ$ & 13.4 & 1.4& 89 & 3.97 & 11.7 & 3.0& 63 & 4.09 & 13.1 & 2.0& 43 & 4.01\\
$-5.3<\delta<-4.9^\circ$ & 12.3 & 1.8& 111 & 3.90 & 10.5 & 2.7& 155 & 3.94 & 11.4 & 2.6& 77 & 3.90\\
$-5.7<\delta<-5.3^\circ$ & 9.6 & 1.9& 190 & 3.93 & 10.0 & 3.4& 258 & 3.90 &10.0 & 2.2& 245 & 3.94\\
$-6.1<\delta<-5.7^\circ$ & 9.2 & 1.0& 139 & 4.12 & 8.6 & 1.6& 65 & 3.95 & 8.3 & 3.4& 40 & 4.02\\
$-6.5<\delta<-6.1^\circ$ & 9.6 & 0.9& 96 & 3.90 & 8.0 & 1.3& 31 & 3.93 & 7.6 & 2.1& 45 & 3.93\\
		\hline
	\end{tabular}
\begin{tablenotes}
   \item[a] Mean (lsr) radial velocity and radial velocity dispersion in \kms\, fitted as a Gaussian to the RV distribution in the slice, ignoring outlying wide wings to exclude likely spectroscopic binaries.
   \item[b] Mean \logg\ of the stars in the slice
  \end{tablenotes}
\end{threeparttable}
\end{table*}

Examining the distribution of the stars in the ONC in the plane of the sky and their proper motions allows for an incredible precision in inferring the dynamical state of the cluster, however, some leverage can be also gained via examining distance of the stars and the radial velocity. However, as even with Gaia EDR3, parallaxes can be very uncertain, resulting in a large spread in the inferred distance, we limit the sample only to the sources with $\sigma_\pi<$0.04 mas, which, at the distance of the ONC translates to $\sim$6 pc, which is still considerable, as it is comparable to the size of the cluster in the plane of the sky, but, nonetheless, allows to resolve some structure along the line of sight. However, imposing this constraint on $\sigma_\pi$ significantly limits the sample to only 533 stars, particularly towards the central cluster due to large degree of nebulosity in the region degrading the quality of the parallaxes.

We examine the distance versus the vertical extent of the cluster in $\delta$ converted to physical units in Figure \ref{fig:rv}, highlighting the velocities of the sources in these two respective dimensions. The assignment of infalling and outflowing sources is retained from the previous figures based on their plane of the sky velocities. On a first glance, the outflowing sources in this plane appear to be somewhat different than in Figure \ref{fig:onc}; we note that this is primarily due to the strict $\sigma_\pi$ cut preferentially excluding sources in the region where expansion is most strongly apparent.

When examining the full sample of infalling and outflowing sources only in the radial velocity space, where the cut on parallax quality is not necessary, we find that the stars that are infalling have a good agreement with RVs of the gas along all $\delta$ in the cluster. On the other hand, the outflowing stars not only often have a larger $\sigma_v$ (Table \ref{tab:stats}, Figure \ref{fig:rv1}), they also may be offset from the gas. This is most strongly apparent at $\delta\sim-5^\circ$ where the peak of the RV distribution of outflowing stars is blueshifted relative to the gas. The origin of this blueshifted stellar component has long since been questioned \citep{furesz2008,tobin2009,kounkel2016}. We can finally offer a partial explanation: radial velocities of these stars could have been dynamically processed by the central cluster. The mean velocity of the expanding stars is more comparable to the mean velocity of the central cluster, on the other hand the gas north of the cluster is intrinsically redshifted. As such, the total distribution of the stars formed from the molecular gas in that region coupled with the stars that most likely originate in the central cluster but since have been scattered to the same line of sight becomes asymmetric relative to the gas.

\section{Discussion}\label{sec:discussion}
\subsection{Distance to the ONC and receding star formation}

\begin{figure}
\includegraphics[width=\columnwidth]{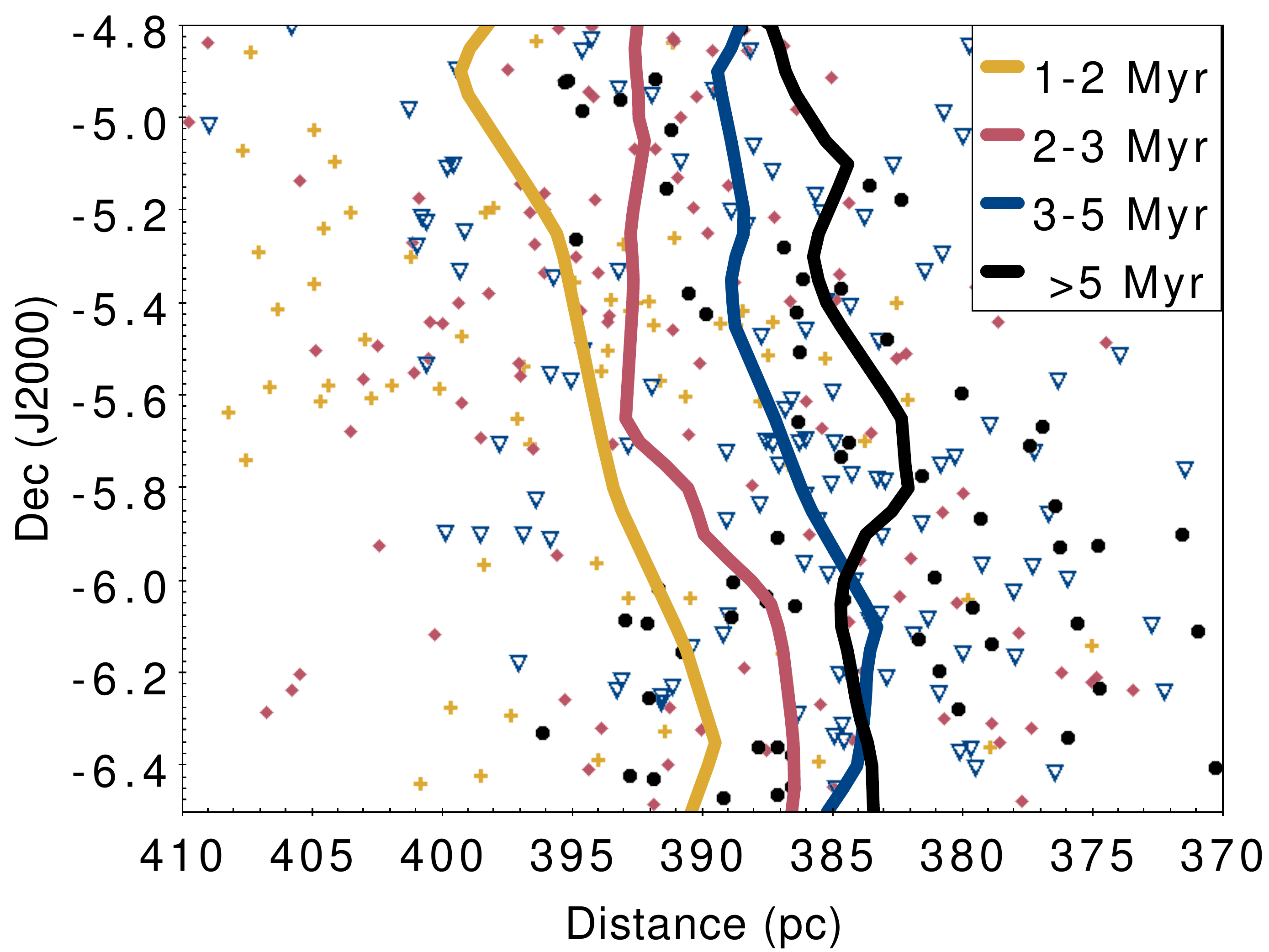}
\caption{Distance to the stars in the ONC separated in different age bins, with different lines showing the interpolated distance along the filament in each bin. The sources shown in the plot have parallax uncertainty $<$0.04 mas, typically $\sim$6 pc.
\label{fig:distance}}
\end{figure}

Previously there has been some contention regarding the orientation of the ONC's parent filament in the plane of the sky: \citet{kounkel2018a} and \citet{grosschedl2018} found it to be largely flat, with a near-constant distance across its length, whereas \citet{stutz2018a} and \citep{getman2019} found a significant variation in distance, with the central cluster being found at a larger distance than everything else in the ONC.

\begin{table}
	\centering
	\caption{Typical distance to the ONC as a function of age
\label{tab:dist}}
	\begin{tabular}{ccc} 
		\hline
		 Age & Parallax & Distance\\
		 (Myr) & (mas) & (pc) \\
		\hline
1--2 & 2.5386$\pm$0.0025 &  393.92$\pm$0.39  \\
2--3 & 2.5567$\pm$0.0019  & 391.12$\pm$0.29  \\
3--5 & 2.5937$\pm$0.0017  & 385.55$\pm$0.26  \\
$>$5 & 2.5995$\pm$0.0022  & 384.68$\pm$0.32  \\
		\hline
	\end{tabular}
\end{table}

Separating the sources into different age bins based on their \logg\ does show a peculiar trend in Figure \ref{fig:rv}: the distance to the ONC does not appear to be constant as a function of age. As previously, we restrict the sample only to the sources with parallax uncertainty $<0.04$ mas, to maximize the resolution in distances. The median parallax with the corresponding reduced uncertainty of all the stars in each age bin is given in Table \ref{tab:dist}. It shows that the younger stars are found at increasingly larger distances. The two-sample KS test rejects the null hypothesis that the distribution of parallaxes in the oldest and in the youngest age bins originate from the same distribution with the $P=2.4\times10^{-9}$, or $\sim6\sigma$.

Figure \ref{fig:distance} shows the interpolated distance of the stars at each age bin, and it accentuates the variance in distance further. The interpolation was performed by a neural net trained to predict distances of the stars as a function of their position in the cluster and their log g. The separation across different age bins appears to hold true across the entire filament, although given that younger stars are far more numerous in the Trapezium than elsewhere, this could result in the apparent asymmetry that \citet{getman2019} has observed. On the other hand, if the sample selection is less sensitive to the youngest stars, this could produce a more uniform distance distribution as in \citet{kounkel2018a} and \citet{grosschedl2018}.

We note that stars with age $<1$ Myr may appear to break this trend of receding star formation, their average distance appears to be centered on 390 pc, closer than stars in 1--2 Myr age bin (Figure \ref{fig:rv}). However, as they are most likely to be heavily extinguished, (both due to being still embedded within the envelope of gas and due to having a protoplanetary disk, many without any optical emission), imposing strict parallax quality cuts, with $\sigma_\pi$ depending on $G$ flux, may result in their census being particularly biased to sources sitting towards the front of the cloud.

The overall progression of distances from 5 Myr stars to 1 Myr stars itself is unlikely to be attributable to such a bias. Older stars tend to have somewhat lower luminosity (by $\sim$0.5 mag) due to having smaller radii, however the difference in the distance modulus between 385 and 395 pc is negligible (0.05 mag). Even considering the difference in extinction increasing with distance traveling through the cloud, it is difficult to explain older and younger stars not necessarily being co-located.

The stars (particularly those that are infalling, regardless of their age) tend to share a common radial velocity with the gas. However, if older stars are physically separated from the younger stars that are currently being formed, this implies that the bulk of the ONC is not necessarily co-located with molecular gas, with stars sitting in front of the reservoir of gas. Indeed, a 3-d model of ionized gas of the nebula assumes that even the central cluster is located in front of the cloud \citep{odell2001}

This, combined with gravitational infall \citep[which has also been previously observed by][]{getman2019}, can in part explain the peculiar RV structure in the ONC towards NGC 1977 ($\delta\sim-4.9^\circ$). The RVs towards it are not only more red-shifted than what is found in the south of the cluster, the stellar RVs (of the infalling stars) are also somewhat more redshifted than the gas. Various scenarios for its motions have been considered by \citet{getman2019}. However, this is likely a signature of gravitational infall, sources in NGC 1977 being attracted to the younger stars in the central cluster that are located further away. This is similar to the scenario considered by \citet{tobin2009} and \citet{proszkow2009}. 

Similarly, \citet{hacar2017a} have observed a blue-shifted ``wedge'' in the N$_2$H+ gas in the inner 1 pc within the central cluster that is less apparent in the more diffuse CO gas. The fresh dense molecular gas that is sitting behind the cluster can nonetheless feel gravitational attraction to the cluster, and it is also infalling as it is pulled towards it.

\subsection{Ejected stars and their effect on the observed boundness of the ONC}

\subsubsection{Framework}

The Orion Nebula is the singularly most massive cluster that is found within more than 1 kpc of the Sun. It has long since been considered that that the ONC is a cluster that may be similar to the progenitor of the Pleiades \citep{kroupa2001b,moraux2004}. 

The comprehensive membership catalog from \citet{mcbride2019}, indicates that the maximum projected central stellar density is $\sim2\times10^5$ stars pc$^{-2}$, well in excess of $\sim$50 stars pc$^{-2}$ in Pleiades \citep{gaia-collaboration2018a}. This does not change significantly even estimating the conversion from surface density to the spatial density. While the entire ONC does extend $\sim$10 pc along the line of sight, in the bulk of the sources towards the central cluster are young, and are concentrated at a much smaller spread of distances \citep[hence the shape of the ONC in the works of ][]{stutz2018a,getman2019}. The Pleiades also contains $\sim$3 times fewer stars than the ONC in a similar volume. The overall density of the ONC increases further if one were to consider a sizable mass of the molecular gas that is still forming stars. Over the course of its evolution, Pleiades has lost a substantial fraction of its stars \citep{dinnbier2020a}, in part due to dynamical heating by binaries \citep{converse2010}.

Given its mass, if ONC isn't bound, no other cluster associated with a star forming regions is expected to be - not just in the Orion Complex, but across the solar neighborhood as a whole, within at least 1 kpc, as no other young cluster can match it in terms of its mass or its density within that volume. Since approximately 16\% of stars appear to form in bound clusters \citep{anders2021}, this seems statistically unlikely.

However, while boundness of the ONC seems to be apparent conceptually, not just in comparison to other clusters, but also based on its morphology, and the effect its presense has on the surrounding stars (e.g., it appears to have substantially large self gravity to attract other parts of the filament), there has been a long-standing difficulty in definitively proving it. Virial parameter \citep{bertoldi1992} is often used to test boundness of a cluster, but while the ONC is a young cluster that is closest to being considered in a virial equillibrium \citep{kuhn2019}, depending on the precise assumptions used to infer the mass distribution, and given the uncertainty in the total mass and radius of the cluster, it may fall just short of being considered unequivocally bound \citep{da-rio2014}.

We propose an method that could help to mitigate this long-standing issue. The $\sigma_v$ of the ONC may be inflated due to ejections of stars in unstable N-body interactions, i.e., incidents where three or more stars are in close proximity exchanging energy and kicking one of the stars outwards. This can occur in primordial unstable triple systems, or a close approach of a binary to another star in a dense cluster. Such ejection events can explain the apparent excess of stars moving on outward trajectories as well as a significantly wider $\sigma_v$ of the outflowing vs infalling or rotating stars. We note that not every star that is moving on the outward trajectory would necessarily have to have been ejected; the excess can be explained by as few $\sim$5\% of stars in the total sample within 0.4$^\circ$ radius of the central clusters being runaways.

It may be difficult to fully disentangle which specific stars have been ejected, and which stars have proper motions that only coincidentally appear to point away from the cluster center. However, stars that are outside of the innermost cluster core, stars that are found in the parameter space not balanced by an equal number of infalling stars, stars that project back directly to the center without any angular offset are particularly likely candidates of bona fide orphans from a past ejection.

\subsubsection{Likelihood of ejection events}

To test the likelihood of the ONC having a suitable number of unstable interactions between its members, we model a simple cluster in GalPy \citep{galpy} with AMUSE framework \citep{amuse}. While this model is not complex enough to accurately reproduce such unstable interactions, in part due to a lack of a prescription for binary stars, it allows to examine how often close encounters are expected to occur, and compare their proximity to the typical orbital separations.

The model cluster consists of 2000 stars within a radius of 2 pc, overall potential of 4500 \msun, and velocity dispersion of $\sim$2.2 \kms. The cluster evolved for 5 Myr, recording position of the stars every 100 years. In this simulation, there are 2700 unique encounters closer than 1000 au (i.e., on average each star has experienced 1.35 such encounters). Of these, 517 encounters (26\%) are closer than 200 au, and even 32 encounters (1.6\%) closer than 30 au.

Approximately 50\% of the field stars are found in multiple systems. In the field, $\sim$50\% of systems have companions with separations $>$30 au, 33\% have separations $>$200 au, and $\sim$20\% of systems have separations $>$1000 au \citep{raghavan2010}. That is to say, $\sim$25\%, 16\%, and 10\% of all stars are expected to have companions with such separations respectively. Considering the the above encounter rate, there is 10\% chance that there will be an encounter $<$1000 au in which at least one of a star has a companion $>$1000 au, 3\% chance for an encounter $<$200 au with $>$200 au companion, and 0.3\% chance for an encounter of an encounter $<30$ au with $>$30 au companion. Integrating across a full range of separations and considering the sizable number of stars within the ONC, this amounts to hundreds of binary stars that could have been involved in a dynamical altercation with another star in a dense cluster.

Given the frequency of encounters with other stars approaching closer than $<$1000 au, from this simulation it is expected that most of binaries with separations $>1000$ au to be largely unstable. This is supported by observations: throughout the ONC the wide binary fraction of sources with companions with separations $>$1000 au is only $\sim$5\% \citep{jerabkova2019}, meaning that the orbits of at least half of wide binaries would have been processed in some degree (either hardened or became unbound). This fraction would grow higher if systems with separations $<$1000 au are also considered. Similarly, the estimates above include only a chance encounter between a binary and a single star; primordial triple systems would increase likelihood of unstable interactions to a degree that is difficult to quantify, but it should be noted that Class 0 YSOs have a much higher multiplicity fraction than the field stars, and that 50\% of multiples among them found in a high order (3+) system in comparison to only $\sim$20\% among the field stars \citep{chen2013}.

In some configurations, the energy exchange in the encounter would lead to unbinding of the binary. Often, one of the stars would form a binary system with an interloper/wider triple star while ejecting the original/closer companion. In other configurations, the binary would grow tighter, giving energy to an interloper/wider triple. As such, even stars for which a wide companion is detected, this does not mean that they weren't participating in a dynamical interaction.

\subsubsection{Properties of runaways}

In dynamical simulations of triple systems, the mean ejection velocity of the unstable N-body interactions is 2.8 \kms\ which later decays to 1.1 \kms\ through intracluster interactions \citep{reipurth2010}. This is similar to the typical velocity of the stars that are expanding away that we observe. Comparatively, true high velocity walkaway and runaway stars are rare, though, there are several dozen that are currently known to be associated with the ONC \citep{mcbride2019,schoettler2020,farias2020}. As such, hundreds of stars with lower ejection velocity are expected. As we have deliberately cut the amplitude of the proper motions to exclude high velocity stars that could be considered as runaways, the bulk of these expanding stars may still remain gravitationally bound to the cluster, and they would be forced to turn around as they climb out the potential well. The typical free-fall time in the cluster is $\sim$0.5--1 Myr, comparable to the time it would take for a star ejected with a speed of 2 \kms\ to fully decelerate. Stars with ejection speed of $>$5 \kms\ are likely to be unbound. The primary reason why so many could be detected can be attributed to the youth of these stars and the recency of their ejection.

Almost all of the expanding stars ($\sim$94\%) have traceback ages that are smaller than the age assigned to a star based on their \logg. This is to be expected, as it would be impossible to eject them from the central cluster otherwise. Interestingly, the candidate ejected stars in the younger age bins appear to have a larger high velocity ($>$4 \kms) tail in comparison to their older counterparts (Figure \ref{fig:onc}, third column). As the catalog on which this analysis is performed consists of sources that have been targeted for spectroscopic observations, older stars moving with comparable speeds are likely to be out of bounds of the targeted area, assuming they were ejected shortly after their formation.

\subsubsection{Implications}

The virial parameter is a somewhat imperfect metric of a cluster boundness, as it is difficult to measure directly, without any assumptions pertaining to e.g., cluster morphology and mass distribution within it. Moreover, it can be overestimated due to the binary stars. For example, $\sigma_v$ can be inflated by the presence of spectroscopic binaries \citep{kouwenhoven2008,Gieles2010}. In the more evolved clusters boundness of which is not in question, mass of a cluster estimated from the virial parameter may be inflated by as much as 8--42\% \citep{rastello2020}.

Here we also introduce another consideration which can inflate the velocity dispersion, and thus affect the virial parameter, namely unstable interactions with multiple systems exchanging orbital energy, resulting in ejection of the individual stars. These individual stars may or may not become unbound depending on their velocity kick they receive relative to the escape velocity of the cluster as a whole. Such sources are able to contribute to the wider wings of $\sigma_v$ (Figure \ref{fig:rv1}), but, they would not affect boundness of the stars that did not experience such interactions, since, at the moment, the energy injection is applied to the individual stars, not the entire cluster as a whole.

Eventually the compounding effect of such ejections will propagate to the other stars. If an ejected star is unbound, it would lead to an incremental mass loss for a cluster. If a star would remain bound, it would eventually be able to equillibrate its velocity with other stars, incrementally heating up the cluster. The final $\sigma_v$ of the cluster will be higher than the initial $\sigma_v$ that were not involved in the aforementioned dynamical interactions, but as all of the excess energy would be shared equally, it would be lower than $\sigma_v$ in the sample in which the ejected stars are dominating the wings of the the velocity dispersion.

But, after some time, a cluster would be able to ``settle'': all the wide binaries that could be disrupted from their primordial separations would already do so, the hard binaries would increasingly become harder. Thus, eventually a close encounter with another star would be less likely to result in an ejection event in more evolved clusters compared to those that are still near their infancy.

Such a state of equilibrium has not yet been achieved in ONC, as can be inferred by a presence of a large number of runaways with an ejection age $<1$ Myr ago \citep{mcbride2019}, including a rather spectacular event mere $\sim$540 years ago that has led to the formation of the BN/KL nebula and singlehandedly ejected at least 4 stars \citep{gomez2008,luhman2017,rodriguez2020}. Arguably, such a state of equillibrium cannot be achieved in a young cluster that is still actively forming stars, and thus is still actively producing binary stars at the full range of their orbital separations.

Orbital evolution of binary systems have often been considered to play a vital role in characterizing the dynamics in young populations as a whole, and young dense clusters such as the ONC in particular \citep{kroupa2001,clarke2015}. However, there is still much regarding young multiples that is not well understood - in part because the observational data towards them is still sparse, in part because multiple systems add a significant degree of complexity that is often disregarded in theoretical framework \citep[e.g., even in simulations that specifically aim to characterize ejected runaways from young clusters][]{schoettler2019}. This also holds true for the studies that have so far applied the virial parameter to test boundness of young clusters, including the ONC \citep[e.g.,][]{da-rio2014,kuhn2014}. More detailed numerical simulations of ONC-like clusters are needed to constrain the role of multiples in general and ejected stars specifically to test the manner in which they challenge the typical approximations of virial equillibrium.

\subsection{Combined model of the star forming history of the ONC}

\begin{figure}
\includegraphics[width=\columnwidth]{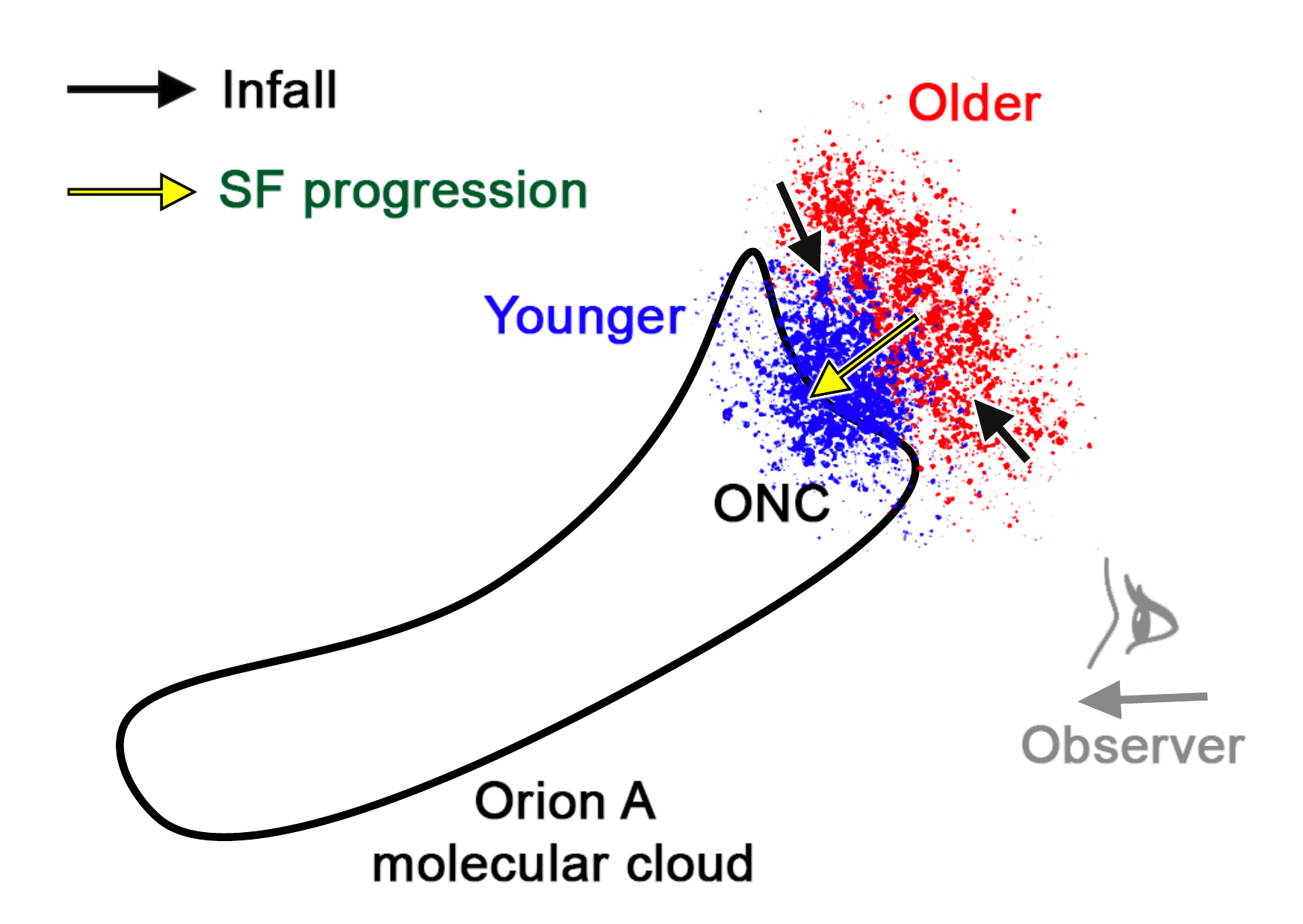}
\caption{A conceptual model showing a side view of the Orion A molecular cloud (corresponding to distance vs $\delta$ projection), and the relation of ONC relative to it. The cluster shows the age gradient with distance, with the younger stars located closer to the cloud; direction of star formation propagation is indicated by a yellow arrow. Meanwhile, northern and southern part of the cluster are contracting towards the central cluster, indicated by black arrows.
\label{fig:model}}
\end{figure}

It has been previously noted by \citet{kounkel2020} and \citet{grossschedl2021} that 6 Myr ago a supernova (or several supernovae) have triggered the global expansion of the Orion Complex. Given that the ONC is being pushed into the Orion A molecular cloud \citep{grosschedl2018} from the direction of the center of the expansion (several pc away from the current position of the ONC), it is likely that the shockwave from a supernova has swept along the gas through the filamentary cloud, compressing the gas, and jump-starting the formation of the cluster in a ``conveyor belt'' like fashion \citep{krumholz2020}.

This shock-driven star formation scenario can account for the variable distance to the ONC as a function of age. If the shockwave is propagating into the cloud at a slightly faster rate (faster by $\sim$2 \kms) than the typical radial velocity of the gas and the stars, this would displace the star forming front.  Such propagation velocity is comparable to the sound speed in the ONC \citep{goicoechea2015}.

Early on, star formation has occurred all throughout the length of the filament at an equal rate, but, eventually, as the cluster grew more and more massive, self-gravity became more important. The molecular gas was pulled increasingly more towards the middle, forming the central cluster, even though the bulk of the stars in the central cluster were formed further away than the initial burst of star formation in the region. As the star forming front of the ONC was being pushed into the Orion A molecular cloud (Figure \ref{fig:model}) and continuing to access fresh gas, this provided sufficient fuel to form multiple generations of stars in the central cluster, increasing its mass and gravitational pull. On the other hand, NGC 1980 to the south and NGC 1977 to the north of it could not sustain star formation beyond the initial burst. To various degree, the molecular gas that was originally there was a) consumed in forming stars, b) attracted towards the central cluster, and c) dissipated through stellar feedback. They could not replenish their gas content with Trapezium hoarding all of the new material.

\citet{getman2019} have noted that the more distant stars in the ONC are receding from the observer at a slower rate than the closer parts of the cluster. This does appear to be the case of the entire population, particularly due to NGC 1977. However, this distance-velocity relation is not immediately apparent in the RV distributions of stars segregated into different age bins along a given line of sight. Nonetheless, this may be a signature of either the shockwave slowing down as it is encountering more and more gas of the filament, or the self-gravity of the ONC is protesting against the sheering of the spatially differential star formation and is attempting to bring the older nearby stars and younger distant stars closer together.

\subsection{Order of magnitude mass estimate of the ONC}

The sources that are infalling may offer a possible way estimate the dynamical mass of both the ONC and the Orion Complex, however, as has been previously stated, such a calculation is difficult as we do not fully know the initial velocity field that was present in the region beforehand. Similarly, as acceleration changes velocity incrementally over time, it is important to consider the time scales over which the force of gravity affects the surroundings, which is non-trivial to estimate, as the conditions in a star-forming cloud can and do drastically change over time. 

However, it is possible to make a rough order of magnitude inference assuming simple conditions to compare the derived mass with what is typically assumed for the population in question.

If we treat ONC as an isolated system and if we assume typical infall velocities of 2 \kms\ at 5 pc, which is a rough order of magnitude of the observed velocities, starting from rest, over $\sim$3 Myr, it would require a cluster mass of 2,500 \msun to achieve such an acceleration. This is well-comparable to the estimate of mass of the ONC from \citet{hillenbrand1998} of 4,500 \msun.

\section{Conclusions}\label{sec:concl}

We analyze the dynamics of young stars within the Orion Nebula and the solar neighborhood in the vicinity of the Orion Complex as a whole. We find that

\begin{itemize}
\item Examining the orientation of the proper motions of Orion members, we detect a significant ($>4 \sigma$) excess of sources whose proper motions are consistent with infall toward the ONC. These infall signatures are most prominent to the north and south of the ONC, within the integral filament where stars are least likely to have been dynamically processed by a passage through the ONC.  We interpret this signature as evidence that the stars in and around the ONC are contracting lengthwise along the integral filament, likely due to the self-gravity of the Trapezium.
\item We also detect a more modest ($\sim3 \sigma$) excess of sources with proper motions consistent with expansion from the ONC; the significance of this excess is largest for sources offset in RA, rather than Dec, from the ONC, placing them outside the integral filament, and thus more likely to be on orbits that have been dynamically processed via interactions in the ONC. We interpret this signature as evidence of dynamical processing of the ONC's stellar population, not due to the cluster itself being unbound. However, dynamical ejections could inflate the estimate of the virial parameter of a young population, which is a common metric that tests for boundness. 
\item Among the youngest stars in the ONC that have purely tangential motion, there is a preferred direction of rotation around the central cluster, there is no such preference among the older stars. As such, the angular momentum of the cluster has either evolved to develop organized rotation after the cluster has accreted enough mass, or early signatures of organized rotation in older stars has since been washed out through the dynamical evolution.
\item The distance to the ONC depends on the age of the sample; it varies from $\sim$385 pc for the oldest stars, to $\sim$395 pc for the younger stars; the star formation is continuously propagating into the Orion A molecular cloud at a faster rate than the typical velocity of the gas and the stars, consuming the outer layers of the gas in the process.
\item These observations reveal that young star forming clusters are highly dynamic entities, and the structure of these clusters as well as velocities of stars within them are affected by various processes occurring across different scales. The ONC is not homogeneous. Its dynamical evolution is influenced in different measure by the gravitational potential of its surroundings, gravitational interactions between the stars, as well as the activity within the rest of the Complex. All these effects leave different kinematical signatures that are difficult to separate when examining the entire cluster as a whole, but they become more apparent through identifying appropriate subsets of stars.
\end{itemize}

\section*{Data Availability}

This work is based on publicly available data. APOGEE Net reduction of APOGEE spectra are available at \url{http://vizier.u-strasbg.fr/viz-bin/VizieR?-source=J/AJ/163/152}; cross-matched with the catalog of members of the ONC \url{http://vizier.u-strasbg.fr/viz-bin/VizieR?-source=J/ApJ/884/6} and with Gaia DR3 photometry and astrometry. No additional data products are computed in this work.

\bibliographystyle{mnras}

\bsp	
\label{lastpage}
\end{document}